\newif\ifsubmission
\submissiontrue

\newif\ifllncs
\llncsfalse
\newif\ifblind
\blindfalse
\ifllncs
\documentclass[runningheads,envcountsame]{llncs}

\usepackage{amssymb}
\usepackage{amsmath}
\usepackage{amsthm}
\else
\documentclass[11pt]{article}
\usepackage{amssymb}
\usepackage{amsmath}
\usepackage{amsthm}
\usepackage{fullpage}

\usepackage{microtype}      % microtypography
\usepackage{libertine}
\usepackage{libertinust1math}

\usepackage{geometry}
\newtheorem{theorem}{Theorem}[section]
\newtheorem{lemma}[theorem]{Lemma}

\newtheorem{corollary}[theorem]{Corollary}

\theoremstyle{definition}
\newtheorem{definition}{Definition} 

\newtheorem*{remark*}{\textbf{Remark}}
\newtheorem{remark}[theorem]{Remark}
\fi

\usepackage[normalem]{ulem}
\usepackage{braket}
\usepackage{url}
\usepackage{algorithm}
\usepackage{algpseudocode}
\usepackage{mathtools}
\usepackage[title]{appendix}
\usepackage[all]{xy}
\usepackage{physics}
\usepackage{cancel}
\usepackage{mdframed}
\usepackage[most]{tcolorbox}
\usepackage{enumitem}
\usepackage{comment}
\usepackage{array}
\usepackage{stackengine,scalerel}
\usepackage[dvipsnames]{xcolor}
\usepackage{booktabs}
\usepackage{tabularx}

\newcommand{\protlabel}[1]{\textbf{#1}}

\definecolor{DarkBlue}{RGB}{0,0,150}
\definecolor{NotSoDarkBlue}{RGB}{15,15,210}
\definecolor{DarkRed}{RGB}{150,0,0}
\definecolor{DarkGreen}{RGB}{0,100,0}
\usepackage[pdfstartview=FitH,pdfpagemode=UseNone,colorlinks,linkcolor=DarkRed,filecolor=blue,citecolor=DarkRed,urlcolor=DarkRed,pagebackref,breaklinks]{hyperref}
\usepackage{cleveref}

\usepackage{multirow} 

\ifsubmission
\newcommand{\authnote}[3]{}
\else
\newcommand{\authnote}[3]{\textcolor{#3}{[{\footnotesize {\bf #1:} { {#2}}}]}}
\fi

\newtheorem*{conjecture*}{\textbf{Conjecture}}

\newcommand{\iter}{\mathsf{iter}}
\newcommand{\negl}{\mathsf{negl}}

\newcommand{\Gen}{\mathsf{Gen}}
\newcommand{\Setup}{\mathsf{Setup}}
\newcommand{\Sign}{\mathsf{Sign}}

\newcommand{\Ver}{\mathsf{Ver}}

\newcommand{\Eval}{\mathsf{Eval}}
\newcommand{\pk}{\mathsf{pk}}
\newcommand{\sk}{\mathsf{sk}}
\newcommand{\eps}{\varepsilon}
\newcommand{\poly}{\mathsf{poly}}

\newcommand{\secp}{\lambda}
\newcommand{\QPT}{\mathsf{QPT}}

\newcommand{\PPT}{\mathsf{PPT}}

\newcommand{\CRS}{\mathsf{CRS}}

\newcommand{\IO}{\mathsf{iO}}
\newcommand{\Hash}{\mathsf{Hash}}
\newcommand{\Punc}{\mathsf{Punc}}
\newcommand{\Perm}{\mathsf{Permute}}
\newcommand{\kin}{k_\mathsf{in}}
\newcommand{\kout}{k_\mathsf{out}}
\newcommand{\klin}{k_\mathsf{lin}}
\newcommand{\LF}{\mathsf{LF}}
\newcommand{\LFKeyGen}{\mathsf{LF}.\mathsf{KeyGen}}
\newcommand{\LFF}{\mathsf{LF}.\mathsf{F}}
\newcommand{\OWF}{\mathsf{OWF}}

\newcommand{\calA}{\mathcal A}
\newcommand{\calB}{\mathcal B}

\newcommand{\calD}{\mathcal D}

\newcommand{\calH}{\mathcal H}

\newcommand{\calM}{\mathcal M}

\newcommand{\calO}{\mathcal O}
\newcommand{\calP}{\mathcal P}

\newcommand{\calS}{\mathcal S}

\newcommand{\calX}{\mathcal X}

\newcommand{\sfF}{\mathsf F}

\newcommand{\sfO}{\mathsf O}
\newcommand{\sfP}{\mathsf P}

\newcommand{\sfW}{\mathsf W}

\newcommand{\bbZ}{\mathbb Z}

\newcommand{\bbE}{\mathbb E}

\newcommand{\bbN}{\mathbb N}

\newcommand{\Z}{\mathbb{Z}}

\def\matA{\mathbf{A}}
\def\matB{\mathbf{B}}
\def\matC{\mathbf{C}}
\def\matD{\mathbf{D}}
\def\matE{\mathbf{E}}
\def\matF{\mathbf{F}}

\def\matI{\mathbf{I}}

\def\matK{\mathbf{K}}
\def\matL{\mathbf{L}}
\def\matM{\mathbf{M}}

\def\matZero{\mathbf{0}}
\def\matOne{\mathbf{1}}

\def\vecb{\mathbf{b}}

\def\vecd{\mathbf{d}}
\def\vece{\mathbf{e}}

\def\vecm{\mathbf{m}}

\def\vecs{\mathbf{s}}

\def\vecu{\mathbf{u}}
\def\vecv{\mathbf{v}}
\def\vecw{\mathbf{w}}
\def\vecx{\mathbf{x}}
\def\vecy{\mathbf{y}}
\def\vecz{\mathbf{z}}
\def\idx{\mathsf{idx}}
\DeclareMathOperator{\spn}{span}
\DeclareMathOperator{\ColSpan}{ColSpan}

\newcommand{\probcond}[2]{\Pr \left[
\begin{array}{l}#1\end{array} \; : \;
\begin{array}{l}#2\end{array} \right]
}

\makeatletter
\newcommand{\customlabel}[2]{%
   \protected@write \@auxout {}{\string \newlabel {#1}{{#2}{\thepage}{#2}{#1}{}} }%
   \hypertarget{#1}{#2}
}
\makeatother

\newcommand{\construction}[3]{
    \stepcounter{figure}
    \vspace{0.05cm}
    { \small
    \begin{tcolorbox}[breakable=true, enhanced]
    \begin{center}
    {\bf \underline{Construction~\customlabel{construction:#2}{\thefigure}: #1}}
    \end{center}
    
    #3
    \end{tcolorbox}
    }
}

\title{A Simple and Efficient \\ One-Shot Signature Scheme}
\ifblind
  \author{Anonymous Submission to Eurocrypt 2026}
  \date{}
  \institute{}
\else
  \ifllncs
    \author{Andrew Huang \and Vinod Vaikuntanathan}
    \institute{MIT, Cambridge, MA, USA\\ 
    \email{\{ahuang,vinodv\}@mit.edu}}
    \date{}
    \index{Huang, Andrew}
    \index{Vaikuntanathan, Vinod}

    \titlerunning{A Simple and Efficient One-Shot Signature Scheme}
    \authorrunning{A. Huang and V. Vaikuntanathan}

  \else
    \author{Andrew Huang\thanks{\texttt{E-mail: ahuang@mit.edu}}\\MIT \and Vinod Vaikuntanathan\thanks{\texttt{E-mail: vinodv@mit.edu}. Research supported in part by NSF CNS-2154149, a Simons Investigator award and a Ford Foundation Chair at MIT.}\\MIT
   }
  \fi
\fi 

\begin{document}
\maketitle
\begin{abstract}
One-shot signatures (OSS) are a powerful and uniquely quantum cryptographic primitive which allows anyone, given common reference string, to come up with a public verification key $\mathsf{pk}$ and a secret signing state $\ket{\mathsf{sk}}$. With the secret signing state, one can produce the signature of {\em any one} message, but no more. In a recent breakthrough work, Shmueli and Zhandry (CRYPTO 2025) constructed one-shot signatures, either unconditionally in a classical oracle model or assuming post-quantum indistinguishability obfuscation and the hardness of Learning with Errors (LWE) in the plain model.

In this work, we address the inefficiency of the Shmueli-Zhandry construction which signs messages bit-by-bit, resulting in signing keys of $\Theta(\lambda^4)$ qubits and signatures of size $\Theta(\lambda^3)$ bits for polynomially long messages, where $\lambda$ is the security parameter. We construct a new, simple, direct, and efficient one-shot signature scheme which can sign messages of any polynomial length using signing keys of $\Theta(\lambda^2)$ qubits and signatures of size $\Theta(\lambda^2)$ bits. We achieve corresponding savings in runtimes, in both the oracle model and the plain model. In addition, unlike the Shmueli-Zhandry construction, our scheme achieves perfect correctness. 

Our scheme also achieves strong signature incompressibility, which implies a public-key quantum fire scheme with perfect correctness among other applications, correcting an error in a recent work of \c{C}akan, Goyal and Shmueli (QCrypt 2025) and recovering their applications.
\end{abstract}

\ifllncs
\else
\newpage
% TABLE OF CONTENTS
\enlargethispage{1cm}
\tableofcontents
\pagenumbering{roman}
\newpage
\pagenumbering{arabic}
\fi 

\section{Introduction}\label{sec:intro}
One-shot signatures (OSS), first defined in the work of Amos, Georgiou, Kiayias and Zhandry~\cite{AGKZ20}, are a uniquely quantum form of digital signatures~\cite{GMR88} where, given some public parameters, anyone can sample a classical verification key together with a quantum signing key. The signing key allows anyone to produce the signature, a classical bit-string, of any message of their choice, but only one. In particular, the signing key self-destructs once a single signature is produced. Security requires that given the public parameter, no quantum polynomial-time algorithm can produce valid signatures for two different messages. (One could strengthen this definition to allow the adversary to produce a different signature of the same message; for more discussion, see Section~\ref{sec:overview}.)

One-shot signatures are very powerful: they immediately imply a number of exotic quantum cryptographic primitives, including quantum money, quantum lightning, non-collapsing hash functions, non-interactive and publicly verifiable proofs of quantumness and certified min-entropy, and more.

In a recent breakthrough work, Shmueli and Zhandry~\cite{SZ25} gave the first construction of one-shot signatures. In a classical oracle model, where all algorithms have oracle access to a (classical) function, their construction is unconditionally secure. In the plain model with no oracles, they show a construction assuming post-quantum secure indistinguishability obfuscation (pqIO), one-way functions (pqOWF), lossy functions (pqLF), and (essentially) a noisy trapdoor claw-free function (pqNTCF), the latter three of which can be instantiated assuming the post-quantum subexponential learning with errors (LWE) assumption.

Shmueli and Zhandry~\cite{SZ25} in fact construct a related primitive called a {\em non-collapsing hash function}, a notion that originates from the works of Ambainis, Rossmanis and Unruh~\cite{ARU14,Unr16}, and use generic reductions from \cite{AGKZ20,DS22} to boost this to a one-shot signature scheme. Unfortunately, this path leads to a large polynomial factor blowup in the sizes of the signing state and the signature itself, and the runtimes of the associated algorithms. For example, as-is, the one-shot signature scheme from \cite{SZ25} has signing keys with $\Theta(\secp^{12})$ qubits, public keys with $\Theta(\secp^{12})$ bits, and signatures with $\Theta(\secp^3)$ bits, where $\secp$ is the security parameter\footnote{As is standard in cryptography, this means that we demand security against $2^{\secp}$-time adversaries that attempt to succeed with better than $2^{-\secp}$ advantage.}. With an improved analysis of their non-collapsing hash function (see Appendix~\ref{appendix:more_details}) and opening up the \cite{DS22} reduction, the signing and public keys can be reduced to $\Theta(\secp^4)$ qubits and $\Theta(\secp^3)$ bits, respectively.

Still, several inefficiencies remain. For one, the \cite{SZ25} construction is seemingly inherently restricted to signing one bit at a time. Secondly, the signing algorithm seems to require a Marriott-Watrous-style~\cite{MW04} alternating projections algorithm.\footnote{The \cite{DS22} reduction implements a more complicated quantum rewinding/quantum singular value transform (QSVT) circuit, but at its core, still involves an alternating projections algorithm.} The first of these issues causes a $\secp$ factor blowup in the signing key size and the public-key and signature size; the second causes a further $\secp$ factor blowup in the signing key size.

In this work, we construct a new one-shot signature scheme which is:
\begin{itemize}
    \item \textbf{Efficient:} Our scheme signs polynomially many bits in one shot. In the classical oracle model, the signing keys have size $O(\secp^2)$ qubits, and the public keys and signatures have size $O(\secp^2)$ bits. Signing and verification also requires significantly fewer oracle queries. We achieve analogous savings in the plain model; for more details, see Table~\ref{fig:OSS}.
    \item \textbf{Simple and Direct:} The algorithms are simply described and, for example, avoid complex Marriott-Watrous/QSVT-style circuits.
    \item \textbf{Perfectly Correct:} The OSS scheme of \cite{SZ25} has a negligible correctness error, again seemingly inherently so, in the sense that the signing algorithm may fail to produce a valid signature. Our scheme, in contrast, enjoys perfect correctness.
    \item \textbf{Incompressible:} A simple modification of our scheme in the oracle model satisfies an incompressibility property, first defined in \cite{CGS25}, which requires that it is impossible to telegraph a classical bit string of any polynomial length which communicates information about too many valid signatures. This in turn leads us to the construction of a quantum fire~\cite{BNZ25} and quantum key-fire~\cite{CGS25} scheme, both with the added advantage of perfect correctness.
    In the process, we identify a gap in the work of \cite{CGS25} who claimed that the $\secp$-bit OSS scheme from \cite{SZ25} satisfies incompressibility; indeed, we show that this scheme, a parallel repetition of a single-bit OSS scheme, cannot be incompressible. We fix this bug by showing that our multi-bit OSS scheme is indeed incompressible, recovering the claim of \cite{CGS25}, albeit for our OSS scheme. (For more details, we refer the reader to Section~\ref{subsec:key-fire}.)
\end{itemize}

%\noindent
%We now describe our results in more detail. 

%\subsection{Our Results}
\begin{figure}[!t]
\centering
\begin{tabular}{|c|c|c|c|c|}
\hline
 &  & \textbf{Signing Key Size} & \textbf{Public Key Size} & \textbf{Signature Size} \\
 &  & {(\# qubits)} & {(\# bits)} & {(\# bits)} \\
\hline
\multirow{2}{*}{Oracle Model} & \cite{SZ25} & $\Theta(\lambda^4)$  & $\Theta(\lambda^3)$  & $\Theta(\lambda^3)$  \\ 
    \cline{2-5} & \textbf{This work} & $\Theta(\lambda^2)$  & $\Theta(\lambda^2)$  & $\Theta(\lambda^2)$   \\
\hline

\multirow{2}{*}{Plain Model} & \cite{SZ25} & $\Theta(\lambda^4)$  & $\Theta(\lambda^4) \cdot \lambda = \Theta(\lambda^5)$  & $\Theta(\lambda^3)$  \\
    \cline{2-5} & \textbf{This work} & $\Theta(\lambda^2)$  & $\Theta(\lambda^4)$  & $\Theta(\lambda^2)$  \\
\hline
\end{tabular}
\caption{OSS parameters for signing polynomially long messages. By standard collision-resistant hashing, it suffices to sign $\secp$-bit messages. The public-key size in the plain model \cite{SZ25} construction consists of $O(\secp)$ repetitions of a base public key of $O(\secp^4)$ bits, whereas our scheme has a single one.}
\label{fig:OSS}
\end{figure}

\noindent
Our results can be summarized by the following theorems:
\begin{theorem}[Informal]
  There exist one-shot signature schemes for polynomially long messages with perfect correctness and strong unforgeability with the following efficiency properties: 
  \begin{enumerate}
      \item In the classical oracle model, unconditionally: the scheme achieves signing key size $O(\secp^2)$ qubits; public-key and signature size $O(\secp^2)$ bits; key generation and verification algorithms make a constant number of oracle calls, and signing makes $O(\secp)$ oracle calls.
      \item In the plain model, assuming the existence of post-quantum secure indistinguishability obfuscation (pqIO) and the post-quantum subexponential learning with errors (LWE) assumption: the scheme achieves signing key size $O(\secp^2)$ qubits; public-key size $O(\secp^4)$ bits; and signature size $O(\secp^2)$ bits.
  \end{enumerate}
\end{theorem}

\noindent
We do not attempt to optimize the runtimes of the algorithms in the plain model construction as the limiting factor there is the use of an expensive indistinguishability obfuscation scheme. Still, we note that the signing and verification algorithms are conceptually simpler than those of \cite{SZ25}. 

\begin{theorem}[Informal]
    There exists an {\em incompressible} OSS scheme for $\secp$-bit messages with perfect correctness, strong signature security and incompressibility. As a corollary of this and \cite{CGS25}, there exists:
    \begin{itemize}
        \item a public-key quantum fire (and key-fire) scheme with perfect verification correctness, perfect strong cloning correctness, and untelegraphability security;
        \item a public-key quantum key-fire scheme for any classically-unlearnable functionality with perfect cloning correctness, perfect evaluation correctness, and key-untelegraphability security; and
        \item a public-key encryption scheme with perfect correctness, perfectly clonable yet unbounded-leakage-resilient secret keys, and classical public keys and ciphertexts.
    \end{itemize}
    Moreover, assuming the existence of post-quantum one-way functions, all of these schemes can be made efficient (with computational security).
\end{theorem}

\section{Technical Overview}\label{sec:overview}
We start by reviewing the one-shot signature (OSS) scheme of Shmueli and Zhandry~\cite{SZ25} (SZ, henceforth), taking care to point out the sources of inefficiency along the way. We then describe our ideas to improve the bandwidth and runtimes of the algorithms.

The SZ work first constructs a primitive called a {\em non-collapsing collision-resistant hash function}~\cite{ARU14,Unr16,SZ25} and invokes the general results of \cite{DS22,AGKZ20} to turn it into a one-shot signature scheme. A non-collapsing hash function is one where the public parameters $\CRS$ define a function $H(\CRS, \cdot)$. The novel and inherently quantum property of this hash function is that there should exist a QPT state sampler that, given $\CRS$, generates a state $\ket{\psi}$ and an auxiliary state $\mathsf{aux}$; and a QPT distinguisher (that we will call the collapsing distinguisher) that distinguishes between the following two experiments. In the first experiment, the challenger measures $\ket{\psi}$ and returns the result to the adversary; in the second, the challenger computes $H(\CRS,\cdot)$ on $\ket{\psi}$ in superposition and measures only the result. In addition, the hash function is collision-resistant in the standard cryptographic sense against quantum polynomial-time adversaries.

To construct such a hash function, SZ uses a bijection (or permutation) $\Pi: \{0, 1\}^n \to \{0,1\}^n$ where $n = \Theta(\secp^2)$, as well as a family of subspaces and cosets $\{(\matA_\vecy, \vecb_\vecy)\}_{\vecy \in \{0, 1\}^r}$ where $\matA_\vecy \in \Z_2^{n\times (n-r)}$ and $\vecb_\vecy \in \Z_2^{n}$. With these in place, SZ defines $$H(x) := \Pi(x)_{[1:r]}~,$$ namely the hash function simply outputs the first $r$ bits of $\Pi(x)$. SZ also  defines the following auxiliary functions, which are treated as oracles in the construction and the proof:
\begin{itemize}
    \item $\calP(\vecx) = (\vecy, \matA_\vecy \cdot \vecw + \vecb_\vecy)$ where $\Pi(x) = \vecy || \vecw$: the function $\calP$ outputs the first $r$ bits of $\Pi(x)$ as-is, and outputs the last $n-r$ bits encoded into the affine subspace indexed by $\vecy$.
    \item $\calP^{-1}(\vecy, \vecu) = \begin{cases} \Pi^{-1}(\vecy || \vecw) & \mbox{if }\exists \vecw \in \bbZ_2^{n-r} \text{ such that } \matA_\vecy \cdot \vecw + \vecb_\vecy = \vecu \\ \bot & \text{otherwise}. \end{cases}$
    \item $\calD$ is the function that checks if a given vector $\vecv$ belongs to the dual of the subspace indexed by $\vecy$. That is,  $$\calD(\vecy, \vecv) = \begin{cases} 1 & \text{if } \vecv^T \cdot \matA_\vecy = 0^{n-r} \\ 0 & \text{otherwise}. \end{cases}$$
\end{itemize}
Given oracle access to $\calP,\calP^{-1}$ and $\calD$, the construction of the collapsing distinguisher is straightforward: the state sampler produces the uniform superposition state $\ket{\psi} = \sum_x \ket{x}$ (no auxiliary information is necessary). In the first experiment, the adversary will have access to the basis state $\ket{x}$ for some $x$, while in the second experiment, the adversary holds $\ket*{\psi_\vecy} := \sum_{x: H(x) = \vecy} \ket{x}$ for some $\vecy$. Distinguishing between the two states is straightforward: one can apply $\calP$ in the forward direction (and use $\calP^{-1}$ to uncompute $x$) to get either 
\[
 \ket{\vecy,\vecz} \hspace{.2in} \mbox{or} \hspace{.2in} \ket{\vecy} \otimes \sum_{\vecw\in \bbZ_2^{n-r}} \ket*{\matA_\vecy \vecw + \vecb_\vecy}
\]
where $\vecz\in \bbZ_2^{n-r}$ is just some string. Doing a Hadamard transform on the second register results in either 
a full superposition over $\bbZ_2^n$ (with phases depending on $\vecz$) or a uniform superposition over the dual subspace of $\matA_\vecy$ (again, with phases depending on $\vecb_\vecy$). Now, applying $\calD$ in superposition on $\vecy$ together with the second register results in $0$ in the first case with overwhelming probability, and $1$ in the second case with probability $1$, distinguishing between the two cases.

Demonstrating collision-resistance of $H$ is significantly more difficult: we do not go into detail about the proof and note only that the proof of collision resistance of our new scheme closely follows the proof strategy of \cite{SZ25}. We now explain how \cite{DS22} and \cite{AGKZ20} produce a one-shot signature scheme from a non-collapsing and collision-resistant hash function. 

The template of the OSS construction, essentially from \cite{AGKZ20}, is to let the public key of the OSS scheme be an element $\vecy \in \bbZ_2^r$ in the image of the hash function $H$. The signature of a bit $b$ with respect to the public key $\vecy$ is $\vecx \in \bbZ_2^n$ such that $H(\vecx) = \vecy$ and $\vecx[1] = b$, i.e. a hash preimage  of $\vecy$ whose first bit is $b$. All one needs now to sign a bit $b$ is to produce a hash preimage of $\vecy$ that starts with the bit $b$; this is where one needs an equivocator for the hash function. The key innovation is a way to use the collapsing distinguisher, together with the pre-image state $\ket*{\psi_\vecy}$ and the oracles, to build an equivocator; this is done using the reduction of Dall'Agnol and Spooner~\cite{DS22}.

While the \cite{DS22} reduction from equivocating to distinguishing only considers non-collapsing commitment schemes, we outline how a similar reduction might work for non-collapsing and collision-resistant hash functions.\footnote{Technically speaking, the \cite{SZ25} distinguisher is not actually enough for the \cite{DS22} reduction to go through. There is an easy fix to this, as we explain in Section \ref{subsec:details}.} The first step of the \cite{DS22} reduction is to use the collapsing distinguisher for the hash function to construct an algorithm that breaks sum binding, that is, an adversary that can produce preimages of a hash that begin with either 0 or 1 on demand. The sum of the probabilities $p_0$ and $p_1$ that it succeeds in producing preimages that start with $0$ and $1$ respectively, should be non-negligibly larger than $1$. This may only succeed with a small, but non-negligible, probability. Therefore, the second step of \cite{DS22} is an amplification step which uses the sum binding algorithm in parallel on many states $\ket*{\psi_{\vecy_i}}$. 

The first step of \cite{DS22}, the reduction from a collapsing distinguisher $D$ (which we will think of as a projective measurement) to sum-binding algorithm $S$,  works roughly as follows: $S$ will measure the first bit of the preimage $x$. Upon being asked to open to bit $b$, if the measured bit is $b$, the algorithm is all set: it can measure all of $x$. Otherwise, it will try to use the projective measurement $(D, I-D)$ before measuring $x$. The hope is that when $x_1 \neq b$, the collapsing distinguisher $D$ will allow $S$ to project the partially measured state back to a subspace with non-trivial weight on preimages with $x_1 = b$. This reduction is mildly lossy: if $D$ distinguishes with probability $\frac{1}{2}+\eps$, then $S$ succeeds with probability roughly $\frac{1}{2}+O(\eps^2)$.

At this stage, one needs to amplify the sum-binding algorithm $S$ into an equivocator $E$, which must succeed with overwhelming probability. A na\"{i}ve attempt might be as follows: after measuring a hash value, $E$ may try to equivocate by running an alternating projectors/Marriott-Watrous style rewinding algorithm using $S$. Unfortunately, for technical reasons that we will explain in more detail in Section~\ref{subsec:details}, this requires multiple copies of the state. Roughly speaking, with only one copy, we may get unlucky and require a super-polynomial amount of time to equivocate.

Dall'Agnol and Spooner~\cite{DS22} deal with this issue by using many preimage states $\ket*{\psi_{\vecy_i}}$ and a QSVT/quantum rewinding circuit to produce an equivocator. This step introduces a fair bit of overhead; in particular, if $S$ succeeds with probability $\frac{1}{2}+\eps$, equivocating to a single bit with probability $1-2^{-\secp}$ requires roughly $\secp/\eps^2$ preimage states, and the same number of hash functions. Overall, the \cite{DS22} reduction will produce an equivocator with $\Theta(\secp)$ overhead in signing.\footnote{One could move this overhead into the key generation algorithm, but ultimately either the key generation algorithm or signing algorithm will incur $\Theta(\secp)$ overhead.}

Finally, to produce a one-shot signature scheme for $\secp$-bit messages, one needs to repeat this scheme in parallel, which introduces another $\secp$ overhead. This again seems unavoidable with the \cite{SZ25} construction as we explain in Section~\ref{subsec:details}. 

Thus, even with a collapsing disguisher that succeeds with probability a constant larger than $1/2$, the signature scheme that results from \cite{SZ25} ends up with $\Theta(\secp) \cdot \secp \cdot \Theta(\secp^2) = \Theta(\secp^4)$ qubit signing keys, $\secp \cdot \Theta(\secp^2) = \Theta(\secp^3)$-bit signatures, and $\secp \cdot \Theta(\secp^2) = \Theta(\secp^3)$-bit public keys. The runtime of the signing algorithm will likely also require close to $\Omega(\secp^4)$ queries, where an $\Omega(\secp^2)$ overhead comes from the QSVT circuit, to the underlying programs/oracles $\calP, \calP^{-1}$ and $\calD$, while the key generation and verification algorithms will require $\Theta(\secp)$ queries.

\subsection{Our Construction}
\label{sec:our-cons}

As noted earlier, \cite{SZ25} technically only provides a distinguisher for the complete measurement of the preimage state, which is not enough to apply the \cite{DS22} reduction. Fortunately, in Appendix~\ref{appendix:more_details}, we observe that their distinguisher has a constant advantage on a known index: the first bit. Using this fact allows us to break sum binding with probability $\frac{1}{2}+c$ for some constant $c > 0$, thus allowing the \cite{DS22} reduction to go through (see Section \ref{subsec:details} for more details). However, the final $\Theta(\secp)$ overhead of the \cite{DS22} reduction seems unavoidable without significant modification. Additionally, there is no clear way to boost the \cite{SZ25} scheme to one which has perfect correctness, since the collapsing distinguisher already begins with imperfect advantage.

Our construction avoids the \cite{DS22} and \cite{AGKZ20} reductions altogether by constructing a one-shot signature scheme in one go. Our main idea is simple: instead of signing a bit $b$ by looking for it in the hash preimage, we will look for $b$ in the \emph{image}. That is to say, to sign a bit $b$ with respect to a public key $\vecy$, we will produce a vector $\vec\sigma$ such that $\vec\sigma[1] = b$ and $\calP(\vecx) = (\vecy, \vec\sigma)$ for some $\vecx$.

To see that this new scheme has an equivocator, we rely on intuition from \cite{AGKZ20}: we can easily implement a reflection oracle across valid coset vectors $\vec\sigma$ such that $\vec\sigma[1] = b$, while the dual oracle/program $\calD$ (essentially) provides a reflection oracle across the state $$\ket*{\psi_{\vecy}} := \sum_{\vecx: H(\vecx) = \vecy}\ket*{\matA_\vecy \cdot \vecw_{\vecx} + \vecb_\vecy}~.$$ where $\vecw_{\vecx}$ is the last $n-r$ bits of $\Pi(\vecx)$. Note that $\ket*{\psi_\vecy}$ is the uniform superposition over the affine subspace $(\matA_\vecy, \vecb_\vecy)$. This naturally leads us to attempt to implement a Grover search algorithm to sign a bit. Because we have switched to the setting where the bit to be signed is in the image of $\calP$ rather than the pre-image, we know that with overwhelming probability, the coset $(\matA_\vecy, \vecb_\vecy)$ contains vectors that begin with $b$, in which case precisely half of them do (unless the first row of $\matA_\vecy$ is the zero vector, a case we will deal with shortly) and so Grover search will succeed. Producing a signature directly is now simple: given $\ket*{\psi_\vecy}$, we can use Grover search using $\calD$ to transform it into $$\ket*{\psi_{\vecy, b}} := \sum_{x: \calP(x)_{[1:r+1]} = \vecy || b} \ket*{\matA_\vecy \cdot \vecw_x + \vecb_\vecy}~.$$ Once we have done so, we can simply measure $\ket*{\psi_{\vecy, b}}$ to get a coset vector $\vec\sigma$ such that $\vec\sigma[1] = b$ and $\calP(\vecx) = (\vecy, \vec\sigma)$ for some $\vecx$. Verification simply consists of applying $\calP^{-1}$ to a supposed coset vector $\vec\sigma$ and checking this condition.

By construction, any two message-signature pairs which are valid with respect to the same public key/hash $\vecy$ (after an invocation of $\calP^{-1}$) give us $\vecx_0 \neq \vecx_1$ such that $\calP(\vecx_0)_{[1:r]} = \calP(\vecx_1)_{[1:r]}$ -- but this is exactly a collision in the hash function of \cite{SZ25}! Thus, in this simple setting, we get collision resistance essentially for free. As a corollary, this also allows us to easily achieve the stronger notion of security for our one-shot signature scheme, where it is not only hard to produce a signature of a different message, but also to produce a new signature of the same message.

There are still two undesirable properties in this basic construction: imperfect correctness and short messages. First, observe that our equivocation/signing algorithm will only work if the coset $(\matA_\vecy, \vecb_\vecy)$ has vectors that start with the bit we want to sign, which does not happen with probability $1$. Our next idea, therefore, is to force this to be the case. Specifically, note that the bad event happens when all the column vectors of $\matA_\vecy$ begin with 0. With this in mind, we will instead sample $\matA_\vecy$ by first sampling a random column vector that has a $1$ in the first coordinate, followed by $n-r-1$ linearly independent vectors with 0's in the first coordinate, thereby guaranteeing that our cosets are ``good''. 

The second undesirable property of this basic scheme is that it only allows us to sign single-bit messages, which is the same issue that we encountered with \cite{SZ25}. Consider for a moment the most natural extension to allow us to sign $\secp$-bit messages: simply equivocate to the first $\secp$ bits of the coset register. Of course, this will break our signing algorithm, because we only have a dual oracle for $\matA_\vecy$! However, there is a crucial difference between our signature scheme and that of \cite{SZ25}: unlike in the \cite{SZ25} construction, where the set of preimages whose first $\secp$ bits have been fixed are essentially a random collection of points, our images are very structured: indeed, the coset $(\matA_\vecy, \vecb_\vecy)$, after fixing the first $i$ bits to arbitrary values, is \emph{still a coset}. Thus, we can modify $\calD$ to provide a series of oracles for the dual of the $(\matA_\vecy, \vecb_\vecy)$ cosets restricted to the first $i$ bits being fixed; note that the duals do not depend on the particular fixing of the bits. Once we have done so, the signing algorithm can once again use $\calD$ to essentially sign bit-by-bit \emph{before} measuring an image, allowing us to compress $\secp$ signatures into a single signature.

With some work, we show that even with this set of oracles/programs, our scheme remains secure. To do so, we prove several lemmas about random subspaces (see Appendix \ref{appendix:helper_lemmas}) and modify the security proof of \cite{SZ25} as necessary (see Appendices \ref{appendix:missing_proofs_oracle} and \ref{appendix:missing_proofs_plain} for more details). This gives us a signature scheme with $\Theta(\secp^2)$ qubit signing keys (which consists of just one preimage state), $\Theta(\secp^2)$-bit public keys and signatures for $\secp$-bit messages, and a simple, iterative signing algorithm that makes one call to $\calD$ per bit of the message being signed. Verification is also more efficient, as we can now verify a signature for a $\secp$-bit message with a single call to $\calP^{-1}$.

To sign arbitrary polynomial-length messages, we use a random oracle to compress the message to $\secp$ bits before using our scheme to sign; this hash-and-sign transformation maintains the stronger security notion we want as well as perfect correctness.

Our construction in the plain model looks largely the same as our oracle model construction, where we replace our permutation with a pseudorandom permutation (PRP) and random functions with pseudorandom functions (PRFs) before obfuscating all our programs instead of providing oracle access, similar to \cite{SZ25}. One difference between the oracle and plain model constructions is that the public key becomes larger due to padding; this is also a feature/bug of the \cite{SZ25} construction. We note that in our construction, the public key used to verify a $\secp$-bit signature is the same size as the public keys used to verify a single bit in \cite{SZ25}, which we estimate to be around $\Theta(\secp^4)$ bits. In the plain model, we can instantiate the random oracle with any arbitrary-domain collision-resistant hash function to extend the scheme to polynomial length messages.

\subsection{Application to Quantum Fire}\label{subsec:key-fire}
Our new OSS construction also allows us to construct public-key quantum fire, a notion first formalized in \cite{BNZ25}. The recent result of \cite{CGS25} constructed such a scheme (along with other applications of quantum fire) with respect to a classical oracle using the OSS scheme of \cite{SZ25}. However, their proof relies on the existence of a one-shot signature scheme with strong incompressibility security, which does \emph{not} hold for the \cite{SZ25} construction.\footnote{It is potentially possible that the \emph{one-bit} \cite{SZ25} construction satisfies a weaker version of incompressibility where the oracles the adversary gets are quantum rather than classical, but we leave the question of whether this is the case to future work.}

Let us briefly recall the strong incompressibility guarantee: we say that a one-shot signature scheme $(\Gen, \Sign, \Ver)$ relative to a tuple of classical oracles $(\calO_{\Gen}, \calO_{\Sign}, \calO_{\Ver})$ respectively, has strong signature incompressibility if for all query-bounded quantum adversaries $\calA_1$, the following is true with overwhelming probability: given oracle access to $(\calO_{\Gen}, \calO_{\Sign}, \calO_{\Ver})$, for any polynomial-sized string $L$ produced by $\calA_1$, there is a list $\mathfrak{L}_L$ of size $|\mathfrak{L}_L| \leq \poly(|L|, \secp)$ such that no query-bounded quantum adversary $\calA_2$ with access to only $\calO_{\Sign}$ and $\calO_{\Ver}$ can produce valid signatures with respect to verification keys outside of $\mathfrak{L}_L$ (with noticeable probability). 

We observe that the \cite{SZ25} construction is {\em not incompresssible} for at least two reasons: 
\begin{enumerate}
    \item First, the signing algorithm uses both the $\calP$ and $\calP^{-1}$ oracles, and given these two oracles, $\calA_2$ can ignore the string $L$ it receives and simply run the key generation algorithm before measuring a random valid key-message-signature tuple. Each verification key appears with equal (but exponentially small) probability, so for any fixed list $\mathfrak{L}_L$, $\calA_2$ will almost always produce signatures with respect to verification keys which are outside of $\mathfrak{L}_L$. \cite{CGS25} incorrectly claim that the \cite{SZ25} signing algorithm does not require access to $\calP$ and $\calP^{-1}$.
    \item Second, the \cite{SZ25} construction for $\secp$-bit messages (which is what is needed for \cite{CGS25}) is simply a parallel repetition of the one-bit OSS scheme. However, schemes based on parallel repetition are generically \emph{highly compressible}: the adversary $\calA_1$ can generate two valid key-message-signature tuples $(\pk_{i, 0}, 0, \sigma_{i, 0}), (\pk_{i, 1}, 1, \sigma_{i, 1})$ for each bit and store this as the string $L$. The adversary $\calA_2$ can then output a valid signature with respect to an exponential number of verification keys, by mixing and matching the signatures for each bit. 
\end{enumerate}

Nevertheless, we show that a simple modification of our many-bit one-shot signature scheme \emph{does} have strong signature incompressibility and thereby recover the results of \cite{CGS25}. In fact, since this modified one-shot signature scheme has perfect correctness, the corresponding public-key quantum fire scheme (and all other related schemes) will also have perfect correctness. In Section \ref{sec:applications}, we elaborate on how to tweak our scheme from Section \ref{sec:long_signatures_oracle} to produce an OSS scheme with signature incompressibility. 

\subsection{On the Use of the \texorpdfstring{\cite{DS22}}{} Reduction in \texorpdfstring{\cite{SZ25}}{}}\label{subsec:details}
As noted earlier, \cite{SZ25} provides a collapsing distinguisher between the case of measuring the hash value alone versus measuring the entire preimage. On the other hand, the \cite{DS22} reduction is only able to construct equivocal hashes/commitments from non-collapsing \emph{bit} commitments/hash functions. Thus, applying their reduction would require a proof that there is a distinguisher with good advantage between measuring the hash value alone versus measuring the hash value together with {\em some specific bit} of the preimage. It turns out that this is not hard to show; we include a proof that the \cite{SZ25} distinguisher itself has this property in Appendix \ref{appendix:more_details}.

We alluded to the possibility of using a simple Mariott-Watrous style algorithm for ``amplifying'' the success probability of a sum-binding adversary and turn it into a equivocal adversary. We argue that this is unlikely to work, at least without significant further analysis of the specifics of the \cite{SZ25} scheme. Recall that to analyze the runtime of the alternating projections algorithm between projections $\Pi_A$ and $\Pi_B$, we need to consider the Jordan decomposition between $\Pi_A$ and $\Pi_B$. Specifically, Jordan's lemma states that there is an orthogonal decomposition of the Hilbert space $\calH$ upon which $\Pi_A$ and $\Pi_B$ act into one-dimensional or two-dimensional subspaces $\calS_j$, such that $\calS_j$ is invariant under both $\Pi_A$ and $\Pi_B$. This decomposition also has the property that on two-dimensional subspaces $\calS_j$, $\Pi_A$ and $\Pi_B$ are rank-one projectors onto the states $\ket*{v_j}$ and $\ket*{w_j}$, respectively, which defines the values $p_j = \bra*{w_j} \Pi_A \ket*{w_j}$, the cosine of the angle between $v_j$ and $w_j$. The runtime of an average execution of the Mariott-Watrous algorithm starting with the state $\ket{\psi} = \sum_j \alpha_j \ket*{w_j}$ can be equivalently described as first sampling $j$ with probability $|\alpha_j|^2$ and then flipping a $p_j$-biased coin repeatedly until a head appears; thus, the expected runtime is $\sum_j |\alpha_j|^2 / {p_j}$. However, the only clear guarantee about the state $\ket{\psi}$ is that the \emph{average} probability $\sum_j |\alpha_j|^2 \cdot p_j$ is good, i.e. some small constant. It might be the case that one of the $p_j$'s is very small (potentially exponentially so), in which case the average runtime may become superpolynomial or exponential in the unlucky scenario where the $j$ sampled corresponds to this bad angle $p_j$. This motivates the need to consider multiple copies of the preimage state, as \cite{DS22} does, which introduces overhead into the key generation and/or signing algorithms. 

While it is potentially possible that a more detailed analysis of the Jordan decomposition could show that one could avoid using a rewinding/QSVT circuit, such an equivocator would still not be able to sign more than $O(\log \secp)$ bits at a time, since after $\omega(\log \secp)$ bits of the preimage state have been measured, the state does not admit any subspace/coset structure, and the advantage of the \cite{SZ25} distinguisher vanishes completely.

\subsection{Open Problems}
We end this section with two open problems. First, the elephant in the room: can we improve the size of all parameters in the oracle model to $\Theta(\secp)$ bits? For the reader familiar with the \cite{SZ25} proof, we offer a brief explanation why this is likely to require completely new techniques: in the oracle model, the bottleneck currently lies in analyzing the dual-free case, which requires collision resistance despite a coset structure. \cite{SZ25} sidesteps this issue by using the parallel repetition of a 2-to-1 function and taking advantage of the fact that any pair of preimages is a coset for the $2$-to-$1$ and noting that repetition preserves coset structure; this causes a $\secp$ factor blowup. A more direct analysis could yield better parameters: in particular, the recent frameworks for compressed permutation oracles \cite{MMW25,Car25} may prove useful in deriving better query lower bounds. This could eliminate the $\secp$ factor blowup and result in (optimal) $\Theta(\secp)$-bit public keys, signatures, and signing states.
 
Secondly, one can ask if the parameters of the plain model construction can be improved. Here, the blowup due to the parallel repetition of a 2-to-1 function shows up as well, and it is less clear how to deal with this issue. Additionally, in the \cite{SZ25} proof, the base 2-to-1 function needs more properties (essentially, the function needs to be a simplified version of a NTCF), which forces a blowup in the public-key size. We suspect that entirely new techniques will be needed to significantly improve the parameters of our construction (and that of \cite{SZ25}) in the plain model.

\section{Preliminaries}\label{sec:prelims}
\paragraph{Notations.}
We will let $\PPT$ denote classical probabilistic polynomial-time algorithms, and $\QPT$ quantum polynomial-time algorithms. 

For a matrix $\matA \in \bbZ_2^{n \times m}$, let $\matA^{[i:j]}$ denote the $i$th through $j$th columns of $\matA$, and let $\matA^{[k]}$ refer to the $k$th column of $\matA$. For a matrix $\matA$ as above, we will let $\ColSpan(\matA)$ denote the subspace generated by the columns of $\matA$. Throughout this work, we will slightly abuse notation and refer to spans (denoted by $\spn\{\cdot\}$) of sets of vectors as well as sets of vectors and subspaces. For example, $\spn\{\ColSpan(\matA), \vece_1\}$ refers to the set of all vectors which can be written as the linear combination of the unit vector $\vece_1$ and the column vectors of $\matA$.

\subsection{One-Shot Signatures}

\begin{definition}[One-Shot Signatures, adapted from \cite{AGKZ20,SZ25}]\label{def:OSS}
    A one-shot signature (OSS) scheme is a tuple of algorithms $(\Setup, \Gen, \Sign, \Ver)$ together with message-space $(\calM_{\secp})_{\secp \in \mathbb{N}}$ with the following interface:
    \begin{itemize}
        \item $\CRS \gets \Setup(1^{\secp})$: A $\PPT$ algorithm that, given the security parameter $\secp$, samples a classical common reference string $\CRS$.
        \item $(\pk, \ket*{\sk}) \gets \Gen(\CRS)$: A $\QPT$ algorithm that takes the classical $\CRS$ and samples a classical public key $\pk$ and a quantum secret key $\ket*{\sk}$.
        \item $\sigma \gets \Sign(\CRS, \ket*{\sk}\!, m)$: A $\QPT$ algorithm that, given $\CRS$, the quantum key $\ket*{\sk}$ and any message $m \in \calM_{\secp}$, produces a classical signature $\sigma$.
        \item $\Ver(\CRS, \pk, m, \sigma) \in \{0, 1\}$: A classical deterministic polynomial-time algorithm which verifies a message $m$ and signature $\sigma$ relative to a public key $\pk$ and a common reference string $\CRS$.
        \end{itemize}
        The algorithms satisfy the following correctness and security properties:

        \begin{itemize}
        \item \textbf{Perfect Correctness:} For any $\secp \in \bbN$ and $m \in \calM_{\secp}$,
           \[ \probcond{
            \Ver(\CRS, \pk, m, \sigma_m) = 1
            }{\CRS \gets \Setup(1^{\secp}) \\ (\pk, \ket{\sk}) \gets \Gen(\CRS) \\ \sigma_m \gets \Sign(\CRS, \ket*{\sk}, m)} = 1. 
            \]
        \item \textbf{Strong Security:} For any $\QPT$ algorithm $\calA$ and for all $\secp \in \bbN$,
            \[ \probcond{(m_0, \sigma_0) \neq (m_1, \sigma_1) \hspace{5pt} \land \\
            \Ver(\CRS, \pk, m_0, \sigma_0) = 1 \hspace{5pt} \land \\ 
            \Ver(\CRS, \pk, m_1, \sigma_1) = 1 } {\CRS \gets \Setup(1^{\secp}) \\ (\pk, m_0, m_1, \sigma_0, \sigma_1) \gets \calA(\CRS)}
            = \negl(\secp).
            \] 
    \end{itemize}
\end{definition}
\begin{remark}
    Both the correctness and the security properties in this definition are stronger than \cite{SZ25}.
    
    Our perfect correctness property guarantees that for any message, the signing algorithm produces a valid signature with probability $1$ (over the CRS and the internal randomness of $\Gen$ and $\Sign$). One could consider a weaker notion of correctness where, for every message $m$, for most $\CRS$, the (strict) quantum polynomial-time signing algorithm produces a correct signature with prob $1-\negl(\secp)$. The construction of \cite{SZ25} only achieves this weaker form of correctness, and seemingly inherently so.

    Additionally, one may consider a weaker security notion, where an adversary may be able to produce multiple valid signatures for the same message, but not for different messages. Both our construction and the construction of \cite{SZ25} (the latter, along with our proof in Appendix~\ref{appendix:more_details}) achieve the strong security notion. We note, however, that a generically non-collapsing and binding commitment scheme would only give rise to a one-shot signature scheme with the weaker form of security via the \cite{DS22,AGKZ20} reduction.
\end{remark}

\section{Our Construction in the Oracle Model}\label{sec:long_signatures_oracle}
We begin by describing our one-shot signature scheme that allows us to sign $\secp$-bit messages in the oracle model.

\subsection{The Construction}
Let $\secp \in \bbN$ be the security parameter and $\ell$ denote the length of the messages we sign. We use the following parameters:
    \[ s := 16 \cdot \secp, r := s \cdot (\secp-1), \ell = \ell(\secp) := \secp, \hspace{.1in} \mbox{and} \hspace{.1in} n := r+\ell+\frac{3}{2} \cdot s~. \]
Let $\Pi: \{0, 1\}^n \to \{0, 1\}^n$ be a random permutation and let $F : \{0, 1\}^r \to \{0, 1\}^{n \cdot (n-r+1)}$ be a random function.  For each $\vecy \in \{0, 1\}^r$, let $\matB_\vecy \in \bbZ_2^{(n-\ell) \times \ell}$ be a random matrix, $\matC_\vecy \in \bbZ_2^{(n-\ell) \times (n-r-\ell)}$ a random matrix with full column-rank, and $\vecb_\vecy \in \bbZ_2^n$ be a uniformly random vector, all generated by the output randomness of $F(\vecy)$. Define $$\matA_\vecy := \begin{bmatrix} \matI_{\ell} & \matZero \\ \matB_\vecy & \matC_\vecy \end{bmatrix} \in \bbZ_2^{n \times (n-r)}~.$$ 
Then, let $\calP,\calP^{-1},\calD$ be functions (treated as oracles by all algorithms) defined as follows: 
    \begin{itemize}
        \item $\calP: \{0, 1\}^n \to (\{0, 1\}^r \times \bbZ_2^n)$ is defined to be $$\calP(x) = (\vecy, \matA_\vecy \cdot \vecw + \vecb_\vecy) \text{ where } \Pi(x) = \vecy || \vecw~.$$
        \item $\calP^{-1}: (\{0, 1\}^r \times \bbZ_2^n) \to \{0, 1\}^n \cup \{ \bot \}$ is defined to be $$\calP^{-1}(\vecy, \vecu) = \begin{cases} \Pi^{-1}(\vecy || \vecw) & \mbox{if } \exists \vecw \in \bbZ_2^{n-r} \text{ such that } \matA_\vecy \cdot \vecw + \vecb_{\vecy} = \vecu \\ \bot & \text{otherwise} \end{cases}$$
        \item $\calD: ([\ell+1] \times \{0, 1\}^r \times \bbZ_2^n) \to \{0, 1\}$ is defined to be $$\calD(j, \vecy, \vecv) = \begin{cases} 1 & \text{if } \vecv^T \cdot \matA_\vecy^{[j:n-r]} = 0^{n-r-j+1} \text{ and } j \in [\ell+1] \\ 0 & \text{otherwise} \end{cases}$$
    \end{itemize}
    The above distribution of oracles is denoted $\calO_{n, r, \ell}$. We define our hash function $H(\vecx)$ to be the first $r$ bits of $\Pi(\vecx)$, so
    $$ H(\vecx) = \Pi(\vecx)_{[1:r]}$$ 
    We will take the message space $\calM_{\secp}$ to be the set of all strings $m \in \{0, 1\}^{\ell(\secp)}$. (In Appendix~\ref{sec:hash_and_sign}, we show how to sign messages of arbitrary length, so we consider for now only messages with one fixed length for a given security parameter.)

\begin{figure}[!ht]
\centering
\begin{minipage}{0.98\linewidth}
\setlength{\parskip}{3pt}
\hrule
\vspace{0.5ex}
\begin{center}
\textbf{\underline{Our OSS Scheme in the Classical Oracle Model}}
\end{center}

\protlabel{\colorbox{lightgray}{$\Gen^{\calP, \calP^{-1}}$:}}
Prepare the state $\ket{+}^{\otimes n} = \frac{1}{{2^{n/2}}} \cdot \sum_{\vecx \in \{0, 1\}^n} \ket{\vecx}$ and apply $\calP$. 
Measure $H$ to get a string $\pk := \vecy$ and the state $$\propto \sum_{\vecx: H(\vecx) = \vecy} \ket{\vecx} \ket*{\matA_\vecy \cdot \vecw_\vecx + \vecb_\vecy}~.$$ 
where $\vecw_\vecx = \Pi(\vecx)_{[r+1:n]}$. 
Apply $\calP^{-1}(\vecy, \cdot)$ to uncompute $\vecx$, resulting in the state $$\ket{\sk} \propto \sum_{\vecx: H(\vecx) = \vecy} \ket*{\matA_\vecy \cdot \vecw_{\vecx} + \vecb_\vecy}~.$$ 
Output $(\pk, \ket{\sk})$.

\medskip
\protlabel{\colorbox{lightgray}{$\Ver^{\calP^{-1}}(\pk, m, \vec\sigma)$:}}
Output 1 if $\vec\sigma_{[1:\ell]} = m$ and $\calP^{-1}(\pk, \vec\sigma) \neq \bot$; else, output 0.

\medskip
\protlabel{\colorbox{lightgray}{$\Sign^{\calD}(\pk, \ket{\sk}, m)$:}}
Initialize $\ket*{\psi_0} := \ket*{\sk}$ and parse $\pk = \vecy$. \\
\\ 
For $\iter = 1$ to $\ell$:
    \begin{itemize}
      \item Define $$O_1(\iter, m, \vecy) := (i-1)\sum_{\vec\sigma[1:\iter] = m[1:\iter]}\ket*{\vec\sigma}\bra*{\vec\sigma}+I$$ to be the first Grover reflection oracle that, on input $\vec\sigma \in \bbZ_2^n$, adds an imaginary phase $i$ to every $\vec\sigma$ such that $\vec\sigma[1:\iter] = m[1:\iter]$.
      \item Define $$O^{\calD}_2(\iter, m, \vecy) := H^{\otimes n}((i-1)\calD(\iter, \vecy, \cdot)+I)H^{\otimes n}$$ to be the second Grover reflection oracle that, on input $\vecv\in\bbZ_2^n$, adds an imaginary phase $i$ to every  $\vecv$ in the dual of the submatrix $\matA_\vecy^{[\iter:n-r]}$ generated by all but the first $\iter-1$ columns of $\matA_\vecy$. 
      \item Compute $$\ket*{\psi_{\iter}} := O^{\calD}_2(\iter, m, \vecy)O_1(\iter, m, \vecy)\ket*{\psi_{\iter-1}}~.$$
    \end{itemize}
Given $\ket*{\psi_\ell}$, measure in the standard basis to get a string $\vec\sigma \in \{0, 1\}^n$. Output $\vec\sigma$.

\vspace{0.5ex}
\hrule
\end{minipage}
\caption{Our OSS scheme in the classical oracle model using oracles $\calP$, $\calP^{-1}$ and $\calD$. For ease of exposition, we choose to give the signing algorithm the public key $\pk$ in addition to the state $\ket{\sk}$; this is without loss of generality.}
\label{fig:oss-scheme-classical}
\end{figure}

\paragraph{Our OSS Scheme.} 
    With this setup, we are ready to construct our one-shot signature scheme; the scheme is described in Figure~\ref{fig:OSS}. Since we are working in the oracle model, we omit $\Setup$ and let $\calO_{n, r, \ell}$ play the role of the CRS. 

    The key generation and verification algorithms are self-explanatory, but the signing algorithm, which follows the general framework of \cite{AGKZ20}, requires a bit of elaboration. For simplicity, consider the process of signing a single bit, say $1$. We start with the signing state $\ket{\sk}$ which is a superposition of all vectors in the affine subspace $(\matA_\vecy, \vecb_\vecy)$. Indeed, $$\ket{\sk} = \frac{1}{\sqrt{2}} (\ket{\sk_0} + \ket{\sk_1})$$ where $\ket{\sk_b}$ is the set of all vectors in the affine subspace $(\matA_\vecy,\vecb_\vecy)$ whose first coordinate is the bit $b$.  Note that the support of $\ket{\sk_b}$ is the set of all valid signature of the bit $b$.  Applying the oracle $\calO_1$ from Figure~\ref{fig:oss-scheme-classical} with $m=1$ to $\ket{\sk}$ results in the state $\frac{1}{\sqrt{2}} (\ket{\sk_0} + i \ket{\sk_1})$.

    Denote the result of the Hadamard transform applied to $\frac{1}{\sqrt{2}} (\ket{\sk_0}+\ket{\sk_1})$ as $\ket*{\overline{\sk_0}}$, and that applied to $\frac{1}{\sqrt{2}} (\ket{\sk_0}-\ket{\sk_1})$ as $\ket*{\overline{\sk_1}}$. The key observation is that the unitary $(i-1)\calD(\iter, \vecy, \cdot)+I$ transforms $\ket*{\overline{\sk_0}}$ to $i\ket*{\overline{\sk_0}}$ and leaves $\ket*{\overline{\sk_1}}$ unchanged. 
    
    Now, applying the oracle $\calO_2$ results in the following sequence of transformations: 
    \begin{align*}
     \frac{1}{\sqrt{2}} (\ket{\sk_0} + i \ket{\sk_1})  \xrightarrow{H^{\otimes n}}  \frac{1}{2} (\ket*{\overline{\sk_0}} &+ \ket*{\overline{\sk_1}}) + \frac{i}{2} (\ket*{\overline{\sk_0}} - \ket*{\overline{\sk_1}})  \\ 
     \xrightarrow{(i-1)\calD(\iter, \vecy, \cdot)+I} \frac{i(1+i)}{2} \ket*{\overline{\sk_0}} + \frac{(1-i)}{2} \ket*{\overline{\sk_1}} &= \frac{(i-1)}{2} (\ket*{\overline{\sk_0}}  - \ket*{\overline{\sk_1}} ) \xrightarrow{H^{\otimes n}} \frac{(i-1)}{\sqrt{2}} \ket{\sk_1}
    \end{align*}
    This final state is simply the superposition over all valid signatures of the bit $1$, so measuring it in the standard basis does the job.
\begin{remark}
    Each iteration of the Grover search will add a global phase to the signing token; this does not affect the correctness of the algorithm as this phase simply accumulates over iterations and the final step of the algorithm is to measure the token state, thereby erasing the phase.
    For our applications, we will not need to implement a controlled version of the signing algorithm, for which a global phase could make a difference~\cite{Zha25a}. However, if that were to be desired in a future application, one can easily eliminate the global phase either by (a) extending our scheme to only sign messages whose lengths are the nearest multiple of 8; this will erase the global phase as $((i-1)/\sqrt{2})^8 = 1$, or (b) applying another phase gate, similar to $\calO_1$, at the end of each iteration.
\end{remark}
\begin{remark}
    We note that in our oracle (and plain model) construction, the signing algorithm only needs to call $\calD$ with indices $j$ where $1 \leq j \leq \ell$, even though we also allow access to $j = \ell+1$. This is purely for notational convenience in the security proof and not necessary for correctness.
\end{remark}

\subsection{Perfect Correctness}
In this section, we show that our signing algorithm always produces a valid signature for any message.
\begin{lemma}\label{lemma:signing_prefix}
    For all security parameters $\secp \in \bbN$ and messages $m \in \calM_{\secp}$,
        \[ \probcond{\Ver^{\calP^{-1}}(\pk, m, \sigma_m) = 1}{(\pk, \ket{\sk}) \gets \Gen^{\calP, \calP^{-1}},\\ \sigma_m \gets \Sign^{\calD}(\ket{\sk}, m) }  = 1. \]
    where the probability is over the experiments and the choice of the oracles $\calO_{n, r, \ell}$.
\end{lemma}
\begin{proof}
    First, observe that for any $\vecy$, the coset $$W_\vecy := \{\matA_\vecy \cdot \vecw + \vecb_\vecy\}_{\vecw \in \{0, 1\}^{n-r}}$$ contains vectors which start with every prefix in $\{0, 1\}^{\ell}$ by construction. For $m \in \{0, 1\}^{\ell}$ and $0 \leq j \leq \ell$, define the sets $$W_{\vecy, m, j} := \{ w \in \{0, 1\}^n : w_{[1:j]} = m_{[1:j]} \land \calP^{-1}(\vecy, w) \neq \bot \}$$  to be the set of all valid signatures of $m$ w.r.t. the public key $\vecy$. 
    We will now fix $m$.
    
    Since $W_\vecy$ contains vectors that have all prefixes, this also means that it is split equally among all prefixes, i.e. $|W_{\vecy, m, j}| = 2^{n-r-j}$ for all $m \in \{0, 1\}^{\ell}, 0 \leq j \leq \ell$. Indeed, this follows from the structure of the first $\ell$ column vectors of $\matA_{\vecy}$ for all $\vecy$. 
    
    For $1 \leq \iter \leq \ell$, the idea is to run a phase-matched variant of Grover search~\cite{BHMT00} to transform $$\ket*{W_{\vecy, m, \iter-1}} := \sum_{w \in W_{\vecy, m, \iter-1}} \ket{w} \longrightarrow \ket*{W_{\vecy, m, \iter}} := \sum_{w \in W_{\vecy, m, \iter}} \ket{w}$$ perhaps up to some global phase, so that we can eventually sign $m$.

    In order to run Grover search, we need two reflection oracles: the first is an oracle which reflects marked items, which we have in $O_1(\iter, m, \vecy)$. The second is $U(\iter, m, \vecy) := (i-1)\ket*{W_{\vecy, m, \iter}}\bra*{W_{\vecy, m, \iter}}+I$; unfortunately, we do not have access to such an oracle. However, as in \cite{AGKZ20}, we do have access to $\calD$, which we claim suffices.
    
    Observe that since the Grover search will operate on the subspace spanned by $\ket*{W_{\vecy, m, \iter}}$ and $\ket*{W_{\vecy, m+\vece_{\iter}, \iter}}$, it suffices to merely implement an oracle which applies a phase of $i$ to $\ket*{W_{\vecy, m, \iter-1}} = \frac{1}{\sqrt{2}}(\ket*{W_{\vecy, m, \iter}}+\ket*{W_{\vecy, m+\vece_{\iter}, \iter}})$ and acts as the identity on $\ket*{-_{\vecy, m, \iter-1}} := \frac{1}{\sqrt{2}}(\ket*{W_{\vecy, m, \iter}}-\ket*{W_{\vecy, m+\vece_{\iter}, \iter}})$. It is not hard to see that $O_2^{\calD}(\iter, m, \vecy)$ is precisely such a reflection oracle. Let us first consider what happens to $\ket*{W_{\vecy, m, \iter-1}}$:
    \begin{enumerate}
        \item We begin with the state $\ket*{W_{\vecy, m, \iter-1}} = \sum_{w \in W_{\vecy, m, \iter-1}} \ket*{w}_\sfW$.
        \item Applying $H^{\otimes n}$ gives the state $\sum_{d \in \left(A^{[\iter:n-r]}_{\vecy}\right)^{\perp}} \ket{d}_\sfW$.
        \item Applying $((i-1)\calD(\iter, \vecy, \cdot)+I)$ adds a phase of $i$ to the state. Applying $H^{\otimes n}$ again results in the state $i\sum_{w \in W_{\vecy, m, \iter-1}} \ket*{w}_\sfW = i \ket*{W_{\vecy, m, \iter-1}}$.
    \end{enumerate}
    On the other hand, we can consider what happens to $\ket*{-_{\vecy, m, \iter-1}}$:
    \begin{enumerate}
        \item We begin with the state 
            \[ \ket*{-_{\vecy, m, \iter-1}} = \sum_{w \in W_{\vecy, m, \iter}} \ket*{w}_\sfW-\sum_{w \in W_{\vecy, m+\vece_{\iter}, \iter}} \ket*{w}_\sfW. \]
        \item Applying $H^{\otimes n}$ gives the state 
            \[ \sum_{d \in \left(A^{[\iter+1:n-r]}_{\vecy}\right)^{\perp} \setminus \left(A^{[\iter:n-r]}_{\vecy}\right)^{\perp}} (-1)^{f(d, m)} \ket{d}_\sfW, \]
        where $f$ is some Boolean function.
        \item Now, applying $((i-1)\calD(\iter, \vecy, \cdot)+I)$ acts as the identity. Thus, applying $H^{\otimes n}$ again results in the state 
            \[ \sum_{w \in W_{\vecy, m, \iter}} \ket*{w}_\sfW-\sum_{w \in W_{\vecy, m+\vece_{\iter}, \iter}} \ket*{w}_\sfW = \ket*{-_{\vecy, m, \iter-1}}. \]
    \end{enumerate}
    
    Since $|W_{\vecy, m, \iter}| = 2^{n-r-\iter}$ is exactly half of the size of $W_{\vecy, m, \iter-1}$, we will succeed in rotating our state with certainty in just one iteration (up to some global phase) by the correctness of Grover search. Iterating for $1 \leq \iter \leq \ell$ results in the state $\ket*{W_{\vecy, m, \ell}} = \sum_{w \in W_{\vecy, m, \ell}} \ket{w}$. All that remains is to measure this state, which gives some string $\sigma_m$ such that $(\sigma_m)_{[1:\ell]} = m$ and $\calP^{-1}(\vecy, \sigma_m) \neq \bot$; this is precisely the condition for the signature verification to succeed.
\end{proof}

\subsection{Strong Security}
We now show that our new one-shot signature scheme is strongly secure. We use the same general proof strategy as \cite{SZ25}, modifying components as needed. As such, we defer most proofs to Appendix \ref{appendix:missing_proofs_oracle}.

We begin by considering what happens when we ``bloat'' the dual oracle. Let $\calO'_{n, r, \ell, s}$ denote the following distribution over oracles $\calP$, $\calP^{-1}$, and $\calD'$. The oracles $\calP$ and $\calP^{-1}$ are defined identically to the oracles in $\calO_{n, r, \ell}$. However, our new dual oracle $\calD'$ will be a relaxed version of the original dual oracle $\calD$:
    \[ \calD'(j, \vecy, \vecv) = \begin{cases} 1 & \text{if } \vecv^T \cdot \begin{bmatrix}
        \matA_\vecy^{[j:\ell]} & \matA_\vecy^{[\ell+1+s:n-r]}
    \end{bmatrix} = 0^{n-r-s-j+1} \text{ and } j \in [\ell+1], \\ 0 & \text{otherwise.} \end{cases} \]
The following lemma states that any adversary $\calA$ that finds a collision given access to $\calP$, $\calP^{-1}$, and $\calD$, is also likely to find a collision\footnote{Technically, the lemma states that $\calA$ produces something which is stronger than a collision.} given access to $\calP$, $\calP^{-1}$, and $\calD'$.
\begin{lemma}[The Bloating Lemma]\label{lemma:bloating_dual}
    Let $n, r, s, \ell \in \bbN^{+}$ such that $r+s+\ell \leq n$. Suppose there is an oracle-aided $q$-query quantum algorithm $\calA$ such that
        \[ \Pr[(\vecy_0 = \vecy_1) \land (\vecx_0 \neq \vecx_1):\begin{aligned} (\calP, \calP^{-1}, \calD) &\gets \calO_{n, r, \ell} \\ (\vecx_0, \vecx_1) &\gets \calA^{\calP, \calP^{-1}, \calD} \\ (\vecy_b, \vecu_b) &\gets \calP(\vecx_b)\end{aligned}] \geq \eps. \]
    Define $t := n-r-s$. If $q^3 \cdot n^3 \cdot \frac{1}{\eps^2} \cdot (2^{t-s}+2^{-t+\ell}) = o(1)$ and $\frac{q^7 \cdot n^{10} \cdot \frac{1}{\eps^4}}{\sqrt{2^{n-r-s-\ell}}} = o(1)$, then 
        \[ \Pr[\begin{array}{l} {\scriptstyle (\vecy_0 = \vecy_1 := \vecy) \hspace{5pt} \land} \\ {\scriptstyle (\vecu_0-\vecu_1) \notin \ColSpan}\left(\begin{bmatrix}{\scriptstyle \matA_\vecy^{[1:\ell]}} & {\scriptstyle \matA_\vecy^{[\ell+s+1:n-r]}}\end{bmatrix}\right) \end{array}:\begin{aligned} {\scriptstyle (\calP, \calP^{-1}, \calD')} &{\scriptstyle \gets \calO'_{n, r, \ell, s}} \\ {\scriptstyle (\vecx_0, \vecx_1)} &{\scriptstyle \gets \calA^{\calP, \calP^{-1}, \calD'}} \\ {\scriptstyle (\vecy_b, \vecu_b)} &{\scriptstyle\gets \calP(\vecx_b)}\end{aligned}] \geq \frac{\eps}{256n^2}. \]
\end{lemma}
We now show that given an adversary which succeeds with a bloated dual oracle, one can actually entirely simulate the dual with only access to the primal oracles. 
\begin{lemma}[The ``Dual is Useless'' Lemma]\label{lemma:simulating_dual}
    Suppose that there is an oracle aided $q$-query algorithm $\calA$ such that
        \[ \Pr[\begin{array}{l} {\scriptstyle (\vecy_0 = \vecy_1 := \vecy) \hspace{5pt} \land} \\ {\scriptstyle (\vecu_0-\vecu_1) \notin \ColSpan}\left(\begin{bmatrix}{\scriptstyle \matA_\vecy^{[1:\ell]}} & {\scriptstyle \matA_\vecy^{[\ell+s+1:n-r]}}\end{bmatrix}\right) \end{array}:\begin{aligned} {\scriptstyle (\calP, \calP^{-1}, \calD')} &{\scriptstyle \gets \calO'_{n, r, \ell, s}} \\ {\scriptstyle (\vecx_0, \vecx_1)} &{\scriptstyle \gets \calA^{\calP, \calP^{-1}, \calD'}} \\ {\scriptstyle (\vecy_b, \vecu_b)} &{\scriptstyle\gets \calP(\vecx_b)}\end{aligned}] \geq \eps. \]
    Then there is an oracle aided $q$-query algorithm $\calB$ such that
        \[ \Pr[(\overline{\vecy}_0 = \overline{\vecy}_1) \land (\overline{\vecx}_0 \neq \overline{\vecx}_1):\begin{aligned} (\overline{\calP}, \overline{\calP}^{-1}, \overline{\calD}) &\gets \calO_{r+s, r, 0} \\ (\overline{\vecx}_0, \overline{\vecx}_1) &\gets \calB^{\overline{\calP}, \overline{\calP}^{-1}} \\ (\overline{\vecy}_b, \overline{\vecu}_b) &\gets \overline{\calP}(\overline{\vecx}_b)\end{aligned}] \geq \eps. \]
\end{lemma}
The proofs of Lemmas \ref{lemma:bloating_dual} and Lemma \ref{lemma:simulating_dual} closely follow the proofs of Lemmas 30 and 31 in \cite{SZ25}, so we include their proofs for completeness but defer them to Appendix \ref{appendix:missing_proofs_oracle}.

Let us now recall the definition of a coset partition function.
\begin{definition}[Coset Partition Functions, \cite{SZ25}]
    For $n, k \in \bbN$ such that $k \leq n$ we say a function $Q : \{0, 1\}^n \to \{0, 1\}^m$ is a $(n, m, k)$-coset partition function if, for each $y$ in the image of $Q$, the pre-image set $Q^{-1}(y)$ has size $2^k$ and is a coset of a linear space of dimension $k$. We allow different preimage sets to be cosets of different linear spaces.
\end{definition}

Having reduced our collision resistance to the dual-free case, we can use the reduction in \cite{SZ25} from the dual-free case to finding collisions in coset-partition functions:
\begin{theorem}[Adapted from Theorem 34, \cite{SZ25}]\label{thm:dual-free-oracle}
    Suppose there is an oracle aided $q$-query quantum algorithm $\calA$ such that
        \[ \Pr[(\vecy_0 = \vecy_1) \land (\vecx_0 \neq \vecx_1):\begin{aligned} (\calP, \calP^{-1}, \calD) &\gets \calO_{r+s, r, 0} \\ (\vecx_0, \vecx_1) &\gets \calA^{\calP, \calP^{-1}} \\ (\vecy_b, \vecu_b) &\gets \calP(\vecx_b)\end{aligned}] \geq \eps. \]        
    Then there is an oracle aided $q$-query quantum algorithm $\calB$ that given any $(r+s, r, s)$-coset partition function $Q$, satisfies
        \[ \Pr[(Q(\vecw_0) = Q(\vecw_1)) \land (\vecw_0 \neq \vecw_1) : (\vecw_0, \vecw_1) \gets \calB^Q] \geq \eps. \]
\end{theorem}

Combining the three results gives us the following corollary:
\begin{corollary}\label{corollary:reduce_to_coset_partition}
    Let $n, r, s, \ell \in \bbN^{+}$ such that $r+s+\ell \leq n$. Suppose there is an oracle-aided $q$-query quantum algorithm $\calA$ such that
        \[ \Pr[(\vecy_0 = \vecy_1) \land (\vecx_0 \neq \vecx_1):\begin{aligned} (\calP, \calP^{-1}, \calD) &\gets \calO_{n, r, \ell} \\ (\vecx_0, \vecx_1) &\gets \calA^{\calP, \calP^{-1}, \calD} \\ (\vecy_b, \vecu_b) &\gets \calP(\vecx_b)\end{aligned}] \geq \eps. \]
    Define $t := n-r-s$ and suppose $q^3 \cdot n^3 \cdot \frac{1}{\eps^2} \cdot (2^{t-s}+2^{-t+\ell}) = o(1)$ and $\frac{q^7 \cdot n^{10} \cdot \frac{1}{\eps^4}}{\sqrt{2^{t-\ell}}} = o(1)$. Then there is an oracle aided $q$-query quantum algorithm $\calB$ that given any $(r+s, r, s)$-coset partition function $Q$, satisfies
        \[ \Pr[(Q(\vecw_0) = Q(\vecw_1)) \land (\vecw_0 \neq \vecw_1) : (\vecw_0, \vecw_1) \gets \calB^Q] \geq \frac{\eps}{256n^2}. \]
\end{corollary}

Finally, we use the collision-resistance of coset partition functions as shown in \cite{SZ25}.
\begin{theorem}[Adapted from Theorem 35, \cite{SZ25}]\label{thm:coset_partition_collision_resistance}
    For any $r, s \in \bbN$ such that $s \mid r$, there exists a distribution over $(r+s, r, s)$-coset partition functions $Q$, such that any algorithm making $q$ queries to $Q$ can only find collisions in $Q$ with probability at most $O(\frac{s^3 \cdot q^3}{2^{(r+s)/s}})$.
\end{theorem}

Combining all these results allows us to conclude that our one-shot signature scheme is collision-resistant:
\begin{theorem}\label{thm:hash_oracle_cr}
    Let $\calA$ be an oracle-aided $q$-query quantum algorithm. Then,
        \[ \Pr[(\vecx_0 \neq \vecx_1) \land (H(\vecx_0) = H(\vecx_1)): \begin{aligned} (\calP, \calP^{-1}, \calD) &\gets \calO_{n, r, \ell} \\ (\vecx_0, \vecx_1) &\gets \calA^{\calP, \calP^{-1}, \calD}\end{aligned}] = O\left(\frac{\secp^3 \cdot q^3 \cdot n^3}{2^{\secp}}\right). \]
\end{theorem}
\begin{proof}
    Suppose for the sake of contradiction that $\calA$ which makes $q$ queries which outputs a collision in $H$ with probability $\eps = \omega\left(\frac{\secp^3 \cdot q^3 \cdot n^3}{2^{\secp}}\right)$. By the choice of our parameters in the OSS Construction (Figure~\ref{fig:oss-scheme-classical}), we have that for $t := n-r-s$,
    \begin{align*}
        q^3 \cdot n^3 \cdot \frac{1}{\eps^2} \cdot (2^{t-s}+2^{-t+\ell}) &= q^3 \cdot n^3 \cdot \frac{1}{\eps^2} \cdot (2^{\ell-s/2}+2^{-s/2})  \\
        &\leq q^3 \cdot n^3 \cdot \frac{2^{2\secp}}{\secp^6 \cdot q^6 \cdot n^6} \cdot (2^{\ell-s/2}+2^{-s/2}) \\
        &\leq \frac{1}{\secp^6 \cdot q^3 \cdot n^3} \cdot (2^{-5\secp}+2^{-6\secp}) = o(1),
    \end{align*}
    and 
        \[ \frac{q^7 \cdot n^{10} \cdot \frac{1}{\eps^4}}{\sqrt{2^{n-r-s-\ell}}} = \frac{q^7 \cdot n^{10} \cdot \frac{1}{\eps^4}}{2^{s/4}} \leq \frac{1}{\secp^{12} \cdot q^5 \cdot n^2} = o(1). \]
    Thus, the conditions of Corollary \ref{corollary:reduce_to_coset_partition} are fulfilled, so there is a $q$-query algorithm $\calB$ that given oracle access to any $(r + s, r, s)$-coset partition function $Q$ (and thus over any distribution of such functions), finds a collision in $Q$ with probability $\frac{\eps}{256n^2}$. Since $s \mid r$, we can consider the distribution over $(r + s, r, s)$-coset partition functions generated by Theorem \ref{thm:coset_partition_collision_resistance}, which implies that
        \[ \frac{\eps}{256n^2} = O\left(\frac{s^3 \cdot q^3}{2^{(r+s)/s}}\right) = O\left(\frac{\secp^3 \cdot q^3}{2^\secp}\right), \]
    and so 
        \[ \eps = O\left(\frac{\secp^3 \cdot q^3 \cdot n^2}{2^\secp}\right), \]
    contradicting $\eps = \omega(\frac{\secp^3 \cdot q^3 \cdot n^3}{2^{\secp}})$.
\end{proof}

We can now see that our scheme is strongly secure:
\begin{corollary}\label{corollary:strong_security_oracle}
    Let $\calA$ be an oracle aided $q$-query quantum algorithm. Then,  
        \[ \Pr[\begin{array}{l} (m_0, \sigma_0) \neq (m_1, \sigma_1) \hspace{3pt} \land \\ \Ver^{\calP^{-1}}(\pk, m_0, \sigma_0) = 1 \hspace{3pt} \land\\ \Ver^{\calP^{-1}}(\pk, m_1, \sigma_1) = 1  \end{array}: \begin{aligned} (\calP, \calP^{-1}, \calD) &\gets \calO_{n, r, \ell} \\ (\pk, m_0, \sigma_0, m_1, \sigma_1) &\gets \calA^{\calP, \calP^{-1}, \calD} \end{aligned}] = O\left(\frac{\secp^3 \cdot q^3 \cdot n^3}{2^{\secp}}\right). \]
\end{corollary}
\begin{proof}
    Fix any adversary $\calA$ that outputs $(\pk, m_0, \sigma_0, m_1, \sigma_1)$ such that $(m_0, \sigma_0) \neq (m_1, \sigma_1)$ and
        \[ \Ver^{\calP^{-1}}(\pk, m_0, \sigma_0) = \Ver^{\calP^{-1}}(\pk, m_1, \sigma_1) = 1. \]
        
    If $\sigma_0 = \sigma_1$, then we must have $(\sigma_i)_{[1:\ell]} = m_i$ for $i \in \{0, 1\}$ and thus $m_0 = m_1$, which is a contradiction. Thus, it must be the case that $\sigma_0 \neq \sigma_1$.

    This means that $\calA$ produces two signatures $\sigma_0 \neq \sigma_1$ and a public key $\pk$ such that $\calP^{-1}(\pk, \sigma_i) \neq \bot$. This means that there exists a $(q+2)$-query algorithm $\calB$ which produces $x_0 \neq x_1$ such that $H(x_0) = H(x_1)$ which first runs $\calA$ and then queries $x_0 \gets \calP^{-1}(\pk, \sigma_0)$ and $x_1 \gets \calP^{-1}(\pk, \sigma_1)$ with the same success probability as $\calA$. As $\calB$ succeeds with probability at most $O\left(\frac{\secp^3 \cdot (q+2)^3 \cdot n^3}{2^{\secp}}\right) = O\left(\frac{\secp^3 \cdot q^3 \cdot n^3}{2^{\secp}}\right)$ by Theorem \ref{thm:hash_oracle_cr}, $\calA$ is only able to produce two valid message-signature pairs with respect to the same verification key with probability $O\left(\frac{\secp^3 \cdot q^3 \cdot n^3}{2^{\secp}}\right)$ as desired.
\end{proof}

\section{Public-Key Quantum Fire and Other Applications}\label{sec:applications}
\newcommand{\QKeyFireSign}{\mathsf{QKeyFireSign}}
\newcommand{\Spark}{\mathsf{Spark}}
\newcommand{\Clone}{\mathsf{Clone}}
\newcommand{\sn}{\mathsf{sn}}
\newcommand{\sig}{\mathsf{sig}}
\newcommand{\flame}{\mathsf{flame}}
\newcommand{\OSS}{\mathsf{OSS}}

To build a public-key quantum fire scheme, we first need a one-shot signature scheme with strong incompressibility security, as defined below:
\begin{definition}[Strong Incompressibility Security, \cite{CGS25}]
     Consider a one-shot signature scheme $\OSS = (\OSS.\Setup, \OSS.\Gen, \OSS.\Sign, \OSS.\Ver)$ relative to a tuple of classical oracles $(\calO_{\OSS.\Gen}, \calO_{\OSS.\Sign}, \calO_{\OSS.\Ver})$ where each algorithm of $\OSS$ uses the related oracle only. 
     
     $\OSS$ is said to satisfy average-case strong (polynomial) incompressibility security if for any query-bounded adversary $\calA_1$ with polynomial size classical output, there is a polynomial $s$ such that
        \[ \Pr[E: \begin{aligned} (\calO_{\OSS.\Gen}, \calO_{\OSS.\Sign}, \calO_{\OSS.\Ver}) &\gets \OSS.\Setup(1^{\secp}) \\ L &\gets \calA_1^{\calO_{\OSS.\Gen}, \calO_{\OSS.\Sign}, \calO_{\OSS.\Ver}} \end{aligned}] \geq 1-\negl(\secp), \]
     where $E$ is the event defined as
     \begin{align*}
         &\exists \mathfrak{L}_L \in (\{0, 1\}^{\poly(\secp)})^{s(\secp)}: \forall \textrm{\ \em query-bounded uniform adversaries } \calA_2, \\
        &\probcond{\pk \notin \mathfrak{L}_L \ \land \\ \OSS.\Ver^{\calO_{\OSS.\Ver}}(\pk, m, \sig) = 1}{(\pk, m, \sig) \gets \calA_2^{\calO_{\OSS.\Sign}, \calO_{\OSS.\Ver}}(L)} = \negl(\secp).
     \end{align*}
\end{definition}

We are now ready to describe the incompressible OSS scheme. It is nearly identical to the construction in Figure~\ref{fig:oss-scheme-classical}, but modified so that the coset vector $\vecb_{\vecy}$ is never a valid signature w.r.t. the public key $\vecy$. This will help with the incompressibility proof, where we will prove a stronger statement for adversaries $\calA_2$ that have $\vecb_{\vecy}$.

Define parameters $s := 16 \cdot (\secp+1)$, $r := s \cdot \secp$, $\ell = \ell(\secp) := \secp+1$, and $n := r+\ell+\frac{3}{2} \cdot s$.
Let the random permutation $\Pi$ and the random function $F$ be as before. For each $\vecy \in \{0, 1\}^r$, let $\matB_\vecy \in \bbZ_2^{(n-\ell) \times \ell}$ be a random matrix, $\matC_\vecy \in \bbZ_2^{(n-\ell) \times (n-r-\ell)}$ a random matrix with full column-rank. Slightly differently from before, we will choose $\vecb_\vecy \in \bbZ_2^n$ to be a uniformly random vector whose $\ell$'th coordinate is 1. All these are generated by the output randomness of $F(\vecy)$. Define as before $$\matA_\vecy := \begin{bmatrix} \matI_{\ell} & \matZero \\ \matB_\vecy & \matC_\vecy \end{bmatrix} \in \bbZ_2^{n \times (n-r)}~.$$
    
Then, let $\calP$, $\calP^{-1}$, $\calD$, and $\calD_0$ be the following oracles:
\begin{itemize}
    \item $\calP, \calP^{-1}, \calD$ are the same as in Figure~\ref{fig:oss-scheme-classical}.
    \item $\calD_0(\vecy, \vecu) = \begin{cases} 1 & \exists \vecz \in \bbZ_2^{n-r} \text{ such that } \matA_\vecy \cdot \vecz + \vecb_{\vecy} = \vecu \\ 0 & \text{otherwise} \end{cases}$
\end{itemize}
We will refer to the distribution of oracles $(\calP, \calP^{-1}, \calD, \calD_0)$ by $\widetilde{\calO}_{n, r, \ell}$. We define our incompressible OSS scheme for signing $\secp$-bit messages in Figure~\ref{fig:oss-scheme-incompressible}.

\begin{figure}[!ht]
\centering
\begin{minipage}{0.98\linewidth}
\setlength{\parskip}{3pt}
\hrule
\vspace{0.5ex}
\begin{center}
\textbf{\underline{Incompressible OSS Scheme in the Classical Oracle Model}}
\end{center}

\protlabel{\colorbox{lightgray}{$\OSS.\Setup(1^{\secp})$:}}
Sample $\calP$, $\calP^{-1}$, $\calD$, $\calD_0$ as above. Set $\calO_{\Ver} = \calD_0, \calO_{\Sign} = \calD, \calO_{\Gen} = (\calP, \calP^{-1}, \calD, \calD_0)$. Output $(\calO_{\Gen}, \calO_{\Sign}, \calO_{\Ver})$.

\medskip
\protlabel{\colorbox{lightgray}{$\OSS.\Gen^{\calO_{\Gen}}$:}}
Same as in Figure~\ref{fig:oss-scheme-classical}.

\medskip
\protlabel{\colorbox{lightgray}{$\OSS.\Ver^{\calO_{\Ver}}(\pk, m, \sigma)$:}}
 Output 1 if $\sigma_{[1:\ell]} = m || 0$ and $\calD_0(\pk, \sigma) = 1$; else, output 0.

\medskip
\protlabel{\colorbox{lightgray}{$\OSS.\Sign^{\calO_{\Sign}}(\ket{\sk}\!, m)$:}}
Run $\OSS.\widetilde{\Sign}^{\calO_{\Sign}}(\ket{\sk}\!, m || 0)$, where $\OSS.\widetilde{\Sign}$ is the signing algorithm in Figure~\ref{fig:oss-scheme-classical}.

\vspace{0.5ex}
\hrule
\end{minipage}
\caption{Incompressible OSS scheme in the classical oracle model using oracles $\calP$, $\calP^{-1}$, $\calD$ and $\calD_0$.}
\label{fig:oss-scheme-incompressible}
\end{figure}

\begin{lemma}[Correctness and Security]\label{lemma:incompressible_OSS_original_properties}
    The OSS scheme in Figure~\ref{fig:oss-scheme-incompressible} has perfect correctness and strong signature security.
\end{lemma}
We defer the proof of this lemma to Appendix \ref{appendix:missing_proofs_oracle} since it is nearly identical to the correctness and security proofs of the construction in Figure~\ref{fig:oss-scheme-classical}.
Before we continue, we introduce a result that we will use in our incompressibility proof.
\begin{definition}[\cite{BBBV97}]
    For an oracle algorithm $A$ making $q$ queries to an oracle $\calO$ with input domain $D$, let $\sum_x \alpha^{(i)}_x \ket{x}$ denote the $i$th oracle query, and let $M_x(i) = |\alpha^{(i)}_x|^2$ be the query mass of $x$ in the $i$'th query. For a subset $V \subseteq [q] \times D$, let $M_V = \sum_{(i, x) \in V} M_x(i)$ be the total query mass of points in $V$.
\end{definition}

\begin{theorem}[\cite{BBBV97}]\label{thm:bbbv}
    Let $A$ be an algorithm which makes $q$ queries to an oracle $\calO$ with input domain $D$ and let $V \subseteq [q] \times D$ be a set of time-input pairs to $\calO$. If we modify $\calO$ into an oracle $\calO'$ which is identical to $\calO$ except possibly on inputs in $V$, then the final quantum states of $A$ given access to $\calO$ or $\calO'$ have trace distance at most $O\left(\sqrt{q \cdot M_V}\right)$.
\end{theorem}

\begin{lemma}[Incompressibility]\label{lemma:incompressibility} 
    The OSS scheme in Figure~\ref{fig:oss-scheme-incompressible} has strong incompressibility security.
\end{lemma}
\begin{proof}
The proof follows the general outline of \cite{CGS25} closely; some parts may be cited verbatim. We point out the key differences which address the issues with the proof of \cite{CGS25}.

First, note that in the construction of \cite{SZ25}, the signing algorithm makes critical use of all three oracles $\calP$, $\calP^{-1}$, and $\calD$. But if $\calA_2$ has access to both $\calP$ and $\calP^{-1}$, then even without $\vecb_{\vecy}$ or $L$, it can trivially produce signatures for many keys outside of any polynomial-sized list of keys $\mathfrak{L}_L$. This is since it can always run the key generation algorithm and produce key-message-signature tuples $(\pk, m, \sigma)$ such that with high probability, $\Ver(\pk, m, \sigma) = 1$ and $\pk \notin \mathfrak{L}_L$. In particular, this means that the \cite{SZ25} OSS construction does \emph{not} have signature incompressibility. On the other hand, our construction only requires $\calD$ to sign and $\calD_0$ to verify, which means it is plausible that our scheme has signature incompressibility (and we will show that this is indeed the case).

We will actually prove a stronger result where $\calA_1$ directly gets the values $\Pi$, $(\matA_{\vecy})_{\vecy}$, $(\vecb_{\vecy})_{\vecy}$ and $\calA_2$ directly gets $\Pi$, $(\vecb_{\vecy})_{\vecy}$, and oracle access to $\calO_{\OSS.\Ver}$ and $\calO_{\OSS.\Sign}$. Additionally, instead of the original game, where we ask $\calA_2$ to output a valid key-message-signature tuple, we will ask $\calA_2$ to output a pair $(\vecy, \vecv)$ such that $\vecv \in \ColSpan(\matA_{\vecy}) \setminus \{\mathbf{0}\}$. We explain why this is at least as easy for $\calA_2$ here as in the original security game.

Observe that in order for $\sig$ to be a valid signature for $m$ with respect to some key $\pk = \vecy$, it must be the case that $\sig - \vecb_{\vecy} \in \ColSpan(\matA_\vecy)$ and that $\sig_{\ell} = 0$. However, since the $\ell$'th coordinate of $\vecb_\vecy$ is always nonzero (by construction), $\sig - \vecb_{\vecy}$ is also always nonzero. Thus, given $\vecb_{\vecy}$, any signature $\sig$ automatically gives rise to a nonzero vector $\sig - \vecb_{\vecy}$ in the column span of $\matA_{\vecy}$. This is the second difference between our construction and the \cite{SZ25} scheme; in the \cite{SZ25} construction, $(\vecy, \vecb_{\vecy})$ (or more accurately, $\vecy$ and the preimage of $\vecb_{\vecy}$) is always a valid signature for some message, so $\calA_2$ can once again trivially produce valid key-message-signature tuples (this time for \emph{any} key) given $\vecb_{\vecy}$. On the other hand, in our construction, $\vecb_{\vecy}$ is never a valid signature for any message since it is (by design) nonzero on the $\ell$'th coordinate.

We now observe that for each $\vecy$, there are
    \[ C_1 = 2^{(n-\ell) \ell} \cdot \prod_{i=0}^{n-r-\ell-1} (2^{n-\ell}-2^i) \]
equally likely matrices for $\matA_{\vecy}$.

Now fix any vector $\vecv \in \bbZ_2^n \setminus \{0\}$. We first consider how many matrices of the form $\matA = \begin{bmatrix} \matI_{\ell} & \matZero \\ \matB & \matC \end{bmatrix} \in \bbZ_2^{n \times (n-r)}$, where $\matC$ is a full-rank matrix, contain $\vecv$ in their column span. We can equivalently consider the probability that a random matrix of this form contains $\vecv$ in its column span and derive the total number of matrices accordingly.

Suppose $\vecv_{[1:\ell]} = 0^{\ell}$; in this case, whether a matrix contains $\vecv$ in its column span or not is independent of $\matB$. Thus, this probability is simply the probability that a random full-rank matrix in $\bbZ_2^{(n-\ell) \times (n-r-\ell)}$ contain a nonzero vector $\vecw \in \bbZ_2^{n-\ell}$ in its column span, which is simply $\frac{2^{n-r-\ell}-1}{2^{n-\ell}-1}$ and so there are 
    \[ C_2 = \frac{2^{n-r-\ell}-1}{2^{n-\ell}-1} \cdot C_1 = 2^{(n-\ell) \ell} \cdot (2^{n-r-\ell}-1) \cdot \prod_{i=1}^{n-r-\ell-1} (2^{n-\ell}-2^i)\]
such matrices.

Now suppose instead that $\vecv_{[1:\ell]} \neq 0^{\ell}$; in this case, $\vecv \in \ColSpan(\matA)$ if and only if $\matB \vecv_{[1:\ell]}+\vecv_{[\ell+1:n]} \in \ColSpan(\matC)$. Since $\vecv_{[1:\ell]}$ is nonzero, for a random matrix $\matB$, $\matB \vecv_{[1:\ell]}+\vecv_{[\ell+1:n]}$ is a uniformly random vector in $\bbZ_2^{n-\ell}$. Thus, the probability that $\vecv \in \ColSpan(\matA)$ is
    \[ \frac{1}{2^{n-\ell}} \cdot 1 + \frac{2^{n-\ell}-1}{2^{n-\ell}} \cdot \frac{2^{n-r-\ell}-1}{2^{n-\ell}-1} = \frac{2^{n-r-\ell}}{2^{n-\ell}}, \]
and so there are 
    \[ C_3 = \frac{2^{n-r-\ell}}{2^{n-\ell}} \cdot C_1 = 2^{(n-\ell) (\ell-1)} \cdot 2^{n-r-\ell} \cdot \prod_{i=0}^{n-r-\ell-1} (2^{n-\ell}-2^i) > C_2 \]
such matrices.

The second thing we consider is how many matrices of the form $\matA = \begin{bmatrix} \matI_{\ell} & \matZero \\ \matB & \matC \end{bmatrix} \in \bbZ_2^{n \times (n-r)}$, where $\matC$ is a full-rank matrix, have the property that $\vecv^T \cdot \matA^{[j:n-r]} = 0^{n-r-j+1}$ for some $j \in [\ell+1]$. It is clear that if this condition holds for any $j$, then it must be the case that $\vecv_{[\ell+1:n]}^T \cdot \matC = 0^{n-r-\ell}$. In addition, for any $j$, if $\vecv_{[\ell+1:n]}^T \cdot \matA^{[j:n-r]} = 0^{n-r-j+1}$, then either $\vecv_{[j:n]} = 0^{n-j+1}$, in which case it is always the case that $\vecv^T \cdot \matA^{[j:n-r]} = 0^{n-r-j+1}$, or $\vecv_{[\ell+1:n]} \neq 0^{n-\ell}$. We will ignore (for now) the case where $\vecv_{[j:n]} = 0^{n-j+1}$ since this reveals no information about $\matA$. We upper bound the fraction of matrices $\matC$ such that $\vecw^T \cdot \matC = 0^{n-r-\ell}$ for some nonzero $\vecw \in \bbZ_2^{n-\ell}$, and thus the fraction of matrices $\matA$ such that $\vecv^T \cdot \matA^{[j:n-r]} = 0^{n-r-j+1}$ assuming $\vecv_{[j:n]} \neq 0^{n-j+1}$.

Clearly the dual of $\matC$ has dimension $(n-\ell)-(n-r-\ell) = r$; thus, the fraction of matrices $\matC$ such that $\vecw^T \cdot \matC = 0^{n-r-\ell}$ is at most 
    \[ \frac{2^r-1}{2^{n-\ell}-1} \leq \frac{2^r}{2^{n-\ell}} = 2^{-(n-\ell)+r}. \]
The number of matrices $\matA$ is thus at most
    \[ C_4 = 2^{-(n-\ell)+r} \cdot C_1 = 2^{-3s/2} \cdot C_1 > C_3. \]

For any string $L^{*}$ from the output support of $\calA_1$, let $S_{L^{*}}$ denote the set consisting of all vectors $\vecy \in \{0, 1\}^r$ such that $H_{\infty}(\matA_{\vecy} \mid L = L^{*}) \leq \log C_4 + \secp$. We will argue that $\calA_2$ cannot succeed with non-negligible probability even when given $\{\matA_{\vecy}\}_{\vecy \in S_{L^{*}}}$.

We begin by considering the special case where $\calA_2$ does not make any oracle queries. We claim that the probability that $\calA_2(L^{*})$ can output either a) $(\vecy, \vecv)$ such that $\vecv \in \ColSpan(\matA_{\vecy}) \setminus \{0\}$ or b) $(j, \vecy, \vecv)$ such that $\vecv^T \cdot \matA_{\vecy}^{[j:n-r]} = 0^{n-r-j+1}$ and $\vecv_{[j:n]} \neq 0^{n-j+1}$ for some $j \in [\ell+1]$ where $\vecy \notin S_{L^{*}}$ is negligible.

Suppose this is not the case; then there is some non-negligible $\eps$ such that either $\calA_2$ can output $(\vecy, \vecv)$ with probability $\eps$ or it can output $(j, \vecy, \vecv)$ with probability $\eps$. Then, whenever $\calA_2$ outputs a valid tuple $(\vecy, \vecv)$ or $(j, \vecy, \vecv)$ for $\vecy \notin S_{L^{*}}$, there are at most $\max\{C_2, C_3, C_4\} = C_4$ options for $\matA_{\vecy}$ and so $\calA_2$ can guess $\matA_{\vecy}$ with probability at least $\frac{\eps}{C_4}$. However, this contradicts the fact that $H_{\infty}(\matA_\vecy \mid L = L^{*}) \leq \log C_4 + \secp$; thus, it must be the case that $\calA_2$ succeeds only with negligible probability.

We now bound the size of $S_{L^{*}}$ in this special case. Observe that
\begin{align*}
    &\Pr[H_{\infty}((\matA_{\vecy})_{\vecy} \mid L = L^{*}) \geq 2^r \cdot \log C_1 - |L| - \secp] \\
    =&\Pr[H_{\infty}((\matA_{\vecy})_{\vecy} \mid L = L^{*}) \geq \sum_{\vecy} H_{\infty}(\matA_{\vecy}) - |L| - \secp] \\
    \geq &\Pr[H_{\infty}((\matA_{\vecy})_{\vecy} \mid L = L^{*}) \geq H_{\infty}((\matA_{\vecy})_{\vecy} | L) - \secp] \geq 1-2^{-\secp}.
\end{align*}
On the other hand, for any string $L^{*}$, we have
\begin{align*}
    H_{\infty}((\matA_{\vecy})_{\vecy} \mid L = L^{*}) &= \sum_{\vecy} H_{\infty}(\matA_{\vecy} \mid L = L^{*}) \\
    &= \sum_{\vecy \in S_{L^{*}}} H_{\infty}(\matA_{\vecy} \mid L = L^{*}) + \sum_{\vecy \notin S_{L^{*}}} H_{\infty}(\matA_{\vecy} \mid L = L^{*}) \\
    &\leq |S_{L^{*}}| \cdot \left(\log C_4 + \secp \right) + (2^r-|S_{L^{*}}|) \cdot \log C_1.
\end{align*}
Thus, with probability $1-2^{-\secp}$, 
\begin{align*}
    |S_{L^{*}}| \cdot \left(\log C_4 + \secp \right) + (2^r-|S_{L^{*}}|) \cdot \log C_1 &\geq 2^r \cdot \log C_1 - |L| - \secp \\
    \implies |L| + \secp \geq |S_{L^{*}}| \cdot \left(\log \frac{C_1}{C_4} - \secp \right) &\geq |S_{L^{*}}| \cdot \left(3s/2 - \secp\right) \geq |S_{L^{*}}|,
\end{align*}
and so $\Pr[|S_{L^{*}}| \leq |L| + \secp] \geq 1-\negl(\secp)$ for all $\secp$, which completes the proof for the special case. 

We now move to the general setting where $\calA_2(L^{*})$ can make any polynomial $q(\secp)$ number of queries. Let $\frac{1}{p(\secp)}$ be the success probability of $\calA_2$. We show how to remove the first query of $\calA_2$ before iterating this process to remove all queries of $\calA_2$. Here, we clarify an ambiguity in the proof of \cite{CGS25}, which does not concretely address queries to $\calD$; this modification is not too hard with additional calculations, but requires some care.

Suppose $\calA_2$'s first query is to $\calD_0$. For just the first query, consider running $\calA_2$ with the oracle $\calD'_0$ which is a modification of $\calD_0$ as follows: on all inputs $(\vecy, \vecv)$ where $\vecy \notin S_{L^{*}}$, output 0 if $\vecv \neq \vecb_{\vecy}$ and 1 otherwise; on inputs $(\vecy, \vecv)$ where $\vecy \in S_{L^{*}}$, behave identically to $\calD_0$. Clearly, it is the case that $\calD_0(\vecy, \vecb_{\vecy}) = \calD'_0(\vecy, \vecb_{\vecy}) = 1$, and so the two oracles differ precisely on input pairs $(\vecy, \vecv)$ such that $\vecy \notin S_{L^{*}}$, $\vecv \neq \vecb_{\vecy}$, and $\calD_0(\vecy, \vecv) = 1$. The total query mass on inputs on which $\calD_0$ and $\calD'_0$ differ is at most
    \[ \sum_{\substack{\vecy \notin S_{L^{*}} \\ \vecv: \vecv \neq \vecb_{\vecy} \land \calD_0(\vecv, \vecy) = 1}} |\alpha_{\vecy, \vecv}|^2. \]
We claim that this query mass is negligible. Suppose for the sake of contradiction that this query mass was non-negligible; consider the adversary $\calB$ which runs $\calA_2$ up until it makes its first query, measures the query that $\calA_2$ would have made to $\calD_0$ and gets the measurement outcome $(\vecy, \vecv)$ before returning $(\vecy, \vecv-\vecb_{\vecy})$. By assumption, with non-negligible probability, $\calB$ outputs a pair $(\vecy, \vecv-\vecb_{\vecy})$ such that $\vecy \notin S_{L^{*}}$, $\vecv-\vecb_{\vecy} \neq 0$, but $\calD_0(\vecy, \vecv) = 1$ and so $\vecv-\vecb_{\vecy} \in \ColSpan(\matA_{\vecy}) \setminus \{0\}$. But this violates the security we proved for the special case since $\calB$ makes no queries! Thus, it must be the case that
    \[ \sum_{\substack{\vecy \notin S_{L^{*}} \\ \vecv: \vecv \neq \vecb_{\vecy} \land \calD_0(\vecv, \vecy) = 1}} |\alpha_{\vecy, \vecv}|^2 = \negl(\secp). \]
By Theorem \ref{thm:bbbv}, $\calA_2$ still succeeds with the reprogrammed oracle $\calD'_0$ for its first query (and $\calD_0$ or $\calD$ for the other queries) with probability 
    \[ \frac{1}{p(\secp)}-\sqrt{q \cdot \negl(\secp)} \geq \frac{1}{p(\secp)}-\negl(\secp) \geq \frac{1}{p(\secp)}-\frac{1}{2p(\secp)q(\secp)}. \]
But now, the query that $\calA_2$ makes to $\calD'_0$ is completely unnecessary: given $\vecb_\vecy$ for all $\vecy$ and $\matA_{\vecy}$ for $\vecy \in S_{L^{*}}$, $\calA_2$ can \emph{itself} simulate access to $\calD'_0$! We can therefore remove/simulate $\calA_2$'s first query altogether and end up with an adversary $\calA'_2$ which makes $q-1$ queries and still succeeds with probability at least $\frac{1}{p(\secp)}-\frac{1}{2p(\secp)q(\secp)}$.

We now consider the case where $\calA_2$'s first query is to $\calD$. For just the first query, consider running $\calA_2$ with the oracle $\calD'$ which is a modification of $\calD$ as follows: on inputs $(j, \vecy, \vecv)$ where $\vecy \notin S_{L^{*}}$, output 0 if $\vecv_{[j:n]} \neq 0^{n-j+1}$ and 1 otherwise; if $\vecy \in S_{L^{*}}$, then behave identically to $\calD$. As argued earlier, if $\vecv_{[j:n]} = 0^{n-j+1}$, then we always have that $\vecv^T \cdot \matA_{\vecy}^{[j:n-r]} = 0^{n-r-j+1}$ and thus $\calD(j, \vecy, \vecv) = \calD'(j, \vecy, \vecv) = 1$. We conclude that $\calD$ and $\calD'$ differ precisely on inputs $(j, \vecy, \vecv)$ such that $\vecy \notin S_{L^{*}}$, $\vecv_{[j:n]} \neq 0^{n-j+1}$, and $\calD(j, \vecy, \vecv) = 1$. The total query mass on inputs on which $\calD$ and $\calD'$ differ is thus at most
    \[ \sum_{\substack{\vecy \notin S_{L^{*}} \\ i, \vecv: \vecv_{[j:n]} \neq 0^{n-j+1} \\ \calD(j, \vecv, \vecy) = 1}} |\alpha_{\vecy, \vecv}|^2. \]
As with the previous case, we claim that this query mass is negligible. Suppose otherwise; consider the adversary $\calB$ which runs $\calA_2$ up until it makes its first query, measures the query that $\calA_2$ would have made to $\calD$ and gets the measurement outcome $(j, \vecy, \vecv)$, which it returns. By assumption, with non-negligible probability, $\calB$ outputs $(j, \vecy, \vecv)$ such that $\vecy \notin S_{L^{*}}$, $\vecv_{[j:n]} \neq 0^{n-j+1}$, but $\calD(j, \vecy, \vecv) = 1$ and so $\vecv^T \cdot \matA_{\vecy}^{[j:n-r]} = 0^{n-r-j+1}$. But this violates the security we proved for the special case since $\calB$ makes no queries. Thus, it must be the case that
    \[ \sum_{\substack{\vecy \notin S_{L^{*}} \\ j, \vecv: \vecv_{[j:n]} \neq 0^{n-j+1} \\ \calD(j, \vecv, \vecy) = 1}} |\alpha_{\vecy, \vecv}|^2 = \negl(\secp). \]
By Theorem \ref{thm:bbbv}, $\calA_2$ still succeeds with the reprogrammed oracle $\calD'$ for its first query (and $\calD$ or $\calD_0$ for the other queries) with probability 
    \[ \frac{1}{p(\secp)}-\sqrt{q \cdot \negl(\secp)} \geq \frac{1}{p(\secp)}-\negl(\secp) \geq \frac{1}{p(\secp)}-\frac{1}{2p(\secp)q(\secp)}. \]
But now, the query that $\calA_2$ makes to $\calD'$ is completely unnecessary: given $\vecb_\vecy$ for all $\vecy$ and $\matA_{\vecy}$ for $\vecy \in S_{L^{*}}$, $\calA_2$ can once again simulate access to $\calD'$. We can therefore remove/simulate $\calA_2$'s first query altogether and end up with an adversary $\calA'_2$ which makes $q-1$ queries and still succeeds with probability at least $\frac{1}{p(\secp)}-\frac{1}{2p(\secp)q(\secp)}$.

Iterating this procedure for each query that $\calA_2$ makes, we end up with an adversary $\calA_2^{*}$ which makes no queries and still succeeds with probability at least
    \[ \frac{1}{p(\secp)} - q(\secp) \cdot \frac{1}{2p(\secp)q(\secp)} \geq \frac{1}{2p(\secp)}, \]
which is non-negligible. But this violates the security of the special case, and so $\calA_2$ cannot succeed with non-negligible probability even when given $\{\matA_{\vecy}\}_{\vecy \in S_{L^{*}}}$. We conclude that without $\{\matA_{\vecy}\}_{\vecy \in S_{L^{*}}}$, $\{\vecb_{\vecy}\}_{\vecy}$, and $\Pi$, and given only $L^{*}$, no (polynomially) query-bounded algorithm $\calA_2$ can succeed with non-negligible probability in producing a valid signature for a verification key outside of $S_{L^{*}}$. Since we argued earlier that $|S_{L^{*}}| \leq |L|+\secp$ with all but negligible probability, the proof is complete.
\end{proof}

With this, we can appeal to the result of \cite{CGS25}:
\begin{theorem}[Theorems 10 and 11, \cite{CGS25}]\label{thm:boost_OSS_to_key_fire}
    Suppose there exists a one-shot signature scheme for $\secp$-bit messages with perfect correctness, (weak) signature security, and strong incompressibility security. Then there exists a quantum key-fire scheme for signing with perfect signing correctness, perfect strong cloning correctness, and key-untelegraphability security. Moreover, assuming the existence of post-quantum one-way functions, the key-fire scheme can be made efficient (with computational security) if the one-shot signature scheme is efficient.
\end{theorem}

\begin{corollary}\label{corollary:key_fire_exists}
    Relative to a classical oracle, there exists a public-key quantum key-fire scheme for signing with perfect signing correctness, perfect strong cloning correctness, and key-untelegraphability security. Moreover, assuming the existence of post-quantum one-way functions, the key-fire scheme can be made efficient (with computational security).
\end{corollary}
\begin{proof}
    The corollary follows from combining Lemmas \ref{lemma:incompressible_OSS_original_properties} and \ref{lemma:incompressibility} with Theorem \ref{thm:boost_OSS_to_key_fire}. The efficient version follows from observing that one can efficiently simulate the OSS scheme by replacing the permutation $\Pi$ and random function $F$ with a post-quantum PRP and PRF, which exist assuming post-quantum one-way functions exist \cite{Zha21a,Zha25b}.
\end{proof}

\begin{corollary}[Combining Corollary \ref{corollary:key_fire_exists} and \cite{CGS25}]
    Relative to a classical oracle, there exists:
    \begin{itemize}
        \item a public-key quantum fire scheme with perfect verification correctness, perfect strong cloning correctness, and untelegraphability security,
        \item a public-key quantum key-fire scheme with untelegraphability security for any classically-unlearnable functionality with perfect cloning correctness, perfect evaluation correctness, and key-untelegraphability security,
        \item a public-key encryption scheme with perfect correctness and perfectly clonable yet unbounded-leakage-resilient secret keys (and classical public keys and ciphertexts).
    \end{itemize}
    Moreover, assuming the existence of post-quantum one-way functions, all of these schemes can be made efficient (with computational security).
\end{corollary}

\bibliographystyle{alpha}
\bibliography{references.bib}
\appendix
\section{Cryptographic Definitions and Lemmas}
\begin{definition}[Puncturable PRFs]\label{def:pprf}
    A puncturable pseudorandom function (P-PRF) is a triple of efficient algorithms $(\sfF, \Punc, \Eval)$ with an associated output-length function $m(\secp)$ with the following interface:
    \begin{itemize}
        \item $\sfF : \{0, 1\}^{\secp} \times \{0, 1\}^{*} \to \{0, 1\}^{m(\secp)}$ is a deterministic polynomial-time algorithm.
        \item $\Punc(k, S)$ is a $\PPT$ algorithm which takes as input a key $k \in \{0, 1\}^{\secp}$ and a set of points $S \subseteq \{0, 1\}^{*}$. It outputs a punctured key $k\{S\}$.
        \item $\Eval(k\{S\}, x)$ is a deterministic polynomial-time algorithm.
    \end{itemize}
    The algorithms satisfy the following correctness and security properties:
    \begin{itemize}
        \item \textbf{Correctness:} For any $\secp \in \bbN$, $S \subseteq \{0,1\}^{*}$, $k \in \{0,1\}^{\secp}$, $x \notin S$, and $k\{S\}$ in the support of $\Punc(k, S)$, we have that $$\Eval(k\{S\}, x) = \sfF(k, x)~.$$
        \item \textbf{(Post-Quantum) Security:} A puncturable PRF is $(f,\epsilon)$-secure for some functions $f: \bbN \to \bbN$ and $\epsilon:\bbN \to [0,1]$ if for every $f(\secp)$-time quantum algorithm $\calA$, the following experiment with $\calA$ outputs $1$ with probability at most $\frac{1}{2}+\eps(\secp)$:
        \begin{itemize}
            \item $\calA(1^{\secp})$ produces a set $S \subseteq \{0, 1\}^{*}$.
            \item The experiment chooses a random $k \gets \{0, 1\}^{\secp}$ and computes $k\{S\} \gets \Punc(k, S)$. For each $x \in S$, it also sets $y_x^0 := \sfF(k, x)$ and samples $y_x^1 \gets \{0, 1\}^{m(\secp)}$ uniformly at random. Then, it chooses a random bit $b$. It finally gives $k\{S\}$ and $\{(x, y_x^b)\}_{x \in S}$ to $\calA$.
            \item $\calA$ outputs a guess $b'$ for $b$. The experiment outputs $1$ if and only if $b' = b$.
        \end{itemize}
    \end{itemize}
    Concretely, a sub-exponentially secure P-PRF scheme would be one for which there exists a positive real constant $c > 0$ such that the scheme is $(2^{\secp^c}, \frac{1}{2^{\secp^c}})$-secure.
\end{definition}

\begin{definition}[Indistinguishability Obfuscation (iO)]
    An indistinguishability obfuscator for Boolean circuits is a probabilistic polynomial-time algorithm $\IO(\cdot, \cdot, \cdot)$ with the following properties:
    \begin{itemize}
        \item \textbf{Correctness:} For all $\secp, s \in \bbN$, Boolean circuits $C$ of size at most $s$, and all inputs $x$, 
            \[ \Pr[\sfO_C(x) = C(x): \sfO_C \gets \IO(1^{\secp}, 1^s, C)] = 1. \]
        \item \textbf{(Post-Quantum) Security:} An iO scheme is $(f,\epsilon)$-secure for some functions $f: \bbN \to \bbN$ and $\epsilon:\bbN \to [0,1]$ if for every $f(\secp)$-time quantum algorithm $\calA$, and every two classical circuits $C_0, C_1$ with the same functionality and size at most $s$,
            \[ \probcond{\calA(1^\secp, \sfO_b)=b}{b \gets \{0,1\} \\ \sfO_{b} \gets \IO(1^{\secp}, 1^s, C_b)} \leq  \frac{1}{2} + \epsilon(\secp)  \]
    \end{itemize}
    Concretely, a sub-exponentially secure iO scheme would be one for which there exists a positive real constant $c > 0$ such that the scheme is $(2^{\secp^c}, \frac{1}{2^{\secp^c}})$-secure. When the upper bound on the size of the circuits is clear from the context, we will omit the size parameter $1^s$ from the input to the obfuscator. 
\end{definition}

\noindent
We will use the following lemma from \cite{CLTV15}:
\begin{lemma}[Adapted from Theorem~2, \cite{CLTV15}]\label{lemma:obfuscating_samplers}
    Let $\calX$ be a finite set and let $\calD_0$ and $\calD_1$ be two distribution samplers that take as input an index $x$ and some random coins $r$, and output a sample.
    Suppose that given the samplers $\mathcal{D}_0$ and $\mathcal{D}_1$, for every $x \in \calX$, the distributions $\calD_0(x; \cdot)$ and $\calD_1(x; \cdot)$ are $(f_D, \delta_D)$-indistinguishable. 
    
    Let $(\sfF, \Punc, \Eval)$ and $(\sfF', \Punc', \Eval')$ be $(f_{\sfF}, \delta_{\sfF})$-secure puncturable PRFs and let $\IO$ be a $(f_{\IO}, \delta_{\IO})$-secure iO scheme.
    Let 
        \[ E_{0,k_0}(x) := \calD_0(x; \sfF(k_0, x)) \hspace{.1in} \mbox{and} \hspace{.1in} E_{1,k_1}(x) := \calD_1(x; \sfF'(k_1, x))~, \]
    and let 
        \[ \widetilde{E}_0 \gets \IO(1^\secp,1^s,E_{0,k_0}) 
        \hspace{.1in} \mbox{and} \hspace{.1in} 
        \widetilde{E}_1 \gets \IO(1^\secp,1^s,E_{1,k_1}) \]
    be the obfuscated programs for a sufficiently large polynomial $s$ and uniformly random keys $k_0$ and $k_1$. Then, $\widetilde{E}_0$ and $\widetilde{E}_1$ are $(f,\delta)$-indistinguishable, where 
        \[ f = \min(f_{\sfF}, f_{\IO}, f_D) \hspace{.1in} \mbox{and} \hspace{.1in} \epsilon = O(|\calX| \cdot (\delta_{\sfF} + \delta_{\IO} + \delta_D))~. \]
\end{lemma}
\noindent
Strictly speaking, \cite{CLTV15,SZ25} only shows such a lemma for one P-PRF and two different keys, but the proof easily extends to the setting of two P-PRFs which use separate keys.

\begin{definition}[Lossy Functions]
A lossy function (LF) family consists of classical algorithms $(\LFKeyGen, \LFF)$ with the following interface:
\begin{itemize}
    \item $\pk \gets \LFKeyGen(1^{\secp}, b, 1^{\ell})$: a PPT algorithm that takes as input the security parameter $\secp \in \bbN$, a bit $b$ and a lossiness parameter $\ell \in \bbN$ with $\ell \leq \secp$. The algorithm outputs a public key $\pk$.
    \item $y \gets \LFF(1^{\secp}, \pk, x)$: a deterministic polynomial-time algorithm that takes as input the security parameter $\secp \in \bbN$, the public key $\pk$ and an input $x \in \{0, 1\}^{\secp}$ and outputs a string $y \in \{0, 1\}^m$ for some $m \geq \secp$.
\end{itemize}
The algorithms satisfy the following guarantees:
\begin{itemize}
    \item \textbf{Statistical Correctness for Injective Mode:} There exists a negligible function $\negl$ such that for every $\secp, \ell \in \bbN$,
        \[ \probcond{|\mathsf{Img}(\LFF(\pk, \cdot))| = 2^{\secp}}{\pk \gets \LFKeyGen(1^{\secp}, 0, 1^{\ell})} \geq 1-\negl(\secp). \]
    \item \textbf{Statistical Correctness for Lossy Mode:} There exists a negligible function $\negl$ such that for every $\secp, \ell \in \bbN$,
        \[ \probcond{|\mathsf{Img}(\LFF(\pk, \cdot))| \leq 2^{\ell}}{\pk \gets \LFKeyGen(1^{\secp}, 1, 1^{\ell})} \geq 1-\negl(\secp). \]
    \item \textbf{(Post-Quantum) Security:} An LF family is $(f,\epsilon)$-secure for some functions $f: \bbN \to \bbN$ and $\epsilon:\bbN \to [0,1]$ if for every $f(\secp)$-time quantum algorithm $\calA$,
     \[ \probcond{\calA(1^\secp, \pk_b)=b}{\pk_b \gets \LFKeyGen(1^{\secp}, b, 1^{\ell})} \leq  \frac{1}{2} + \epsilon(\secp)  \]
\end{itemize}
Concretely, a sub-exponentially secure LF scheme would be for which there exists a positive real constant $c > 0$ such that the scheme is $(2^{\secp^c}, \frac{1}{2^{\secp^c}})$-secure.
\end{definition}

\begin{definition}[Permutable PRPs, \cite{SZ25}]
    Let $G = \{G_N\}_{N \in \bbN}$ be a collection where each $G_N$ is a set of permutations over $[N]$. An output-permutable PRP (OP-PRP) for $G$ is a tuple of algorithms $(\Pi, \Pi^{-1}, \Perm, \Eval, \Eval^{-1})$ with the following interface:
    \begin{itemize}
        \item \textbf{Efficient Permutations:} For any key $k \in \{0, 1\}^{\secp}$ and any desired ``block size'' $N$, $\Pi(k, \cdot)$ is an efficiently computable permutation on $[N]$ with $\Pi^{-1}(k, \cdot)$ being its efficiently computable inverse. The same is true of $\Eval$ and $\Eval^{-1}$. 
        \item \textbf{Output Permuting:} $\Perm(k, \Gamma, c)$ is a deterministic polynomial-time procedure which takes as input a key $k \in \{0, 1\}^{\secp}$, the circuit description of a permutation $\Gamma \in G_N$, and a bit $c$. It outputs a permuted key $k^{\Gamma, c}$.
        \end{itemize}
        The algorithms satisfy the following guarantees:
        \begin{itemize}
        \item \textbf{Output Permuted Correctness:} For all $\secp \in \bbN$, $k \in \{0, 1\}^{\secp}$, $c \in \{0,1\}$, $\Gamma \in G_N$, and $x, z \in [N]$,
        \begin{align*}
            \Eval(k^{\Gamma, c}, x) &= \begin{cases} \Pi(k, x) & \text{if $c = 0$} \\ \Gamma(\Pi(k, x)) & \text{if $c = 1$}\end{cases} \\
            \Eval^{-1}(k^{\Gamma, c}, z) &= \begin{cases} \Pi^{-1}(k, z) & \text{if $c = 0$} \\ \Pi^{-1}(k, \Gamma^{-1}(z)) & \text{if $c = 1$}\end{cases}.
        \end{align*}
        with probability $1$. We call $k^{\Gamma, c} \gets \Perm(k, \Gamma, c)$ a permuted key.
        \item \textbf{(Post-Quantum) Security:} 
        An OP-PRP family is $(f,\epsilon)$-secure for some functions $f: \bbN \to \bbN$ and $\epsilon:\bbN \to [0,1]$ if every $f(\secp)$-time quantum interactive algorithm $\calA$ outputs $1$ with probability at most $\frac{1}{2}+\eps(\secp)$ in the following experiment:
        \begin{itemize}
            \item $\calA(1^{\secp})$ produces a block size $N$ (in binary) and the description of a permutation $\Gamma \in G_N$.
            \item The experiment chooses a random $k \gets \{0, 1\}^{\secp}$ and a random bit $c \in \{0, 1\}$. It gives $k^{\Gamma, c} \gets \Perm(k, \Gamma, c)$ to $\calA$.
            \item $\calA$ produces a guess $c'$ for $c$. The experiment outputs $1$ if $c' = c$.
        \end{itemize}
    \end{itemize}
    Concretely, a sub-exponentially secure OP-PRP scheme would be one such that there exists a positive real constant $c > 0$ such that the scheme is $(2^{\secp^c}, \frac{1}{2^{\secp^c}})$-secure.
\end{definition}
\begin{remark}[From \cite{SZ25}]
    We will often overload notation, and use the same symbol for both $\Pi$ and $\Eval$ (and corresponding symbols for $\Pi^{-1}$ and $\Eval^{-1}$), when clear from context whether we are using a permuted or normal key. In this case, an OP-PRP would be just a triple $(\Pi, \Pi^{-1}, \Perm)$.
    
    We say that a PRP is an input permutable (IP-)PRP if we apply $\Gamma$ to the inputs rather than the outputs of $\Pi$ (and accordingly apply $\Gamma^{-1}$ to the output, and not the input, of $\Pi^{-1}$). A PRP which is simultaneously an IP-PRP and OP-PRP is called a permutable PRP (P-PRP).
\end{remark}

\begin{definition}[Decomposable Permutations, \cite{SZ25}]
    Let $N \in \bbN$, $\Gamma$ a permutation on $[N]$, and let $T, s : \bbN \to \bbN$. We say that $\Gamma$ is $(T(N), s(N))$-decomposable if there exists a sequence of permutations $\Gamma_0, \ldots, \Gamma_{T(N)}$ such that:
    \begin{enumerate}
        \item $\Gamma_0$ is the identity,
       and $\Gamma_{T(N)} = \Gamma$.
        \item Each $\Gamma_i, \Gamma_i^{-1}$ has circuit size at most $s(N)$.
        \item For each $i$, either (a) $\Gamma_i = \Gamma_{i-1}$ or (b) there exists a $z_i \in [N]$ such that $\Gamma_i = \Gamma_{i-1} \circ (z_i \enspace z_i+1)$. Here $(z_i \enspace z_i+1)$ is the neighbor-swap permutation for $z_i$, which swaps between $z_i$ and $z_i + 1 \pmod{N}$, and acts as the identity on all other elements in $[N]$.
    \end{enumerate}
    In the uniform setting, we will additionally ask that there is a uniform polynomial-time (quantum) algorithm which, given the description of $\Gamma$ and $i$, constructs both $z_i$ (if it exists) and the circuits for $\Gamma_i, \Gamma_i^{-1}$.
\end{definition}

We refer the reader to \cite{SZ25} for examples of decomposable permutations; we will only need to consider two such classes of efficiently decomposable permutations.
\begin{lemma}[\cite{SZ25}]\label{lemma:decomposable_perms}
The following classes of permutations are decomposable:
\begin{enumerate}
    \item Affine permutations $\vecx \to \matA \vecx + \vecb$, where $\matA \in \bbZ_2^{n \times n}$ is invertible and $\vecb \in \bbZ_2^n$, are $(O(2^n \times n^2), \poly(n))$-decomposable.
    \item If $N = N_0 \cdot N_1$, and for every $v \in [N_0]$, $\Gamma_v$ is a $(T, s)$-decomposable permutation on domain $[N_1]$, the controlled permutation $\Gamma$ with domain $[N] := [N_0] \times [N_1]$ where $\Gamma$ applies $\Gamma_v$ to $[N_1]$ conditioned on the element in $[N_0]$ being $v$ is $(T \cdot N_0, s+\poly\log(N))$-decomposable.
\end{enumerate}
\end{lemma}

\begin{theorem}[Theorem 39, \cite{SZ25}]\label{theorem:OP_PRPs}
    Let $T$ be any exponential function and $p$ any polynomial. Assuming the existence of sub-exponentially-secure one-way functions and sub-exponentially-secure iO, there exists an P-PRP for the class of $(T, p)$-decomposable permutations. Moreover, the P-PRP is itself $(T, p)$-decomposable.
\end{theorem}

We also use the following lemma from \cite{SZ25} which allows us to use OP-PRP's in conjunction with indistinguishability obfuscation:
\begin{lemma}[Lemma 40, \cite{SZ25}]\label{lemma:PRPs}
    Let $\Gamma$ be a permutation, let $(\Pi, \Pi^{-1}, \Perm)$ be an $(f_{\sfP}, \delta_{\sfP})$-secure OP-PRP for a class of permutations which includes $\Gamma$, and let $\IO$ be a $(f_{\IO}, \delta_{\IO})$-secure iO scheme. Then for a sufficiently large polynomial $s$,  
        \[ \IO(1^{\secp}, 1^s, P^{\Pi(k, \cdot), \Pi^{-1}(k, \cdot)}) \approx_{\min(f_{\sfP}, f_{\IO}), O(\delta_{\sfP}+\delta_{\IO})} \IO(1^{\secp}, 1^s, P^{\Gamma(\Pi(k, \cdot)), \Pi^{-1}(k, \Gamma^{-1}(\cdot))}), \]
    where $k \gets \{0,1\}^{\secp}$ is uniformly random.
\end{lemma}

\section{Our Construction in the Plain Model}\label{sec:long_signatures_plain}
Like in \cite{SZ25}, to instantiate Construction \ref{fig:oss-scheme-plain}, we replace the random permutation and functions with pseudorandom versions, and provide an obfuscation of the programs.

Let $\secp \in \bbN$ be the statistical security parameter. We use the following parameters:
    \[ s := 16 \cdot \secp, r := s \cdot (\secp-1), \ell = \ell(\secp) := \secp, \hspace{.1in} \mbox{and} \hspace{.1in} n := r+\ell+\frac{3}{2} \cdot s~. \]
Let $d := \poly(\secp)$ be the expansion parameter, $\kappa := \poly(\secp) \geq d$ be the cryptographic security parameter, and $p := \poly(\kappa) = \poly(\secp)$ be the size parameter, for some sufficiently large polynomials. 

Let $\IO$ be an iO scheme, $(\sfF, \Punc, \Eval)$ a puncturable PRF, and $(\Pi, \Pi^{-1}, \Perm)$ a permutable PRP for the class of all $(2^{\poly(\kappa)}, \poly(\kappa))$-decomposable permutations.

\begin{figure}[!ht]
\centering
\begin{minipage}{0.98\linewidth}
\setlength{\parskip}{3pt}
\hrule
\vspace{0.5ex}
\begin{center}
\textbf{\underline{Incompressible OSS Scheme in the Plain Model}}
\end{center}

\protlabel{\colorbox{lightgray}{$\OSS.\Setup(1^{\secp})$:}}
Sample $\kin, \kout, \klin \gets \{0, 1\}^{\kappa}$. $\Pi(\kin, \cdot)$ is a permutation with domain $\{0, 1\}^n$, $\Pi(\kout, \cdot)$ is a permutation with domain $\{0, 1\}^d$, and $\sfF(\klin, \cdot)$ is a PRF with inputs in $\{0, 1\}^d$ that outputs some polynomial number of bits. Let $H(\cdot)$ denote the first $r$ output bits of $\Pi(\kin, \cdot)$ and $J(\cdot)$ denote the remaining $n-r$ bits. For each $\vecy \in \{0, 1\}^d$, let $\matB_\vecy \in \bbZ_2^{(n-\ell) \times \ell}$ be a matrix, $\matC_\vecy \in \bbZ_2^{(n-\ell) \times (n-r-\ell)}$ a matrix with full column-rank, and $\vecb_\vecy \in \bbZ_2^n$ be a vector, all generated pseudorandomly by the output of $\sfF(\klin, \vecy)$. Define $\matA_\vecy := \begin{bmatrix} \matI_\ell & \matZero \\ \matB_\vecy & \matC_\vecy \end{bmatrix} \in \bbZ_2^{n \times (n-r)}$.
        
        As the common reference string output $\CRS = (\calP, \calP^{-1}, \calD)$ where $\calP \gets \IO(1^{\kappa}, 1^p, P)$, $\calP^{-1} \gets \IO(1^{\kappa}, 1^p, P^{-1})$, and $\calD \gets \IO(1^{\kappa}, 1^p, D)$ such that
        \begin{itemize}
            \item $P(x) = (\vecy, \matA_\vecy \cdot J(x) + \vecb_\vecy)$ where $\vecy \gets \Pi^{-1}(\kout, H(x)||0^{d-r})$,
            \item $P^{-1}(\vecy, \vecu) = \begin{cases} \Pi^{-1}(\kin, \vecw || \vecz) & \exists \vecw, \vecz: (\Pi(\kout, \vecy) = \vecw||0^{d-r}) \\
            &\land \hspace{.05in} (\matA_\vecy \cdot \vecz + \vecb_\vecy = \vecu) \\ \bot & \text{otherwise} \end{cases}$
            \item $D(j, \vecy, \vecv) = \begin{cases} 1 & \text{if } \vecv^{T} \cdot \matA_\vecy^{[j:n-r]} = 0^{n-r-j+1} \text{ and } j \in [\ell+1] \\ 0 & \text{otherwise} \end{cases}$
        \end{itemize}

Let $\Hash(\CRS, x)$ refer to the first $d$ bits of $\calP(x)$. We will take the message space $\calM_{\secp}$ to be the set of all strings $m \in \{0, 1\}^{\ell(\secp)}$.

\medskip
\protlabel{\colorbox{lightgray}{$\OSS.\Gen(\CRS)$:}}
Prepare the state $\ket{+}^{\otimes n} = \sum_{x \in \{0, 1\}^n} \ket{x}$ and apply $\calP$. Measure $\Hash$ to get a string $\pk := \vecy$ and state $\sum_{x: \calP(x)_{[1:d]} = \vecy} \ket{x} \ket*{\matA_\vecy \cdot \vecw_{\vecx} + \vecb_\vecy}$, where $\vecw_{\vecx}$ is $J(\vecx)$ interpreted as a vector in $\bbZ_2^{n-r}$. Apply $\calP^{-1}(\vecy, \cdot)$ to uncompute $x$, resulting in the state $\ket{\sk} := (\vecy, \sum_{x: \calP(x)_{[1:d]} = \vecy} \ket*{\matA_\vecy \cdot \vecw_{\vecx} + \vecb_\vecy})$. Output $(\pk, \ket{\sk})$.

\medskip
\protlabel{\colorbox{lightgray}{$\OSS.\Sign(\CRS, \ket{\sk}\!, m)$:}}
Identical to the signing algorithm of Section \ref{sec:long_signatures_oracle}.

\medskip
\protlabel{\colorbox{lightgray}{$\OSS.\Ver(\CRS, \pk, m, \sigma)$:}}
Output 1 if $\sigma_{[1:\ell]} = m$ and $\calP^{-1}(\pk, \sigma) \neq \bot$; else, output 0.

\vspace{0.5ex}
\hrule
\end{minipage}
\caption{Incompressible OSS scheme in the plain model.}
\label{fig:oss-scheme-plain}
\end{figure}

\subsection{Perfect Correctness}
\begin{lemma}\label{lemma:signing_prefix_plain}
    For all security parameters $\secp \in \bbN$ and messages $m \in \calM_{\secp}$,
        \[ \Pr_{\substack{\CRS \gets \Setup(1^{\secp}) \\ (\pk, \ket{\sk}) \gets \Gen(\CRS)}}\left[\Ver(\CRS, \pk, m, \sigma_m) = 1 : \sigma_m \gets \Sign(\CRS, \ket{\sk}, m) \right] = 1. \]
\end{lemma}
\noindent The perfect correctness of the signing algorithm follows from an identical argument to Section \ref{sec:long_signatures_oracle}.

\subsection{Strong Security}
We now show that our one-shot signature scheme is strongly secure. We use the same proof strategy as the plain model proof in \cite{SZ25}, modifying components as needed. As such, we defer most proofs to Appendix \ref{appendix:missing_proofs_plain}.

We begin by considering what happens when we ``bloat'' the dual oracle. Let $\widetilde{\Setup}(1^{\secp}, n, r, \ell, s)$ denote the following modified setup protocol. It samples a distribution over $\calP, \calP^{-1}, \calD$. $\calP$ and $\calP^{-1}$ are defined identically; however, our new dual program $\calD' := \IO(1^{\kappa}, 1^p, D')$ will be a relaxed version of the original dual program $\calD := \IO(1^{\kappa}, 1^p, D)$:
    \[ D'(j, \vecy, \vecv) = \begin{cases} 1 & \text{if } \vecv^T \cdot \begin{bmatrix}
        \matA_\vecy^{[j:\ell]} & \matA_\vecy^{[\ell+1+s:n-r]}
    \end{bmatrix} = 0^{n-r-s-j+1} \text{ and } j \in [\ell+1], \\ 0 & \text{otherwise.} \end{cases} \]
Observe that $\widetilde{\Setup}(1^{\secp}, n, r, \ell, 0)$ is precisely the distribution of programs in Construction \ref{fig:oss-scheme-plain}. The following lemma states that any adversary $\calA$ that finds a collision given access to $\calP$, $\calP^{-1}$, and $\calD$, is also likely to find a collision\footnote{Technically, the lemma states that $\calA$ produces something which is stronger than a collision.} given access to $\calP$, $\calP^{-1}$, and $\calD'$.
\begin{lemma}\label{lemma:bloating_dual_crypto}
    Let $\secp, \kappa, n, r, s, \ell \in \bbN^{+}$ such that $r+s+\ell \leq n$, $\kappa \geq n-r-s-\ell$, and $n = \poly(\secp)$. Suppose there is a quantum algorithm $\calA$ with complexity $T_{\calA}$ such that
        \[ \Pr[(y_0 = y_1) \land (x_0 \neq x_1):\begin{aligned} (\calP, \calP^{-1}, \calD) &\gets \widetilde{\Setup}(1^{\secp}, n, r, \ell, 0) \\ (x_0, x_1) &\gets \calA(\calP, \calP^{-1}, \calD) \\ (y_b, \vecu_b) &\gets \calP(x_b)\end{aligned}] \geq \eps, \]
    for some non-negligible function $\eps := \eps(\secp)$.
    
    Suppose that the primitives used in Construction \ref{fig:oss-scheme-plain} are $(f(\kappa), \frac{1}{f(\kappa)})$-secure for some sub-exponential $f(\kappa) := 2^{\kappa^{\delta}}$ for some constant real number $\delta > 0$. Suppose there exists a constant $c > 0$ such that for $w := \kappa^{c \cdot \delta}$ and $t := (n-r-s)$, all of the following are true:
    \begin{enumerate}
        \item $\frac{(2^r+T_{\calA}) \cdot n^2/\eps}{f(\kappa)} \leq o(1)$,
        \item $2^w \cdot \frac{t}{\eps} \cdot (2^{t-s}+2^{-t+\ell}) \leq o(1)$, and 
        \item $\frac{2^w \cdot \poly(n, \kappa) \cdot T_{\calA}/\eps}{f(n-r-s-\ell)} \leq o(1)$.
    \end{enumerate}
    Then, 
        \[ \Pr[\begin{array}{l} (y_0 = y_1 := \vecy) \hspace{5pt} \land \\ {\scriptstyle (\vecu_0-\vecu_1) \notin \ColSpan}\left(\begin{bmatrix}{\scriptstyle \matA_\vecy^{[1:\ell]}} & {\scriptstyle \matA_\vecy^{[\ell+s+1:n-r]}}\end{bmatrix}\right) \end{array}:\begin{aligned} {\scriptstyle (\calP, \calP^{-1}, \calD')} &{\scriptstyle \gets \widetilde{\Setup}(1^{\secp}, n, r, \ell, s)} \\ {\scriptstyle (x_0, x_1)} &{\scriptstyle \gets \calA(\calP, \calP^{-1}, \calD')} \\ {\scriptstyle (y_b, \vecu_b)} &{\scriptstyle \gets \calP(x_b)}\end{aligned}] \geq \frac{\eps}{512n^2}. \]
\end{lemma}

We now show that given an adversary which succeeds with a bloated dual program, one can actually entirely simulate the dual with only access to the primal program obfuscations.
\begin{lemma}\label{lemma:simulating_dual_crypto}
    Let $\secp, \kappa, n, r, s, \ell \in \bbN^{+}$ such that $r+s+\ell \leq n$, $\kappa \geq n-r-s-\ell$, and $n = \poly(\secp)$. Suppose there is a quantum algorithm $\calA$ with complexity $T_{\calA}$ such that
        \[ \Pr[\begin{array}{l} (y_0 = y_1 := \vecy) \hspace{5pt} \land \\ {\scriptstyle (\vecu_0-\vecu_1) \notin \ColSpan}\left(\begin{bmatrix}{\scriptstyle \matA_\vecy^{[1:\ell]}} & {\scriptstyle \matA_\vecy^{[\ell+s+1:n-r]}}\end{bmatrix}\right) \end{array}:\begin{aligned} {\scriptstyle (\calP, \calP^{-1}, \calD')} &{\scriptstyle \gets \widetilde{\Setup}(1^{\secp}, n, r, \ell, s)} \\ {\scriptstyle (x_0, x_1)} &{\scriptstyle \gets \calA(\calP, \calP^{-1}, \calD')} \\ {\scriptstyle (y_b, \vecu_b)} &{\scriptstyle \gets \calP(x_b)}\end{aligned}] \geq \eps. \]
    Suppose that the primitives used in Construction \ref{fig:oss-scheme-plain} are $(f(\kappa), \frac{1}{f(\kappa)})$-secure for some sub-exponential $f(\kappa) := 2^{\kappa^{\delta}}$ for some constant real number $\delta > 0$, $\frac{2^r/\eps}{f(\kappa)} \leq o(1)$, and $\frac{T_{\calA}+\poly(\kappa)}{f(\kappa)} \leq o(1)$. Then there is a quantum algorithm $\calB$ running in time $T_{\calA}+\poly(\kappa)$ such that
        \[ \Pr[(\overline{y}_0 = \overline{y}_1) \land (\overline{x}_0 \neq \overline{x}_1):\begin{aligned} (\overline{\calP}, \overline{\calP}^{-1}, \overline{\calD}) &\gets \widetilde{\Setup}(1^{\secp}, r+s, r, 0, 0) \\ (\overline{x}_0, \overline{x}_1) &\gets \calB^{\overline{\calP}, \overline{\calP}^{-1}} \\ (\overline{y}_b, \overline{\vecu}_b) &\gets \overline{\calP}(\overline{x}_b)\end{aligned}] \geq \eps/2. \]
\end{lemma}
The proofs of Lemmas \ref{lemma:bloating_dual_crypto} and \ref{lemma:simulating_dual_crypto} closely follow the proofs of Lemmas 54 and 55 in \cite{SZ25}, so we defer their proofs to Appendix \ref{appendix:missing_proofs_plain}.

Having reduced to the dual-free case, we can now directly use the collision resistance argument of \cite{SZ25}.

\begin{theorem}[Adapted from Theorem 56, \cite{SZ25}]\label{theorem:dual_free_collision_plain}
Let $r, s \in \bbN$ such that $s \mid r$. Suppose the post-quantum sub-exponential hardness of LWE and that $d$ is a sufficiently large polynomial in $\secp$.

Then, for every $\QPT$ algorithm $\calA$, the probability that $\calA(\calP, \calP^{-1})$ outputs a collision in $\Hash$, where $(\calP, \calP^{-1})$ is sampled from $(\calP, \calP^{-1}, \calD) \gets \widetilde{\Setup}(1^{\secp}, r+s, r, 0, 0)$, is $\negl((r+s)/s)$.
\end{theorem}

By combining Lemma \ref{lemma:bloating_dual_crypto}, Lemma \ref{lemma:simulating_dual_crypto}, and Theorem \ref{theorem:dual_free_collision_plain}, we see that our hash function is collision-resistant. 

\begin{corollary}\label{corollary:hash_plain_cr}
    Let $\secp \in \bbN$ and set all other parameters accordingly to Construction \ref{fig:oss-scheme-plain}. Then, for every $\QPT$ algorithm $\calA$, 
        \[ \Pr_{\substack{\CRS \gets \Setup(1^{\secp}) \\ (x_0, x_1) \gets \calA(\CRS)}}\left[(x_0 \neq x_1) \land (\Hash(\CRS, x_0) = \Hash(\CRS, x_1))\right] \leq \negl(\secp). \]
\end{corollary}

We can now conclude that our one-shot signature scheme is strongly secure, from an identical argument as in the proof of Corollary \ref{corollary:strong_security_oracle}.
\begin{corollary}\label{corollary:strong_security_plain}
    Let $\secp \in \bbN$ and set all other parameters accordingly to Construction \ref{fig:oss-scheme-plain}. Then, for every $\QPT$ algorithm $\calA$,
    \[ \Pr_{\substack{\CRS \gets \Setup(1^{\secp}) \\ (\pk, m_0, \sigma_0, m_1, \sigma_1) \gets \calA(\CRS)}}\left[\begin{array}{l} (m_0, \sigma_0) \neq (m_1, \sigma_1) \hspace{5pt} \land \\ \forall b \in \{0, 1\}, \Ver(\CRS, \pk, m_b, \sigma_b) = 1 \end{array}\right] \leq \negl(\secp). \]
\end{corollary}

\section{Signing Polynomial-Length Messages}\label{sec:hash_and_sign}
With the ability to sign $\secp$-sized messages with signatures of size $O(\secp^2)$ and a single signing key, we show how to extend this to messages of arbitrary polynomial length without blowing up the signature size. The idea is simple (in both the oracle and plain models): we will use the classic \emph{hash-and-sign} approach, either with a random oracle or with a collision-resistant hash function.

\subsection{Oracle Model Construction}
In this section, we describe how to extend Construction \ref{fig:oss-scheme-classical} to sign arbitrary polynomial length messages using a random oracle.
\construction{Oracle Model Hash-and-Sign OSS}{OracleHashOSS}{
    Let $\secp \in \bbN$ be the security parameter, and let $(\Gen, \Sign, \Ver)$ be any oracle-based one-shot signature scheme for messages of length $\secp$ which uses the oracle distribution $\calO \gets \calO_{\secp}$. Let $\calH: \{0, 1\}^{*} \to \{0, 1\}^{\secp}$ be a random oracle drawn from the distribution $\calH_\secp$. We omit $\Setup$ and let a random sample from $\calH_\secp, \calO_{\secp}$ play the role of the CRS. Our other algorithms are as follows:
    \begin{itemize}
        \item $\widetilde{\Gen}^{\calH, \calO}$: Output $(\pk, \ket{\sk}) \gets \Gen^{\calO}$.
        \item $\widetilde{\Sign}^{\calH, \calO}(\ket{\sk}, m)$: Apply $\calH(m)$ to get a message $m' \in \{0, 1\}^{\secp}$. Output $\sigma \gets \Sign^{\calO}(\ket{\sk}, m')$.
        \item $\widetilde{\Ver}^{\calH, \calO}(\pk, m, \sigma)$: Apply $\calH(m)$ to get a message $m' \in \{0, 1\}^{\secp}$. Output $b \gets \Ver^{\calO}(\pk, m', \sigma)$.
    \end{itemize}
}

Correctness follows directly from the equivocality of the base OSS scheme. We now show that the hash-and-sign scheme is still strongly secure.

\begin{lemma}\label{lemma:hash_and_sign_oracle}
    Let $(\Gen, \Sign, \Ver)$ be any one-shot signature scheme which uses the oracle distribution $\{\calO_{\secp}\}_{\secp \in \bbN}$ with collision-resistance $\eps_{\mathsf{OSS}}(q, \secp)$ against $q$-query quantum algorithms for a security parameter $\secp$. Let $\calA$ be an oracle aided $q$-query quantum algorithm. Then, for all $\secp$,
        \[ \Pr[\begin{array}{l} (m_0, \sigma_0) \neq (m_1, \sigma_1) \hspace{5pt} \land \\
       \widetilde{\Ver}^{\calH, \calO}(\pk, m_0, \sigma_0) = 1 \ \land \\
       \widetilde{\Ver}^{\calH, \calO}(\pk, m_1, \sigma_1) = 1
       \end{array}: \begin{aligned} \calH \gets \calH_{\secp}, \calO &\gets \calO_{\secp} \\ (\pk, m_b, \sigma_b) &\gets \calA^{\calH, \calO} \\ \mbox{for $b \in$} & \mbox{$\{0,1\}$} \end{aligned}] \leq \eps_{\mathsf{OSS}}(q, \secp) + O\left(\frac{q^3}{2^{\secp}}\right), \]
    where $\calH \gets \calH_{\secp} : \{0, 1\}^{*} \to \{0, 1\}^{\secp}$ is a random oracle. 
\end{lemma}
\begin{proof}
    The lemma follows from a straightforward hybrid argument. Suppose that $\calA$ produces a collision $(\pk, m_0, \sigma_0, m_1, \sigma_1)$. If $\calH(m_0) = \calH(m_1)$ and $m_0 \neq m_1$, this produces a collision in $\calH$. Otherwise, either $\calH(m_0) \neq \calH(m_1)$ or $m_0 = m_1$ and thus $\sigma_0 \neq \sigma_1$; in either case, $\calA$ produces two valid message-signature pairs $(\calH(m_0), \sigma_0) \neq (\calH(m_0), \sigma_1)$ for the underlying one-shot signature scheme. Since $\calA$ makes $q$ total queries, it makes at most $q$ queries to $\calO$ and $\calH$ each, and so it can produce two such message-signature pairs with probability at most $\eps_{\mathsf{OSS}}(q, \secp)$ by assumption and a collision in $\calH$ with probability at most $O\left(\frac{q^3}{2^{\secp}}\right)$ by standard collision lower bounds \cite{Zha15}.
\end{proof}

Instantiating this hash-and-sign construction with the base one-shot signature scheme from Section \ref{sec:long_signatures_oracle} gives a strongly secure scheme for signing arbitrary polynomial-length messages:
\begin{corollary}
    Let $(\Gen, \Sign, \Ver)$ be the one-shot signature scheme from Section \ref{sec:long_signatures_oracle}, and define the one-shot signature scheme $(\widetilde{\Gen}, \widetilde{\Sign}, \widetilde{\Ver})$ accordingly. Let $\calA$ be an oracle aided $q$-query quantum algorithm. Then,
        \[ {\scriptstyle \Pr[\begin{array}{l} {\scriptstyle (m_0, \sigma_0) \neq (m_1, \sigma_1) \hspace{5pt} \land} \\
        {\scriptstyle \forall b \in \{0, 1\}, \widetilde{\Ver}^{\calH, \calP, \calP^{-1}, \calD}(\pk, m_b, \sigma_b) = 1} \end{array}: \begin{aligned} {\scriptstyle \calH \gets \calH_{\secp}, (\calP, \calP^{-1}, \calD)} &{\scriptstyle \gets \calO_{n, r, \ell}} \\ {\scriptstyle (\pk, m_0, \sigma_0, m_1, \sigma_1)} &{\scriptstyle \gets \calA^{\calH, \calP, \calP^{-1}, \calD}} \end{aligned}] \leq O\left(\frac{\secp^3 \cdot q^3 \cdot n^3}{2^{\secp}}\right),} \]
    where $\calH \gets \calH_{\secp} : \{0, 1\}^{*} \to \{0, 1\}^{\secp}$ is a random oracle. 
\end{corollary}
Perfect correctness of $(\widetilde{\Gen}, \widetilde{\Sign}, \widetilde{\Ver})$ follows immediately from the perfect correctness of $(\Gen, \Sign, \Ver)$.
\subsection{Plain Model Construction}
In this section, we describe how to extend Construction \ref{fig:oss-scheme-plain} to sign arbitrary polynomial length messages by instantiating the random oracle with an arbitrary domain collision-resistant hash function.
\construction{Plain Model Hash-and-Sign OSS}{PlainHashOSS}{
    Let $\secp$ be the security parameter. Let $(\Setup_\secp, \Gen_\secp, \Sign_\secp, \Ver_\secp)$ be any one-shot signature scheme which signs messages of length $\ell(\secp)$, and let $\{H_{\secp}\}_{\secp \in \bbN}: \{0, 1\}^{*} \to \{0, 1\}^{\ell(\secp)}$ be any family of collision-resistant hash functions. Our hash-and-sign based one-shot signature scheme is as follows:
    \begin{itemize}
        \item $\widetilde{\Setup}(1^{\secp})$: Sample $\CRS \gets \Setup(1^{\secp})$ and hash function $h \gets H_\secp$. Output $\CRS' = (h, \CRS)$.
        \item $\widetilde{\Gen}(\CRS' = (h, \CRS))$: Output $(\pk, \ket{\sk}) \gets \Gen(\CRS)$.
        \item $\widetilde{\Sign}(\CRS' = (h, \CRS), \ket{\sk}, m)$: Compute $m' := h(m) \in \{0, 1\}^{\ell(\secp)}$. Output $\sigma \gets \Sign(\CRS, \ket{\sk}, m')$.
        \item $\widetilde{\Ver}(\CRS' = (h, \CRS), \pk, m, \sigma)$: Compute $m' := h(m) \in \{0, 1\}^{\ell(\secp)}$. Output $b \gets \Ver(\CRS, \pk, m', \sigma)$.
    \end{itemize}
}

The proof of the following lemma is identical to the proof of Lemma \ref{lemma:hash_and_sign_oracle} and follows from a straightforward hybrid argument.
\begin{lemma}\label{lemma:hash_and_sign_plain}
    Let $(\Setup_\secp, \Gen_\secp, \Sign_\secp, \Ver_\secp)$ and $\{H_{\secp}\}_{\secp \in \bbN}: \{0, 1\}^{*} \to \{0, 1\}^{\ell(\secp)}$ be any one-shot signature scheme which signs messages of length $\ell(\secp)$ and any family of hash functions with collision resistance $\eps_{\mathsf{OSS}}(\secp)$ and $\eps_{\mathsf{CRHF}}(\secp)$, respectively, against $\QPT$ algorithms for a security parameter $\secp$. Then, for all $\QPT$ algorithms $\calA$,
        \[ \Pr_{\substack{\CRS' \gets \widetilde{\Setup}(1^{\secp}) \\ (\pk, m_0, m_1, \sigma_0, \sigma_1) \gets \calA(\CRS')}}\bigg[\begin{array}{l} {\scriptstyle (m_0, \sigma_0) \neq (m_1, \sigma_1) \hspace{5pt} \land} \\ {\scriptstyle \forall b \in \{0, 1\}, \widetilde{\Ver}(\CRS', \pk, m_b, \sigma_b) = 1}\end{array} \bigg] \leq \eps_{\mathsf{OSS}}(\secp) + \eps_{\mathsf{CRHF}}(\secp). \]
\end{lemma}

Instantiating the one-shot signature scheme with the scheme in Section \ref{sec:long_signatures_plain} and any LWE-based collision-resistant hash function gives the following corollary:
\begin{corollary}
    Assuming (1) subexponentially-secure indistinguishability obfuscation, (2) subexponentially-secure one-way functions, and (3) (polynomially-secure) LWE with a subexponential noise-modulus ratio, there exists a secure OSS scheme with perfect correctness and $O(\secp^2)$-bit signatures and $O(\secp^2)$-qubit signing keys for $\poly(\secp)$-bit messages.
\end{corollary}

\section{Subspace Hiding Lemmas}\label{appendix:helper_lemmas}
We begin by stating two lemmas on equivalent ways to sample superspaces which will be used throughout this work.
\begin{lemma}\label{lemma:sampling_superspaces}
    Let $n, r, s, \ell \in \bbN^{+}$ such that $r+s+\ell \leq n$, and let $S_1 \subseteq \ldots \subseteq S_{\ell+1} \subseteq \bbZ_2^n$ be subspaces where $S_j$ has dimension $r+j-1$. Fix any full column-rank matrix $\matA \in \bbZ_2^{n \times (n-r)}$ such that $S_j^{\perp} := \ColSpan(\matA^{[j:n-r]})$.

    \noindent Then, the following distributions are statistically equivalent:
    \begin{enumerate}
        \item Sample a random vector $\vecv \in \bbZ_2^n \setminus S_{\ell+1}$ and output $(T_{1, 1}, \ldots, T_{\ell+1, 1})$, where $T_{j, 1} := S_j \cup (S_j+\vecv)$ for all $j$. \label{distribution:dual_bloating_1}
        \item Sample a random vector $\vecv \in \bbZ_2^n \setminus S_{\ell+1}$. Compute $\vecy = \vecv^T \cdot \matA \in \bbZ_2^{n-r}$. Output $(T_{1, 2}, \ldots, T_{\ell+1, 2})$, where $T_{j, 2} := \{\vecw \in \bbZ_2^n \mid (\vecw^T \cdot \matA^{[j:n-r]} = 0^{n-r-j+1}) \lor (\vecw^T \cdot \matA^{[j:n-r]} = \vecy_{[j:n-r]}) \}$ for all $j$. \label{distribution:dual_bloating_2}
        \item Sample a random matrix $\matB \in \bbZ_2^{(n-r-\ell) \times \ell}$ and compute $\matC := \matA \cdot \begin{bmatrix} \matI_{\ell} & \matZero \\ \matB & \matI_{n-r-\ell} \end{bmatrix}$. Sample a random vector $\vecz \in \bbZ_2^{n-r-\ell} \setminus \{0\}$. Output $(T_{1, 3}, \ldots, T_{\ell+1, 3})$, where $T_{j, 3} := \{\vecw \in \bbZ_2^n \mid (\vecw^T \cdot \matC^{[j:n-r]} = 0^{n-r-j+1}) \lor (\vecw^T \cdot \matC^{[j:n-r]} = 0^{\ell-j+1} || \vecz) \}$ for all $j$. \label{distribution:dual_bloating_3}
    \end{enumerate}
\end{lemma}
\begin{proof}
    We first show that Distributions \ref{distribution:dual_bloating_1} and \ref{distribution:dual_bloating_2} are equivalent distributions. By hypothesis, we have that 
        \[ \{\vecw \in \bbZ_2^n \mid \vecw^T \cdot \matA^{[j:n-r]} = 0^{n-r-j+1}\} = \{\vecw \in \bbZ_2^n \mid \vecw \in \ColSpan\left(\matA^{[j:n-r]}\right)^{\perp}\} = S_j. \]
    Fix any vector $\vecv$ in $\bbZ_2^n \setminus S_{\ell+1}$. Consider any index $j$ and any vector $\vecx \in S_j + \vecv$. By definition, $\vecx = \vecs_j + \vecv$ for some vector $\vecs_j \in S_j$. Thus,
        \[ \vecx^T \cdot \matA^{[j:n-r]} = (\vecs_j + \vecv)^T \cdot \matA^{[j:n-r]} = \vecs_j^T \cdot \matA^{[j:n-r]} + \vecv^T \cdot \matA^{[j:n-r]} = \vecv^T \cdot \matA^{[j:n-r]} = \vecy_{[j:n-r]}, \]
    so $\vecx \in \{\vecw \in \bbZ_2^n \mid \vecw^T \cdot \matA^{[j:n-r]} = \vecy_{[j:n-r]}\}$. In other words, $T_{j, 1} \subseteq T_{j, 2}$ for all $j$. But observe that since $\matA$ has full column-rank, $T_{j, 2}$ is a $(r+j)$-dimensional subspace, while $\vecv \in \bbZ_2^n \setminus S_{\ell+1} \subseteq \bbZ_2^n \setminus S_j$, so $T_{j, 1}$ is also a $(r+j)$-dimensional subspace. This means that for a given $\vecv$, $(T_{1, 1}, \ldots, T_{\ell+1, 1}) = (T_{1, 2}, \ldots, T_{\ell+1, 2})$, so Distributions \ref{distribution:dual_bloating_1} and \ref{distribution:dual_bloating_2} are equivalent distributions (over random $\vecv$).

    We now show that Distributions \ref{distribution:dual_bloating_2} and \ref{distribution:dual_bloating_3} are equivalent distributions. First, observe that for \emph{any} matrix $\matB$, we have that
    \begin{align*}
        \vecw^T \cdot \matC^{[j:n-r]} &= \vecw^T \cdot \left(\matA \cdot \begin{bmatrix} \matI_{\ell} & \matZero \\ \matB & \matI_{n-r-\ell} \end{bmatrix}\right)^{[j:n-r]} \\
        &= \vecw^T \cdot \left(\begin{bmatrix} (\matA^{[1:\ell]}+\matA^{[\ell+1:n-r]} \cdot \matB) & \matA^{[\ell+1:n-r]} \end{bmatrix}\right)^{[j:n-r]} \\
        &= \vecw^T \cdot \begin{bmatrix} (\matA^{[j:\ell]}+\matA^{[\ell+1:n-r]} \cdot \matB^{[j:\ell]}) & \matA^{[\ell+1:n-r]} \end{bmatrix} \\
        &= \vecw^T \cdot \matA^{[j:n-r]} + \begin{bmatrix} (\vecw^T \cdot \matA^{[\ell+1:n-r]} \cdot \matB^{[j:\ell]}) & 0^{n-r-\ell} \end{bmatrix}.
    \end{align*}
    Thus, if $\vecw^T \cdot \matA^{[j:n-r]} = 0^{n-r-j+1}$, then
    \begin{align*}
        \vecw^T \cdot \matC^{[j:n-r]} &= \vecw^T \cdot \matA^{[j:n-r]} + \begin{bmatrix} (\vecw^T \cdot \matA^{[\ell+1:n-r]} \cdot \matB^{[j:\ell]}) & \matZero \end{bmatrix} \\
        &= 0^{n-r-j+1} + \begin{bmatrix} (0^{n-r-\ell} \cdot \matB^{[j:\ell]}) & 0^{n-r-\ell} \end{bmatrix} = 0^{n-r-j+1}.
    \end{align*}
    But the matrix $\begin{bmatrix} \matI_{\ell} & \matZero \\ \matB & \matI_{n-r-\ell} \end{bmatrix}$ is its own inverse, so we also have that $\matA = \matC \cdot \begin{bmatrix} \matI_{\ell} & \matZero \\ \matB & \matI_{n-r-\ell} \end{bmatrix}$ and thus
        \[ \vecw^T \cdot \matA^{[j:n-r]} = \vecw^T \cdot \matC^{[j:n-r]} + \begin{bmatrix} (\vecw^T \cdot \matC^{[\ell+1:n-r]} \cdot \matB^{[j:\ell]}) & 0^{n-r-\ell} \end{bmatrix}. \]
    Therefore, if $\vecw^T \cdot \matC^{[j:n-r]} = 0^{n-r-j+1}$, it must also be the case that $\vecw^T \cdot \matA^{[j:n-r]} = 0^{n-r-j+1}$.

    Since $\matA$ is full column-rank and $\begin{bmatrix} \matI_{\ell} & \matZero \\ \matB & \matI_{n-r-\ell} \end{bmatrix}$ is full-rank for all $\matB$, $\matC$ is always full column-rank. Thus, for all $\vecv$, $T_{j, 2}$ and $T_{j, 3}$ are $(r+j)$-dimensional subspaces. It therefore suffices to show the following distributions are equivalent:
    \begin{itemize}
        \item Sample a random vector $\vecv \in \bbZ_2^n \setminus S_{\ell+1}$. Compute $\vecy = \vecv^T \cdot \matA \in \bbZ_2^{n-r}$. Output $(U_1, \ldots, U_{\ell+1})$, where $U_j := \{\vecw \in \bbZ_2^n \mid \vecw^T \cdot \matA^{[j:n-r]} = \vecy_{[j:n-r]} \}$ for all $j$.
        \item Sample a random matrix $\matB \in \bbZ_2^{(n-r-\ell) \times \ell}$ and compute $\matC := \matA \cdot \begin{bmatrix} \matI_{\ell} & \matZero \\ \matB & \matI_{n-r-\ell} \end{bmatrix}$. Sample a random vector $\vecz \in \bbZ_2^{n-r-\ell} \setminus \{0\}$. Output $(U'_1, \ldots, U'_{\ell+1})$, where $U'_j := \{\vecw \in \bbZ_2^n \mid \vecw^T \cdot \matC^{[j:n-r]} = 0^{\ell-j+1} || \vecz \}$ for all $j$.
    \end{itemize}
    Note that since $\matA$ has full column-rank, the multiplication-by-$\matA$ map $L: \bbZ_2^n \to \bbZ_2^{n-r}$, defined as taking $\vecv \mapsto \vecv \cdot \matA$, is $2^r$-to-1. Since $L(S_{\ell+1}) = \spn\{\vece_1, \ldots, \vece_{\ell}\}$, we see that $L(\bbZ_2^n \setminus S_{\ell+1}) = \bbZ_2^{n-r} \setminus \spn\{\vece_1, \ldots, \vece_{\ell}\}$. In other words, in the first distribution, $\vecy$ is a uniformly random vector in $\bbZ_2^{n-r}$ whose last $n-r-\ell$ coordinates are not all zero.

    We now consider the second distribution: for a fixed matrix $\matB$ and index $j$, we have that
    \begin{align*}
        \vecw^T \cdot \matC^{[j:n-r]} &= \begin{bmatrix} 0^{\ell-j+1} & \vecz \end{bmatrix} \\
        \Leftrightarrow \exists \vecs_j \in \bbZ_2^{j-1}: \vecw^T \cdot \matC &= \begin{bmatrix} \vecs_j & 0^{\ell-j+1} & \vecz \end{bmatrix}  \\
        \Leftrightarrow \exists \vecs_j \in \bbZ_2^{j-1}: \vecw^T \cdot \matA \cdot \begin{bmatrix} \matI_{\ell} & \matZero \\ \matB & \matI_{n-r-\ell} \end{bmatrix} &= \begin{bmatrix} \vecs_j & 0^{\ell-j+1} & \vecz \end{bmatrix} \\
        \Leftrightarrow \exists \vecs_j \in \bbZ_2^{j-1}: \vecw^T \cdot \matA &= \begin{bmatrix} \vecs_j & 0^{\ell-j+1} & \vecz \end{bmatrix} \cdot \begin{bmatrix} \matI_{\ell} & \matZero \\ \matB & \matI_{n-r-\ell} \end{bmatrix} \\
        \Leftrightarrow \exists \vecs_j \in \bbZ_2^{j-1}: \vecw^T \cdot \matA &= \begin{bmatrix} (\begin{bmatrix} \vecs_j & 0^{\ell-j+1} \end{bmatrix}+ \vecz \cdot \matB) & \vecz \end{bmatrix} \\
        \Leftrightarrow \vecw^T \cdot \matA^{[j:n-r]} &= \left(\begin{bmatrix} (\vecz \cdot \matB) & \vecz \end{bmatrix}\right)_{[j:n-r]}.
    \end{align*}
    For any fixed nonzero vector $\vecz$, over the randomness of $\matB$, each column of $\vecz \cdot \matB$ is a i.i.d. random bit, so the joint distribution of $\begin{bmatrix} (\vecz \cdot \matB) & \vecz \end{bmatrix}$ for a random nonzero $\vecz$ and random matrix $\matB$ is precisely a uniformly random vector in $\bbZ_2^{n-r}$ whose last $n-r-\ell$ coordinates are not all zero. Thus, $(U_1, \ldots, U_{\ell+1}) \equiv (U'_1, \ldots, U'_{\ell+1})$ which concludes the proof.
\end{proof}

\begin{lemma}\label{lemma:sampling_superspaces_matrices}
    Let $n, r, s, \ell \in \bbN^{+}$ such that $r+s+\ell \leq n$. Fix any matrix $\matB \in \bbZ_2^{(n-\ell) \times \ell}$ and any full column-rank matrix $\matC \in \bbZ_2^{(n-\ell) \times (n-r-\ell)}$; this defines $\matA := \begin{bmatrix} \matI_{\ell} & \matZero \\ \matB & \matC \end{bmatrix}$ and $S_j := \ColSpan(\matA^{[j:n-r]})^{\perp}$.

    \noindent Then, the following distributions over subspaces are statistically equivalent:
    \begin{enumerate}
        \item Sample $s$ random linearly independent vectors $\vecv_1, \ldots, \vecv_s$ such that 
            \[ \spn\{\vecv_1, \ldots, \vecv_s\} \cap S_{\ell+1} = \{0\}. \]
        Compute $T_j := \spn\{S_j, \vecv_1, \ldots, \vecv_s\}$ for all $j$. Output $(T_1, \ldots, T_{\ell+1})$. \label{distribution:superspace_1}
        \item Sample a random full-rank matrix $\matM \in \bbZ_2^{(n-r-\ell) \times (n-r-\ell)}$ and random matrix $\matM' \in \bbZ_2^{(n-r-\ell) \times \ell}$. Compute $\overline{\matA} := \matA \cdot \begin{bmatrix} \matI_\ell & \matZero \\ \matM' & \matM \end{bmatrix}$ and 
            \[ T_j := \ColSpan\left(\begin{bmatrix}\overline{\matA}^{[j:\ell]} & \overline{\matA}^{[\ell+s+1:n-r]} \end{bmatrix}\right)^{\perp}. \]
        Output $(T_1, \ldots, T_{\ell+1})$. \label{distribution:superspace_2}
    \end{enumerate}
\end{lemma}
\begin{proof}
    We first prove the lemma for the case of $s = 1$. This means that Distribution \ref{distribution:superspace_1} consist of sampling any vector $\vecv \notin S_{\ell+1}$ and setting $T_j = \spn\{S_j, \vecv\}$. We claim that for fixed $S_j$, there are exactly $2^{n-r}-2^{\ell}$ possible sets of subspaces $T_1 \subseteq \ldots \subseteq T_{\ell+1}$. This follows from two facts:
    \begin{itemize}
        \item For any two vectors $\vecv, \vecw \notin S_{\ell+1}$ such that $\spn\{S_1, \vecv\} = \spn\{S_1, \vecw\}$, we have $\vecv - \vecw \in S_1 \subseteq S_j$ for all $j$ and so $\spn\{S_j, \vecv\} = \spn\{S_j, \vecw\}$. Thus, determining $T_1$ fixes all other subspaces $T_2, \ldots, T_{\ell+1}$. It is also not hard to see that among all vectors $\vecv \notin S_{\ell+1}$ consistent with a superspace $T_{j+1}$, there are exactly two possible superspaces $T_j := \spn\{S_j, \vecv\}$.
        \item $T_1$ is always a dimension $(r+1)$ superspace of $S_1$ such that $T_1 \subsetneq S_{\ell+1}$. There are at most $2^{n-r}-2^{\ell}$ such superspaces. On the other hand, there are $2^n-2^{r+\ell}$ vectors $\vecv \notin S_{\ell+1}$, and at most $|S_1| = 2^r$ will result in the same $T_1$, so there are at least $\frac{2^n-2^{r+\ell}}{2^r} = 2^{n-r}-2^{\ell}$ such superspaces. 
    \end{itemize}
    Moreover, these two facts imply that Distribution \ref{distribution:superspace_1} is the uniform distribution over all possible sets of subspaces $T_1 \subseteq \ldots \subseteq T_{\ell+1}$ such that $T_j = \spn\{S_j, \vecv\}$ for some vector $\vecv \notin S_{\ell+1}$.

    We now take a look at Distribution \ref{distribution:superspace_2}. We claim that for any set of subspaces $T_1 \subseteq \ldots \subseteq T_{\ell+1}$ produced in Distribution \ref{distribution:superspace_2}, there exists some vector $\vecv \notin S_{\ell+1}$ such that $T_j = S_j \cup (S_j+\vecv)$. Note that this is equivalent to the condition that there is some vector $\vecv \in \bigcap_{j = 1}^{\ell+1} T_j \setminus S_j$.
    
    Fix any full-rank matrix $\matM$ and matrix $\matM'$. We have that
        \[ \overline{\matA} = \matA \cdot \begin{bmatrix} \matI_\ell & \matZero \\ \matM' & \matM \end{bmatrix} = \begin{bmatrix} \matI_\ell & \matZero \\ \matB & \matC \end{bmatrix} \cdot \begin{bmatrix} \matI_\ell & \matZero \\ \matM' & \matM \end{bmatrix} = \begin{bmatrix} \matI_\ell & \matZero \\ \matB + \matC \cdot \matM' & \matC \cdot \matM \end{bmatrix}, \]
    and $T_j = \ColSpan\left(\begin{bmatrix} \overline{\matA}^{[j:\ell]} & \overline{\matA}^{[\ell+2:n-r]}\end{bmatrix}\right)^{\perp}$. This means that $\vecv \in T_j$ if and only if $\vecv^T \cdot \begin{bmatrix} \overline{\matA}^{[j:\ell]} & \overline{\matA}^{[\ell+2:n-r]}\end{bmatrix} = 0^{n-r-j}$. On the other hand,
    \begin{align*}
        \ColSpan\left(\overline{\matA}^{[j:n-r]}\right) &= \ColSpan\left(\begin{bmatrix} \matI_\ell^{[j:\ell]} & \matZero \\ (\matB + \matC \cdot \matM')^{[j:\ell]} & \matC \cdot \matM \end{bmatrix}\right) \\
        &= \ColSpan\left(\begin{bmatrix} \matI_\ell^{[j:\ell]} & \matZero \\ \matB^{[j:\ell]} & \matC \end{bmatrix}\right) = \ColSpan\left(\matA^{[j:n-r]}\right),
    \end{align*}
    since $\matM$ is a full-rank matrix. Thus, $\vecv \in S_j$ if and only if $\vecv^T \cdot \overline{\matA}^{[j:n-r]} = 0^{n-r-j+1}$. Therefore, an equivalent condition to $\vecv \in \bigcap_{j = 1}^{\ell+1} T_j \setminus S_j$ is if there exists a vector $\vecv$ such that $\vecv^T \cdot \overline{\matA} = 0^{\ell} || 1 || 0^{n-r-\ell-1}$. But since $\matA$ and $\matM$ are both full column-rank matrices, so must $\overline{\matA}$, and so the multiplication-by-$\overline{\matA}$ is surjective and thus such a vector $\vecv$ must exist. In particular, this establishes that the support of Distribution \ref{distribution:superspace_2} is a subset of the support of Distribution \ref{distribution:superspace_1}.

    We move to considering the dual spaces $T_{\ell+1}^{\perp} \subseteq \ldots \subseteq T_1^{\perp}$. Note that 
        \[ T_{\ell+1}^{\perp} = \ColSpan\left(\overline{\matA}^{[\ell+2:n-r]}\right) = \ColSpan\left(\begin{bmatrix} \matZero \\ \matC \cdot \matM^{[2:n-r-\ell]} \end{bmatrix}\right), \]
    while 
        \[ S_{\ell+1}^{\perp} = \ColSpan\left(\matA^{[\ell+1:n-r]}\right) = \ColSpan\left(\begin{bmatrix} \matZero \\ \matC \end{bmatrix}\right),\]
    and since $\matM^{[2:n-r-\ell]}$ is a random full column-rank matrix (as $\matM$ is random and full-rank), $T_{\ell+1}^{\perp}$ is a uniformly random $(n-r-\ell-1)$-dimension subspace of $S_{\ell+1}^{\perp}$; note that there are $2^{n-r-\ell}-1$ such subspaces. 

    Now let us fix any $\matM$, which thereby fixes the last $n-r-\ell$ columns of $\overline{\matA}$ and $T_{\ell+1}^{\perp}$. We claim that over the randomness of $\matM'$, $(T_1^{\perp}, \ldots, T_{\ell}^{\perp})$ take on one of $2^{\ell}$ possible sets of subspaces with equal probability. To see why this is the case, it suffices to show that for any index $j \in [\ell]$, after fixing the $(j+1)$'th through $\ell$'th column of $\matM'$ (which fixes $T_{j+1}^{\perp}, \ldots, T_{\ell}^{\perp}$), picking the $j$'th column $\vecm'_j$ of $\matM'$ at random results in one of two possible $T_j^{\perp}$'s with equal probability.

    How is $T_j^{\perp}$ defined with respect to $T_{j+1}^{\perp}$? Recall that 
    \begin{align*}
        T_j^{\perp} &:= \ColSpan\left(\begin{bmatrix} \overline{\matA}^{[j:\ell]} & \overline{\matA}^{[\ell+2:n-r]} \end{bmatrix}\right) = T_{j+1}^{\perp} \cup \left(T_{j+1}^{\perp} + \overline{\matA}^{[j]}\right) \\
        &= T_{j+1}^{\perp} \cup \left(T_{j+1}^{\perp} + \begin{bmatrix} \matI_{\ell}^{[j]} \\ \matB^{[j]} \end{bmatrix} + \begin{bmatrix} \matZero \\ \matC \cdot \vecm'_j \end{bmatrix} \right),
    \end{align*}
    where there are $2^{n-r-\ell}$ possible vectors $\matC \cdot \vecm'_j$. For any two vectors $\vecm'_{j, 1}, \vecm'_{j, 2}$, the subspaces $T_{j, 1}^{\perp}$ and $T_{j, 2}^{\perp}$ are equal if and only if 
    \begin{align*}
        \begin{bmatrix} \matZero \\ \matC \cdot (\vecm'_{j, 1}-\vecm'_{j, 2}) \end{bmatrix} &\in T_{j+1}^{\perp} \\ \Leftrightarrow \begin{bmatrix} \matZero \\ \matC \cdot (\vecm'_{j, 1}-\vecm'_{j, 2}) \end{bmatrix} &\in T_{\ell+1}^{\perp} = \ColSpan\left(\begin{bmatrix} \matZero \\ \matC \cdot \matM^{[2:n-r-\ell]} \end{bmatrix}\right) \\
        \Leftrightarrow \vecm'_{j, 1} - \vecm'_{j, 2} &\in \ColSpan\left(\matM^{[2:n-r-\ell]}\right),
    \end{align*}
    since columns $j+1$ through $\ell$ of $\overline{\matA}$ have a 1 somewhere in the first $\ell$ coordinates and $\matC$ has full column-rank. Since $\matM$ has full rank and thus $\ColSpan(\matM^{[2:n-r-\ell]})$ has dimension $n-r-\ell-1$, this partitions the vectors $\vecm'_j \in \bbZ_2^{n-r-\ell}$ evenly into exactly two cosets. Thus, $T_j^{\perp}$ takes on one of two possible values over the randomness of the $j$th column of $\matM'$.

    Thus, $(T_1^{\perp}, \ldots, T_{\ell+1}^{\perp})$ take on one of $2^{\ell} \cdot (2^{n-r-\ell}-1) = 2^{n-r}-2^{\ell}$ possible sets of subspaces with equal probability. Consequently, the output distribution $(T_1, \ldots, T_{\ell+1})$ in Distribution \ref{distribution:superspace_2} is also the uniform distribution over a support of size $2^{n-r}-2^{\ell}$. But we argued earlier that the support of Distribution \ref{distribution:superspace_2} is a subset of the support of Distribution \ref{distribution:superspace_1}, which has size $2^{n-r}-2^{\ell}$. Thus Distribution  \ref{distribution:superspace_2} is also the uniform distribution over the same $2^{n-r}-2^{\ell}$ possible subspaces, which concludes the proof for the case of $s = 1$.
    
    For $s > 1$, we assume the lemma is true for all smaller $s$. Consider the following recursive process:
    \begin{enumerate}
        \item First, sample a random matrix $\matD_1 \in \bbZ_2^{(n-r-\ell) \times \ell}$ and a random full-rank matrix $\matE_1 \in \bbZ_2^{(n-r-\ell) \times (n-r-\ell)}$. Compute the matrix $\matA_1 := \matA \cdot \begin{bmatrix} \matI_\ell & \matZero \\ \matD_1 & \matE_1 \end{bmatrix}$ and subspaces $S'_j := \ColSpan\left(\begin{bmatrix}\matA_1^{[j:\ell]} & \matA_1^{[\ell+2:n-r]} \end{bmatrix}\right)^{\perp}$.
        \item Continue by sampling a new random matrix $\matD_2 \in \bbZ_2^{(n-r-\ell-1) \times \ell}$ and random full-rank matrix $\matE_2 \in \bbZ_2^{(n-r-\ell-1) \times (n-r-\ell-1)}$. Compute the matrix 
            \[ \matA_2 := \begin{bmatrix} \matA_1^{[1:\ell]} & \matA_1^{[\ell+2:n-r]} \end{bmatrix} \cdot \begin{bmatrix} \matI_\ell & \matZero \\ \matD_2 & \matE_2 \end{bmatrix} \]
        and subspaces 
            \[ T_j := \ColSpan\left(\begin{bmatrix} \matA_2^{[j:\ell]} & \matA_2^{[\ell+s:n-r-1]} \end{bmatrix}\right)^{\perp}. \]
        Output $(T_1, \ldots, T_{\ell+1})$.
    \end{enumerate}
    After step 1, we compute the matrix 
        \[ \matA_1 = \matA \cdot \begin{bmatrix} \matI_\ell & \matZero \\ \matD_1 & \matE_1 \end{bmatrix} = \begin{bmatrix} \matI_\ell & \matZero \\ \matB & \matC \end{bmatrix} \cdot \begin{bmatrix} \matI_\ell & \matZero \\ \matD_1 & \matE_1 \end{bmatrix} = \begin{bmatrix} \matI_\ell & \matZero \\ \matB+\matC \cdot \matD_1 & \matC \cdot \matE_1 \end{bmatrix}. \]
    Thus, $S'_j = \ColSpan\left(\widetilde{\matA}_1^{[j:(n-1)-r]}\right)^{\perp}$ for 
        \[ \widetilde{\matA}_1 = \begin{bmatrix} \matA_1^{[j:\ell]} & \matA_1^{[\ell+2:n-r]} \end{bmatrix} = \begin{bmatrix} \matI_\ell & \matZero \\ \matB+\matC \cdot \matD_1 & (\matC \cdot \matE_1)^{[2:n-r-\ell]} \end{bmatrix}. \]
    Since $\matC$ and $\matE_1$ have full column-rank, so does $\matC \cdot \matE_1$ and therefore $\left(\matC \cdot \matE_1\right)^{[2:n-r-\ell]}$. By our inductive hypothesis, as $r+(s-1)+\ell \leq n-1$, we have that the distribution of $(T_1, \ldots, T_{\ell+1})$ can be equivalently sampled as follows:
    \begin{enumerate}
        \item Sample a random vector $\vecv_1$ such that $\vecv_1 \notin S_{\ell+1}$. Compute $S'_j := \spn\{S_j, \vecv_1\}$ for all $j$.
        \item Sample $s-1$ random linearly independent vectors $\vecv_2, \ldots, \vecv_s$ such that $\spn\{\vecv_2, \ldots, \vecv_s\} \cap S'_{\ell+1} = \{0\}$. Compute $T_j := \spn\{S'_j, \vecv_2, \ldots, \vecv_s\}$ for all $j$. Output $(T_1, \ldots, T_{\ell+1})$.
    \end{enumerate}
    It is easy to see that this is statistically identical to sampling $\vecv_1, \ldots, \vecv_s$ simultaneously, as in Distribution \ref{distribution:superspace_1}. What remains is to show that the distribution produced by the recursive process is also statistically equivalent to Distribution \ref{distribution:superspace_2}.

    Let us compute the matrix $\matA_2$:
    \begin{align*}
        \matA_2 &= \begin{bmatrix} \matA_1^{[1:\ell]} & \matA_1^{[\ell+2:n-r]} \end{bmatrix} \cdot \begin{bmatrix} \matI_\ell & \matZero \\ \matD_2 & \matE_2 \end{bmatrix} \\
        &= \begin{bmatrix} \matI_\ell & \matZero \\ \matB+\matC \cdot \matD_1 & (\matC \cdot \matE_1)^{[2:n-r-\ell]} \end{bmatrix} \cdot \begin{bmatrix} \matI_\ell & \matZero \\ \matD_2 & \matE_2 \end{bmatrix} \\
        &= \begin{bmatrix} \matI_\ell & \matZero \\ \matB + \matC \cdot \matD_1 + (\matC \cdot \matE_1)^{[2:n-r-\ell]} \cdot \matD_2 & (\matC \cdot \matE_1)^{[2:n-r-\ell]} \cdot \matE_2 \end{bmatrix}.
    \end{align*}
    Observe that $\matA_2$ is closely related to the following matrix:
    \begin{align*}
        \widetilde{\matA}_2 &:= \matA \cdot \begin{bmatrix} \matI_\ell & \matZero \\ \matD_1 & \matE_1 \end{bmatrix} \cdot \begin{bmatrix} \matI_\ell & \matZero & \matZero \\ \matZero & \matOne & \matZero \\ \matD_2 & \matZero & \matE_2 \end{bmatrix} = \begin{bmatrix} \matI_\ell & \matZero \\ \matB+\matC \cdot \matD_1 & \matC \cdot \matE_1 \end{bmatrix} \cdot \begin{bmatrix} \matI_\ell & \matZero & \matZero \\ \matZero & \matOne & \matZero \\ \matD_2 & \matZero & \matE_2 \end{bmatrix} \\
        &= \begin{bmatrix} \matI_\ell & \matZero & \matZero \\ \matB + \matC \cdot \matD_1 + (\matC \cdot \matE_1)^{[2:n-r-\ell]} \cdot \matD_2 & (\matC \cdot \matE_1)^{[1]} & (\matC \cdot \matE_1)^{[2:n-r-\ell]} \cdot \matE_2 \end{bmatrix}.
    \end{align*}
    Thus, $\begin{bmatrix} \widetilde{\matA}_2^{[j:\ell]} & \widetilde{\matA}_2^{[\ell+s+1:n-r]} \end{bmatrix} = \begin{bmatrix} \matA_2^{[j:\ell]} & \matA_2^{[\ell+s:n-r-1]} \end{bmatrix}$. So we can equivalently sample $\matD_1, \matD_2, \matE_1, \matE_2$ together and compute $\widetilde{\matA}_2$ accordingly before defining $T_j := \ColSpan\left(\begin{bmatrix} \widetilde{\matA}_2^{[j:\ell]} & \widetilde{\matA}_2^{[\ell+s+1:n-r]} \end{bmatrix}\right)^{\perp}$. 

    If we now compute the following matrix:
        \[ \matF := \begin{bmatrix} \matI_\ell & \matZero \\ \matD_1 & \matE_1 \end{bmatrix} \cdot \begin{bmatrix} \matI_\ell & \matZero & \matZero \\ \matZero & \matOne & \matZero \\ \matD_2 & \matZero & \matE_2 \end{bmatrix} = \begin{bmatrix} \matI_\ell & \matZero \\ \matD_1 + \matE_1^{[2:n-r-\ell]} \cdot \matD_2 & \matE_1 \cdot \begin{bmatrix} \matOne & \matZero \\ \matZero & \matE_2 \end{bmatrix} \end{bmatrix}, \]
    we observe that (1) $\matE_2$ is full-rank and $\matE_1$ is a random full-rank matrix and (2) for each fixed $\matE_1$ and $\matD_2$, $\matD_1$ is a random matrix. Thus, $\matF$ has the same distribution as $\begin{bmatrix} \matI_\ell & \matZero \\ \matM' & \matM \end{bmatrix}$ in Distribution \ref{distribution:superspace_2}, and since $\widetilde{\matA}_2 = \matA \cdot \matF$ and $T_j := \ColSpan\left(\begin{bmatrix} \widetilde{\matA}_2^{[j:\ell]} & \widetilde{\matA}_2^{[\ell+s+1:n-r]} \end{bmatrix}\right)^{\perp}$, this distribution is equivalent to Distribution \ref{distribution:superspace_2}.
\end{proof}

We will also use variations of the information-theoretical and cryptographic subspace-hiding lemmas of \cite{SZ25}, stating proofs for completeness; many of these proofs will closely track the proof outlines of \cite{SZ25}.

\begin{lemma}[Adapted from \cite{SZ25}]\label{lemma:subspace_hiding_info}
    Let $n, r, s, \ell \in \bbN^{+}$ such that $r+s+\ell \leq n$, and let $S_1 \subseteq \ldots \subseteq S_{\ell+1} \subseteq \bbZ_2^n$ be subspaces where $S_j$ has dimension $r+j-1$. Let $\calS_s$ be the distribution over subspaces $T_j$ defined as follows: pick $s$ random linearly independent vectors $v_1, \ldots, v_s \in \bbZ_2^n$ such that $\spn\{v_1, \ldots, v_s\} \cap S_{\ell+1} = \{0\}$ and let $T_j := \spn\{S_j, v_1, \ldots, v_s\}$ for all $j$. For any subspace $S' \subseteq \bbZ_2^n$ let $\calO_{S'}$ be the oracle that checks membership in $S'$ (outputs 1 if and only if the input is inside $S'$).
    
    Then, for every oracle-aided quantum algorithm $\calA$ making at most $q$ quantum queries, we have the following indistinguishability over oracle distributions:
        \[ \{ \calO_{S_1}, \ldots, \calO_{S_{\ell+1}} \} \approx_{O\left(\frac{q \cdot \ell \cdot s}{\sqrt{2^{n-r-s-\ell}}}\right)} \{ \calO_{T_1}, \ldots, \calO_{T_{\ell+1}} : (T_1, \ldots, T_{\ell+1}) \gets \calS_s \}. \]
\end{lemma}
\begin{proof}
    We prove the claim by a hybrid argument, increasing the dimension of the random super-spaces $T_j$ one at a time.

    First, note that since $S_1 \subseteq \ldots \subseteq S_{\ell+1}$, there is some full-column-rank matrix $\matA \in \bbZ_2^{n \times (n-r)}$ such that $\calO_{S_j}$ can be described as accepting $\vecx \in \bbZ_2^n$ if and only if $\vecx^T \cdot \matA^{[j:n-r]} = 0^{n-r-j+1}$; fix any such matrix $\matA$. 

    We now sample a random matrix $\matB \in \bbZ_2^{(n-r-\ell) \times \ell}$ and compute $\matC := \matA \cdot \begin{bmatrix} \matI_{\ell} & \matZero \\ \matB & \matI_{n-r-\ell} \end{bmatrix}$. We switch to the oracles $\calO'_{S_j}$ which accept $\vecx \in \bbZ_2^n$ if and only if $\vecx^T \cdot \matC^{[j:n-r]} = 0^{n-r-j+1}$. As shown in the proof of Lemma \ref{lemma:sampling_superspaces}, $(\calO_{S_1}, \ldots, \calO_{S_{\ell+1}})$ and $(\calO_{S'_1}, \ldots, \calO_{S'_{\ell+1}})$ are functionally identical for all $\matB$, so this change is indistinguishable. We can also indistinguishably switch to the oracles $\calO''_{S_j}$ which accept $\vecx \in \bbZ_2^n$ if and only if $\vecx^T \cdot \matC^{[\ell+1:n-r]} = 0^{n-r-\ell}$ and $\vecx^T \cdot \matC^{[j:\ell]} = 0^{\ell-j+1}$.
    
    Next, we additionally sample a uniformly random vector $\vecz \in \bbZ_2^{n-r-\ell} \setminus \{0\}$. Consider the oracles $\calO_{S'_j}$ which accept $\vecx \in \bbZ_2^n$ if and only if (1) $\vecx^T \cdot \matA^{[\ell+1:n-r]} = 0^{n-r-\ell}$ or $\vecx^T \cdot \matA^{[\ell+1:n-r]} = \vecz$ and (2) $\vecx^T \cdot \matA^{[j:\ell]} = 0^{\ell-j+1}$. By a simulation argument, it is easy to see that we are essentially giving $\ell+1$ copies of an oracle which in one case accepts $\vecx$ if $\vecx^T \cdot \matA^{[\ell+1:n-r]} = 0^{n-r-\ell}$ and in the other accepts if $\vecx^T \cdot \matA^{[\ell+1:n-r]} = 0^{n-r-\ell}$ or $\vecx^T \cdot \matA^{[\ell+1:n-r]} = \vecz$. By the randomness of $\vecz$ and a standard quantum lower bound, we have that 
        \[ \{ \calO''_{S_1}, \ldots, \calO''_{S_{\ell+1}} \} \approx_{O\left(\frac{q \cdot \ell}{\sqrt{2^{n-r-\ell}}}\right)} \{ \calO_{S'_1}, \ldots, \calO_{S'_{\ell+1}} \}. \]
    Finally, we move from $\calO_{S'_j}$ to the functionally equivalent oracles $\calO'_{S'_j}$ which accept $\vecx \in \bbZ_2^n$ if and only if $\vecx^T \cdot \matA^{[j:n-r]} = 0^{n-r-j+1}$ or $\vecx^T \cdot \matA^{[j:n-r]} = 0^{\ell-j+1} || \vecz$; this change is indistinguishable.
    
    By Lemma \ref{lemma:sampling_superspaces}, $\calO'_{S'_j}$ are precisely membership oracles for the distribution of superspaces produced by adding a random vector outside $S_{\ell+1}$ to all $S_j$'s, thereby proving the claim for $s = 1$. For arbitrary $s$, iterating the argument $s$ times in total gives a distinguishing advantage of
        \[ \sum_{j = 1}^{s} O\left(\frac{q \cdot \ell}{\sqrt{2^{n-r-\ell-(j-1)}}}\right) = O\left(\frac{q \cdot \ell \cdot s}{\sqrt{2^{n-r-s-\ell}}}\right), \]
    as desired.
\end{proof}

A straightforward combination of a simulation argument and a hybrid argument gives the following corollary of Lemma \ref{lemma:subspace_hiding_info}.
\begin{corollary}\label{corollary:superspace_indistinguishability_info}
    Let $n, r, s, \ell \in \bbN^{+}$ such that $r+s+\ell \leq n$, and let $S_1 \subseteq \ldots \subseteq S_{\ell+1} \subseteq \bbZ_2^n$ as in Lemma \ref{lemma:subspace_hiding_info}. Define the distribution $\calS_s$ and membership oracles $\calO_{S'}$ for subspaces $S'$ similarly.
    
    Then, for every oracle-aided quantum algorithm $\calA$ making at most $q$ quantum queries, we have the following indistinguishability over oracle distributions:
    \begin{align*}
        \{ (\calO_{T_{1, 1}}, \ldots, \calO_{T_{1, \ell+1}}), \ldots, (\calO_{T_{m, 1}}, \ldots, \calO_{T_{m, \ell+1}}) : \forall i \in [m], (T_{i, 1}, \ldots, T_{i, \ell+1}) \gets \calS_s \}& \\
        \approx_{O\left(\frac{q \cdot \ell \cdot m \cdot s}{\sqrt{2^{n-r-s-\ell}}}\right)} \{ (\calO^1_{T_1}, \ldots, \calO^1_{T_{\ell+1}}), \ldots, (\calO^m_{T_1}, \ldots, \calO^m_{T_{\ell+1}}) : (T_1, \ldots, T_{\ell+1}) \gets \calS_s \}&.
    \end{align*}
\end{corollary}

Using Corollary \ref{corollary:superspace_indistinguishability_info}, we can now show the following subspace anti-concentration result. 
\begin{lemma}\label{lemma:info_anti_concentration}
    Let $n, r, s, \ell \in \bbN^{+}$ such that $r+s+\ell \leq n$, and let $S_1 \subseteq \ldots \subseteq S_{\ell+1} \subseteq \bbZ_2^n$ as in Lemma \ref{lemma:subspace_hiding_info}. Define the distribution $\calS_s$  and membership oracles $\calO_{S'}$ for subspaces $S'$ similarly.
    
    Assume there is an oracle-aided quantum algorithm $\calA$ making at most $q$ quantum queries and outputting a vector $u \in \bbZ_2^n$ at the end of its execution, such that
        \[ \Pr[\calA^{\calO_{T_1}, \ldots, \calO_{T_{\ell+1}}} \in (T_1^{\perp} \setminus \{0\}) : (T_1, \ldots, T_{\ell+1}) \gets \calS_s] \geq \eps. \]
    Denote $t := n-r-s, m := \frac{n(t+1)}{\eps}$ and assume (1) $\frac{t}{\eps} \cdot \left(2^{t-s}+2^{-t+\ell}\right) \leq o(1)$ and (2) $\frac{q \cdot \ell \cdot m^2 \cdot s}{2^{t/2}} \leq o(1)$. Then,
        \[ \Pr[\calA^{\calO_{T_1}, \ldots, \calO_{T_{\ell+1}}} \in (S_1^{\perp} \setminus T_1^{\perp}) : (T_1, \ldots, T_{\ell+1}) \gets \calS_s] \geq \frac{\eps}{16n \cdot (t+1)}. \]
\end{lemma}
\begin{proof}
    Our proof follows \cite{SZ25} closely, but with a few modifications. We first construct a reduction $\calB$ that uses $\calA$ as a subroutine; $\calB$ has access to $m := n(t+1)/\eps$ samples of oracles $\left(\calO^{(1, 1)}, \ldots, \calO^{(1, \ell+1)}\right)$, $\ldots$, $\left(\calO^{(m, 1)}, \ldots, \calO^{(m, \ell+1)}\right)$, for $t := n-r-s$. $\calB$ executes $\calA^{\calO^{(i, 1)}, \ldots, \calO^{(i, \ell+1)}}$ for every $i \in [m]$ and obtains $m$ vectors $\{u_1, \ldots, u_m\}$, before taking only the vectors $\{v_1, \ldots, v_j \}$ that are inside $S_1^{\perp}$. Finally, it computes $D := \dim(\mathsf{span}\{v_1, \ldots, v_j\})$; note that $\calB$ only makes $q \cdot m$ queries.

    Consider the following oracle distribution $\calD_1$: sample $(T_{i, 1}, \ldots, T_{i, \ell+1}) \gets \calS_s$ i.i.d. $m$ times and give $\calB$ access to all of the associated membership oracles $(\calO_{T_{i, 1}}, \ldots, \calO_{T_{i, \ell+1}})$. Our goal is to show that $D \geq t+1$ with high probability over $\calD_1$.

    For every $i \in [m]$ we define the probability $p_i$ as follows. For indices $i \in [n/\eps]$ in the first bucket, $p_i$ is the probability that given access to $\calO_{T_{i, 1}}, \ldots, \calO_{T_{i, \ell+1}}$, the output of $\calA$ is a vector $u_i \in T_{i, 1}^{\perp} \setminus \{0\} \subseteq S_1^{\perp}$, in which case we call the $i$'th execution \emph{successful}. We denote by $T_{(1, 1)}, \ldots, T_{(1, \ell+1)}$ the first set of subspaces in the first bucket where a successful execution happens (and set $\left(T_{(1, 1)}, \ldots, T_{(1, \ell+1)}\right) := (\bot, \ldots, \bot)$ if no execution in the bucket is successful). For any $i$ inside buckets $2 \leq b \leq t+1$, let $p_i$ be the probability that (1) $u_i \in T_{i, 1}^{\perp} \setminus \{0\}$ and (2) the intersection between $T_{i, 1}^{\perp}$ and each of $T_{(1, 1)}^{\perp}, \ldots, T_{(b-1, 1)}^{\perp}$ contains only $\{0\}$. Similarly to the first bucket, we denote by $T_{(b, 1)}, \ldots, T_{(b, \ell+1)}$ the first set of subspaces in bucket $b$ with a successful execution. Observe that a successful execution in all buckets immediately implies that $\calB$ finds at least $t+1$ linearly independent vectors which are in $S_1^{\perp}$ and thus that $D \geq t+1$.
    
    Consider the probability $p'$ defined as follows. Let $T_{1, 1}, \ldots, T_{t, 1}$ be any $t$ subspaces, each of dimension $r+s$, and hence their duals $T_{1, 1}^{\perp}, \ldots, T_{t, 1}^{\perp}$ have dimension $t$. $p'_{T_{1, 1}, \ldots, T_{t, 1}}$ is the probability that (1) when sampling $T_1^{\perp}$, the intersection of $T_1^{\perp}$ with each of the $t$ dual subspaces $T_{1, 1}^{\perp}, \ldots, T_{t, 1}^{\perp}$ is only the zero vector and (2) the output of $\calA$ is inside $T_1^{\perp} \setminus \{0\}$. $p'$ is defined to be the minimal probability taken over all possible choices of $t$ subspaces $T_{1, 1}, \ldots, T_{t, 1}$. It is easy to see that $p' \leq p_i$ for all $i$ and thus it suffices to give a lower bound on $p'$. 
    
    We first show that the probability for a nontrivial intersection is small. Fix any index $\idx$ and consider the probability that $T_1^{\perp}$ intersects with $T_{\idx, 1}^{\perp}$. How is $T_1^{\perp}$ (and in particular $T_1$) sampled? A random set of linearly independent vectors $v_1, \ldots, v_s$ is chosen such that $\spn\{v_1, \ldots, v_s\} \cap S_{\ell+1} = \{0\}$, and $T_1$ is computed as $\spn\{S_1, v_1, \ldots, v_s\}$. Of course, one alternative way to think of this process is to first pick random linearly independent vectors $v_1, \ldots, v_s$ such that $\spn\{v_1, \ldots, v_s\} \cap S_1 = \{0\}$, compute $T_1 := \spn\{S_1, v_1, \ldots, v_s\}$, and resample if $\spn\{v_1, \ldots, v_s\} \cap S_{\ell+1} \neq \{0\}$. In other words, if we let $A$ denote the event that a random dimension $(r+s)$-superspace $T'_1$ of $S_1$ can be written as $T'_1 = \spn\{S_1, v'_1, \ldots, v'_s\}$ for $v'_1, \ldots, v'_s \in \bbZ_2^n$ such that $\spn\{v'_1, \ldots, v'_s\} \cap S_{\ell+1} = \{0\}$, and $B$ the event that a random superspace $T'_1$ of $S_1$ satisfies $(T'_1)^{\perp} \cap T_{\idx, 1}^{\perp} = \{0\}$, then we want a lower bound on $\Pr[B \mid A]$.

    We begin by bounding $\Pr[\neg A]$. We can think of the process of picking a random superspace of $S_1$ as a $s$-step process of picking vectors $\vecv_z \in \bbZ_2^n$ for $z = 1$ to $s$, where each time, we select $\vecv_z$ such that $\vecv_z \notin \spn\{S_1, \vecv_1, \ldots, \vecv_{z-1}\}$. We observe that the probability that $A$ does not occur is the sum over $z \in [s]$ of the following event: the $z$'th step is the first step such that $A$ becomes false. This is at most the sum over $z \in [s]$ of the probability that in the $z$'th step, (1) $\vecv_z \in \spn\{S_{\ell+1}, \vecv_1, \ldots, \vecv_{z-1}\}$, conditioned on (2) $\vecv_k \notin \spn\{S_{\ell+1}, \vecv_1, \ldots, \vecv_{k-1}\}$ for all $k < z$ and (3) $\vecv_z \notin \spn\{S_1, \vecv_1, \ldots, \vecv_{z-1}\}$. Observe that given condition (2), there are only $2^n-2^{r+z-1}$ vectors which also satisfy condition (3). On the other hand, given condition (2), only $2^{r+\ell+z-1}-2^{r+z-1}$ vectors also satisfy conditions (1) and (3). Thus, the probability for a given $z$ is at most
        \[ \frac{2^{r+\ell+z-1}-2^{r+z-1}}{2^n-2^{r+z-1}} = \frac{2^\ell-1}{2^{n-r-z+1}-1} < \frac{2^\ell}{2^{n-r-z+1}} = 2^{-(n-\ell-r-z+1)}. \]
    We therefore conclude that $\Pr[\neg A]$ is at most
        \[ \sum_{z=1}^{s} 2^{-(n-\ell-r-z+1)} < 2^{-(n-\ell-r-s)} = 2^{-t+\ell}. \]

    Now, let us bound $\Pr[\neg B]$. Recall that if $T'_1$ is a random superspace of $S_1$, then $(T'_1)^{\perp}$ is a random dimension $t$ subspace of $S_1^{\perp}$, which has dimension $n-r$. On the other hand, $T_{\idx, 1}^{\perp}$ is a fixed dimension $t$ subspace of $S_1^{\perp}$. We want to determine the probability that $(T'_1)^{\perp} \cap T_{\idx, 1}^{\perp} = \{0\}$. Like before, we can now think of the process of sampling a basis for $(T'_1)^{\perp}$ as a $t$-step process, where in the $z$'th step, a vector $\vecv_z \in S_1^{\perp}$ is chosen such that $\vecv_z \notin \spn\{\vecv_1, \ldots, \vecv_{z-1}\}$. We again bound the sum over $z \in [t]$ of the following event: conditioned on (1) $\vecv_k \notin \spn\{T_{\idx, 1}^{\perp}, \vecv_1, \ldots, \vecv_{k-1}\}$ for all $k < z$ and (2) $\vecv_z \notin \spn\{\vecv_1, \ldots, \vecv_{z-1}\}$, (3) $\vecv_z \in \spn\{T_{\idx, 1}^{\perp}, \vecv_1, \ldots, \vecv_{z-1}\}$. Observe that condition (1) obviously implies that $\vecv_1, \ldots, \vecv_{z-1}$ are linearly independent and thus given condition (1), there are precisely $2^{n-r}-2^{z-1}$ vectors in $S_1^{\perp}$ which satisfy condition (2). On the other hand, given condition (1), there are only $2^{t+z-1}-2^{z-1}$ vectors which satisfy both conditions (2) and (3). Thus, the probability for a given $z$ is at most
        \[ \frac{2^{t+z-1}-2^{z-1}}{2^{n-r}-2^{z-1}} = \frac{2^t-1}{2^{n-r-z+1}-1} < \frac{2^t}{2^{n-r-z+1}} = 2^{-(n-r-t-z+1)}. \]
    We therefore conclude that $\Pr[\neg B]$ is at most 
        \[ \sum_{z=1}^t 2^{-(n-r-t-z+1)} < 2^{t-s}. \]
    
    Thus, we have that
        \[ \Pr[B \mid A] \geq \Pr[B]-\Pr[\neg A] = 1-\Pr[\neg B]-\Pr[\neg A] \geq 1-2^{t-s}-2^{-t+\ell}. \]
    By a union bound, this means the probability that there exists $\idx$ such that $T_1^{\perp}$ has a nontrivial intersection with $T_{\idx, 1}^{\perp}$ is at most $t(2^{t-s}+2^{-t+\ell})$.
    
    If we now let $C$ denote the event that $\calA$ outputs a vector in $T_1^{\perp} \setminus \{0\}$ and $D$ the event that $T_1^{\perp} \cap T_{\idx, 1}^{\perp} = \{0\}$ for all $\idx \in [t]$, we have that 
        \[ \Pr[C \mid D] \geq \Pr[C]-\Pr[\neg D] \geq \eps - t(2^{t-s}+2^{-t+\ell}). \]
    As $\frac{t}{\eps}(2^{t-s}+2^{-t+\ell}) \leq o(1)$ by assumption, 
    \begin{align*}
        p' &= \Pr[C \land D] = \Pr[C \mid D] \cdot \Pr[D] \\
        &\geq \left(\eps - t\left(2^{t-s}+2^{-t+\ell}\right)\right) \cdot \left(1-t\left(2^{t-s}+2^{-t+\ell}\right)\right) \\
        &\geq \eps/4.
    \end{align*}
    Since every bucket has $n/\eps$ indices, each of which succeeds with probability at least $\eps/4$, the overall success probability in a bucket is at least $1-e^{-\Omega(n)}$. By a union bound, $D \geq t+1$ occurs with probability at least $1-(t+1)e^{-\Omega(n)} = 1-e^{-\Omega(n)}$.

    We now consider a different oracle distribution $\calD_2$: sample $(T_1, \ldots, T_{\ell+1}) \gets \calS_s$ once, then allow an $m$-oracle access to it. Note that $\calB$ is a $(q \cdot m)$-query algorithm and thus by Corollary \ref{corollary:superspace_indistinguishability_info}, we have that
        \[ \calD_1 \approx_{O\left(\frac{q \cdot \ell \cdot m^2 \cdot s}{\sqrt{2^{n-r-s-\ell}}}\right)} \calD_2. \]
    Thus, when we execute $\calB$ on a sample from $\calD_2$, with probability at least 
        \[ 1-e^{-\Omega(n)}-O\left(\frac{q \cdot \ell \cdot m^2 \cdot s}{\sqrt{2^{n-r-s-\ell}}}\right) \geq 1-O\left(\frac{q \cdot \ell \cdot m^2 \cdot s}{\sqrt{2^{n-r-s-\ell}}}\right), \]
    $\calB$ will compute $D \geq t+1$.
    As we assumed that $\frac{q \cdot \ell \cdot m^2 \cdot s}{\sqrt{2^{n-r-s-\ell}}} = O(\frac{q \cdot \ell \cdot m^2 \cdot s}{2^{t/2}}) \leq o(1)$, this means that $\calB$ computes $D \geq t+1$ with probability at least $1/2$ over $\calD_2$. By a simple averaging argument, this means that with probability at least $1/4$ over the sampling of $T_1, \ldots, T_{\ell+1}$, $\calB$ computes $D \geq t+1$ when given access to this set of superspaces with probability at least $1/4$; we refer to such sets of superspaces as ``good'' sets. But when $D \geq t+1$, this means that $\calB$ has computed at least $t+1$ linearly independent vectors in $S_1^{\perp} \setminus \{0\}$ despite the fact that $T_1^{\perp}$ has dimension $t$, and thus $\calB$ produces at least one vector in $S_1^{\perp} \setminus T_1^{\perp}$. Since $\calB$ uses $m$ i.i.d executions of $\calA$, for every $(T_1, \ldots, T_{\ell+1})$ which is ``good'', when we prepare an oracle access to $(T_1, \ldots, T_{\ell+1})$ and execute $\calA$, we will get $\calA^{\calO_{T_1}, \ldots, \calO_{T_{\ell+1}}} \in S_1^{\perp} \setminus T_1^{\perp}$ with probability at least $\frac{1}{4m}$.

    Thus, over random $(T_1, \ldots, T_{\ell+1})$, we have that 
        \[ \Pr[\calA^{\calO_{T_1}, \ldots, \calO_{T_{\ell+1}}} \in S_1^{\perp} \setminus T_1^{\perp} : (T_1, \ldots, T_{\ell+1}) \gets \calS_s] \geq \frac{1}{4} \cdot \frac{1}{4m} = \frac{\eps}{16 \cdot n \cdot (t+1)}, \]
    as desired.
\end{proof}

We now consider a cryptographic version of Lemma \ref{lemma:subspace_hiding_info}.
\begin{lemma}\label{lemma:subspace_hiding_crypto}
    Let $n, r, s, \ell \in \bbN^{+}$ such that $r+s+\ell \leq n$, and let $S_1 \subseteq \ldots \subseteq S_{\ell+1} \subseteq \bbZ_2^n$ be subspaces where $S_j$ has dimension $r+j-1$. Let $\calS_s$ be the distribution over subspaces $T_j$ defined as follows: pick $s$ random linearly independent vectors $v_1, \ldots, v_s \in \bbZ_2^n$ such that $\spn\{v_1, \ldots, v_s\} \cap S_{\ell+1} = \{0\}$ and let $T_j := \spn\{S_j, v_1, \ldots, v_s\}$ for all $j$. For any subspace $S' \subseteq Z_2^n$ let $C_{S'}$ some canonical classical circuit that checks membership in $S'$, say by Gaussian elimination. Let $\IO$ be an indistinguishability obfuscation scheme that is $(f(\secp), \eps(\secp))$-secure, and assume that $(f(\secp), \eps(\secp))$-secure injective one-way functions exist, where $f$ and $1/\eps$ are sub-exponential functions.
    
    Then, for every security parameter $\secp$ such that $\secp \leq n-r-s-\ell$ and sufficiently large $p := p(\secp)$ polynomial in the security parameter, we have the following indistinguishability:
    \begin{align*}
        &\{ \sfO_{S_1}, \ldots, \sfO_{S_{\ell+1}}: \sfO_{S_j} \gets \IO(1^{\secp}, 1^p, C_{S_j}) \} \approx_{(f(\secp)-\ell \cdot \poly(\secp), \poly(n) \cdot \eps(\secp))} \\
        &\{ \sfO_{T_1}, \ldots, \sfO_{T_{\ell+1}}: (T_1, \ldots, T_{\ell+1}) \gets \calS_s,  \sfO_{T_j} \gets \IO(1^{\secp}, 1^p, C_{T_j}) \}.
    \end{align*}
\end{lemma}
\begin{proof}
    The proof of Lemma \ref{lemma:subspace_hiding_crypto} is a simple extension of the proof of Lemma \ref{lemma:subspace_hiding_info} to the cryptographic setting using the same proof outline as in \cite{Zha21b}. We will use an induction argument, first considering the case where $s = 1$.
    
    Consider the following series of hybrids:
    \begin{itemize}
        \item $\mathsf{Hyb}_0$: In this hybrid, $\calA$ receives $\sfO_{S_1}, \ldots, \sfO_{S_\ell}$. All programs are appropriately padded before obfuscating so that all the programs received by $\calA$ in the following hybrids have the same length.
        \item $\mathsf{Hyb}_1$: Since $S_1 \subseteq \ldots \subseteq S_{\ell+1}$, we also have that $S_{\ell+1}^{\perp} \subseteq \ldots \subseteq S_1^{\perp}$. Fix any full column-rank matrix $\matA \in \bbZ_2^{n \times (n-r)}$ such that $S_j^{\perp} = \ColSpan(\matA^{[j:n-r]})$; such a matrix can be computed by Gaussian elimination. Sample a random matrix $\matB \in \bbZ_2^{(n-r-\ell) \times \ell}$ and compute $\matC := \matA \cdot \begin{bmatrix} \matI_{\ell} & \matZero \\ \matB & \matI_{n-r-\ell} \end{bmatrix}$. Let $\hat{P}$ be an obfuscation under $\IO$ of the program $Z$ that always outputs $0$ on inputs in $\bbZ_2^{n-r-\ell}$. Then $\sfO_{S_j}$ is the obfuscation under $\IO$ of the function
        \begin{align*}
            Q_j(\vecx) := \begin{cases} 1 & \text{if } \vecx^T \cdot \matC^{[j:n-r]} = 0^{n-r-j+1} \\
            1 & \text{if } \vecx^T \cdot \matC^{[j:\ell]} = 0^{\ell-j+1} \text{ and } \hat{P}\left(\vecx^T \cdot \matC^{[\ell+1:n-r]}\right) = 1 \\
            0 & \text{otherwise.} \end{cases}
        \end{align*}
        As shown in the proof of Lemma \ref{lemma:sampling_superspaces}, the condition $\vecx^T \cdot \matC^{[j:n-r]} = 0^{n-r-j+1}$ is equivalent to checking that $\vecx \in S_j$ for all matrices $\matB$. Since $\hat{P}$ always outputs $0$, $Q_j$ is a membership program for $S_j$ and thus is functionally equivalent to $C_{S_j}$. Thus $\mathsf{Hyb}_0$ and $\mathsf{Hyb}_1$ are $(f(\secp), (\ell+1) \cdot \eps(\secp))$-computationally indistinguishable by the security of the $\IO$.
        \item $\mathsf{Hyb}_2$: Let $\hat{P}$ now be the obfuscation under $\IO$ of the function
        \begin{align*}
            P_y(\vecz) := \begin{cases} 1 & \text{if } \OWF(\vecz) = y \\
            0 & \text{otherwise,} \end{cases}
        \end{align*}
        where $\OWF$ is an injective one-way function and $y = \OWF(\vecz^{*})$ for a random $\vecz^{*} \in \bbZ_2^{n-r-\ell}$. Since $n-r-\ell \geq \secp$, we can invoke the security of $\OWF$. Specifically, the only point on which $Z$ and $P_y$ differ is $\vecz^{*}$, and finding $\vecz^{*}$ requires inverting $\OWF$; thus, $\IO$ is also a differing-inputs obfuscator for these circuits \cite{BCP14}. Since the reduction will need to embed either an obfuscation of $P_y$ or $Z$ in $Q_1, \ldots, Q_{\ell+1}$ before running $\calA$, it has overhead $\ell \cdot \poly(\secp)$. Therefore, $\mathsf{Hyb}_1$ and $\mathsf{Hyb}_2$ are $(f(\secp)-\ell \cdot \poly(\secp), O(\ell \cdot \eps(\secp)))$-computationally indistinguishable by the security of $\OWF$ and $\IO$. Note that $Q_j(\vecx)$ decides membership in the subspace $S'_j$ of vectors $\vecx$ such that $\vecx^T \cdot \matC^{[j:n-r]} \in \spn\{0^{\ell-j+1}||\vecz^{*}\}$.
        \item $\mathsf{Hyb}_3$: In this hybrid, a random $\matB$ and $\vecz^{*}$ are chosen which fixes $S'_1, \ldots, S'_{\ell+1}$; the canonical membership programs $C_{S'_1}, \ldots, C_{S'_{\ell+1}}$ are then obfuscated. Since all programs in both hybrids still decide membership in $S'_1, \ldots, S'_{\ell+1}$, the programs being obfuscated are functionally equivalent. Thus, $\mathsf{Hyb}_2$ and $\mathsf{Hyb}_3$ are $(f(\secp), (\ell+1) \cdot \eps(\secp))$-computationally indistinguishable by the security of the $\IO$. 
        \item $\mathsf{Hyb}_4$: Now $\vecz^{*}$ is chosen at random, but restricted to being nonzero. Since $\secp \leq n-r-\ell$, $\mathsf{Hyb}_3$ and $\mathsf{Hyb}_4$ are $O\left(\frac{1}{2^{\secp}}\right)$-statistically close. Observe that by Lemma \ref{lemma:sampling_superspaces}, $\mathsf{Hyb}_4$ now contains precisely the obfuscations of $C_{T_j}$.
    \end{itemize}
    The proof follows by iterating over $s$ and observing that $\secp \leq n-r-\ell-s$ and $\ell, s \leq n$.
\end{proof}
We remark that the lemma also trivially extends (with the same base security parameter $\secp$) to the setting where we choose stronger parameters $\kappa \geq n-r-s-\ell, \poly(\kappa) \geq p(n-r-s-\ell)$ to obfuscate the circuits.

We will also need cryptographic versions of Corollary \ref{corollary:superspace_indistinguishability_info} and Lemma \ref{lemma:info_anti_concentration}.
\begin{lemma}\label{lemma:double_obfuscation}
    Let $n, r, s, \ell \in \bbN^{+}$ such that $r+s+\ell \leq n$, and let $S_1 \subseteq \ldots \subseteq S_{\ell+1} \subseteq \bbZ_2^n$ be the same subspaces as in Lemma \ref{lemma:subspace_hiding_crypto}. Define the distribution $\calS_s$ and membership circuits $C_{S'}$ for subspaces $S'$ similarly. Let $\IO$ an indistinguishability obfuscation scheme that is $(f(\secp), \eps(\secp))$-secure, and assume that $(f(\secp), \eps(\secp))$-secure injective one-way functions exist, where $f$ and $1/\eps$ are sub-exponential functions.
    
    Then, for every security parameter $\secp$ such that $\secp \leq n-r-s-\ell$ and sufficiently large $p := p(\secp)$ polynomial in the security parameter, we have the following indistinguishability:
    \begin{align*}
        \{ (\sfO_{S_1}^1, \ldots, \sfO_{S_{\ell+1}}^1), \ldots, (\sfO_{S_1}^m, \ldots, \sfO_{S_{\ell+1}}^m):& \forall i \in [m], \sfO_{S_j}^i \gets \IO(1^{\secp}, 1^p, C_{S_j}) \} \\
        \approx&_{(f(\secp) - m \cdot n \cdot \poly(\secp), m \cdot \poly(n) \cdot \eps(\secp))} \\
        \{ (\sfO_{T_1}^1, \ldots, \sfO_{T_{\ell+1}}^1), \ldots, (\sfO_{T_1}^m, \ldots, \sfO_{T_{\ell+1}}^m) :& (T_1, \ldots, T_{\ell+1}) \gets \calS_s, \\
        &\forall i \in [m], \sfO_{T_j}^i \gets \IO(1^{\secp}, 1^p, C_{T_j}) \}.
    \end{align*}
\end{lemma}
\begin{proof}
    The proof is nearly identical to \cite{SZ25}, but we include one here for completeness.
    
    As long as $p$ is sufficiently large, then it is indistinguishable to tell whether a given circuit is obfuscated under one or two layers of obfuscation. Thus, for a sufficiently large parameter $p$, the distribution 
        \[ \{ (\sfO_{S_1}^1, \ldots, \sfO_{S_{\ell+1}}^1), \ldots, (\sfO_{S_1}^m, \ldots, \sfO_{S_{\ell+1}}^m): \forall i \in [m], \sfO_{S_j}^i \gets \IO(1^{\secp}, 1^p, C_{S_j}) \}, \]
    is $(f(\secp), O(\ell \cdot m) \cdot \eps(\secp))$-computationally indistinguishable from a distribution where $C_{S_j}$ are swapped with obfuscations of $C_{S_j}$ (denote this modified distribution by $\calD_S$). Similarly, the distribution 
    \begin{align*}
        \{ (\sfO_{T_1}^1, \ldots, \sfO_{T_{\ell+1}}^1), \ldots, (\sfO_{T_1}^m, \ldots, \sfO_{T_{\ell+1}}^m) : &(T_1, \ldots, T_{\ell+1}) \gets \calS_s, \\
        &\forall i \in [m], \sfO_{T_j}^i \gets \IO(1^{\secp}, 1^p, C_{T_j}) \},
    \end{align*}
    is $(f(\secp), O(\ell \cdot m) \cdot \eps(\secp))$-computationally indistinguishable from a distribution where $C_{T_j}$ are swapped with obfuscations of $C_{T_j}$ (denote this modified distribution by $\calD_T$). 

    The indistinguishability between $\calD_S$ and $\calD_T$ follows from Lemma \ref{lemma:subspace_hiding_crypto}: consider the reduction $\calB$ that gets a sample $(\sfO_1, \ldots, \sfO_{\ell+1})$ which is either from $\calD_0 := \{ (\sfO_{S_1}, \ldots, \sfO_{S_{\ell+1}}):  \sfO_{S_j} \gets \IO(1^{\secp}, 1^p, C_{S_j}) \}$ or from $\calD_1 := \{ (\sfO_{T_1}, \ldots, \sfO_{T_{\ell+1}}) : (T_1, \ldots, T_{\ell+1}) \gets \calS_s, \sfO_{T_j} \gets \IO(1^{\secp}, 1^p, C_{T_j}) \}$ for random superspaces $(T_1, \ldots, T_{\ell+1})$ of $(S_1, \ldots, S_{\ell+1})$. $\calB$ then generates $m$ i.i.d sets of obfuscations 
        \[ \{(\sfO_1^{(1)}, \ldots, \sfO_{\ell+1}^{(1)}), \ldots, (\sfO_1^{(m)}, \ldots, \sfO_{\ell+1}^{(m)})\} \]
    of $(\sfO_1, \ldots, \sfO_{\ell+1})$ and executes $\calA$ on the $m \cdot (\ell+1)$ obfuscations. When $(\sfO_1, \ldots, \sfO_{\ell+1})$ comes from $\calD_0$ then the output sample of the reduction comes from the distribution $\calD_S$, and when $(\sfO_1, \ldots, \sfO_{\ell+1})$ comes from $\calD_1$ then the output sample of the reduction comes from the distribution $\calD_T$. Since the reduction executes in complexity $\ell \cdot m \cdot \poly(\secp) \leq m \cdot n \cdot \poly(\secp)$, this means that
        \[ \calD_S \approx_{(f(\secp) - m \cdot n \cdot \poly(\secp), \poly(n) \cdot \eps(\secp))} \calD_T. \]
    By transitivity of computational indistinguishability, the two distributions in the statement of the lemma are $(f(\secp) - m \cdot n \cdot \poly(\secp), m \cdot \poly(n) \cdot \eps(\secp))$-computationally indistinguishable.
\end{proof}
\begin{corollary}\label{corollary:superspace_indistinguishability_crypto}
    Let $n, r, s, \ell \in \bbN^{+}$ such that $r+s+\ell \leq n$, and let $S_1 \subseteq \ldots \subseteq S_{\ell+1} \subseteq \bbZ_2^n$ be the same subspaces as in Lemma \ref{lemma:subspace_hiding_crypto}. Define the distribution $\calS_s$ and membership circuits $C_{S'}$ for subspaces $S'$ similarly. Let $\IO$ an indistinguishability obfuscation scheme that is $(f(\secp), \eps(\secp))$-secure, and assume that $(f(\secp), \eps(\secp))$-secure injective one-way functions exist, where $f$ and $1/\eps$ are sub-exponential functions.
    
    Then, for every security parameter $\secp$ such that $\secp \leq n-r-s-\ell$ and sufficiently large $p := p(\secp)$ polynomial in the security parameter, we have the following indistinguishability:
    \begin{align*}
        \{ &(\sfO_{T_1}^1, \ldots, \sfO_{T_{\ell+1}}^1), \ldots, (\sfO_{T_1}^m, \ldots, \sfO_{T_{\ell+1}}^m) : \\
        &(T_1, \ldots, T_{\ell+1}) \gets \calS_s, \forall i \in [m], \sfO_{T_j}^i \gets \IO(1^{\secp}, 1^p, C_{T_j}) \} \\
        &\approx_{(f(\secp) - m \cdot n \cdot \poly(\secp), m \cdot \poly(n) \cdot \eps(\secp))} \\
        \{ &(\sfO_{T_{1, 1}}, \ldots, \sfO_{T_{1, \ell+1}}), \ldots, (\sfO_{T_{m, 1}}, \ldots, \sfO_{T_{n, \ell+1}}) : \\
        &\forall i \in [m], (T_{i, 1}, \ldots, T_{i, \ell+1}) \gets \calS_s, \sfO_{T_{i, j}} \gets \IO(1^{\secp}, 1^p, C_{T_{i, j}}) \}.
    \end{align*}
\end{corollary}
\begin{lemma}\label{lemma:crypto_anti_concentration}
    Let $n, r, s, \ell \in \bbN^{+}$ such that $r+s+\ell \leq n$, and let $S_1 \subseteq \ldots \subseteq S_{\ell+1} \subseteq \bbZ_2^n$ be the same subspaces as in Lemma \ref{lemma:subspace_hiding_crypto}. Define the distribution $\calS_s$ and membership circuits $C_{S'}$ for subspaces $S'$ similarly. Let $\IO$ an indistinguishability obfuscation scheme that is $(f(\cdot), \frac{1}{f(\cdot)})$-secure, and assume that $(f(\cdot), \frac{1}{f(\cdot)})$-secure injective one-way functions exist, where $f$ is a sub-exponential function.
    
    Let $\secp$ be any security parameter such that $\secp \leq n-r-s-\ell$ and $p := p(\secp)$ a sufficiently large polynomial in the security parameter. Denote by $\calO_{\secp, p, s}$ the distribution over obfuscated circuits that samples $(T_1, \ldots, T_{\ell+1}) \gets \calS_s$ and then $O_{T_j} \gets \IO(1^{\secp}, 1^p, C_{T_j})$.
    
    Assume there is a quantum algorithm $\calA$ of complexity $T_{\calA}$ such that, 
        \[ \Pr[\calA(\sfO_{T_1}, \ldots, \sfO_{T_{\ell+1}}) \in T_1^{\perp} \setminus \{0\} : (\sfO_{T_1}, \ldots, \sfO_{T_{\ell+1}}) \gets \calO_{\secp, p, s}] \geq \eps. \]
    Also, denote $t := n-r-s$, $m := \frac{n(t+1)}{\eps}$, and assume (1) $\frac{t}{\eps} \cdot \left(2^{t-s}+2^{-t+\ell}\right) \leq o(1)$ and (2) $\frac{m \cdot \poly(n) \cdot T_{\calA}}{f(\secp)} \leq o(1)$.
    Then, 
        \[ \Pr[\calA(\sfO_{T_1}, \ldots, \sfO_{T_{\ell+1}}) \in \left(S_1^{\perp} \setminus T_1^{\perp} \right) : (\sfO_{T_1}, \ldots, \sfO_{T_{\ell+1}}) \gets \calO_{\secp, p, s}] \geq \frac{\eps}{16 \cdot n \cdot (t+1)}. \]
\end{lemma}
\begin{proof}
The proof of Lemma \ref{lemma:crypto_anti_concentration} is nearly identical to that of Lemma \ref{lemma:info_anti_concentration}; we will point out where the main differences lie.

As before, we consider a reduction $\calB$ which uses $\calA$ as a subroutine. $\calB$ has access to $m := n(t+1)/\eps$ samples of obfuscations 
    \[ ((\sfO^{(1, 1)}, \ldots, \sfO^{(1, \ell+1)}), \ldots, (\sfO^{(m, 1)}, \ldots, \sfO^{(m, \ell+1)})), \]
for $t := n-r-s$. $\calB$ executes $\calA(\sfO^{(i, 1)}, \ldots, \calO^{(i, \ell+1)})$ for every $i \in [m]$ and obtains $m$ vectors $\{ u_1, \ldots, u_m \}$, before taking only the vectors $\{ v_1, \ldots, v_j \}$ that are inside $S_1^{\perp}$. Finally, it computes $D := \dim(\mathsf{span}\{v_1, \ldots, v_j\})$.

The running time of $\calB$ is $O(m \cdot T_{\calA} + m \cdot n^3)$, where the first term arises from running $m$ executions of $\calA$ and the second term comes from running Gaussian elimination to check whether the $i$'th vector $u_i$ produced by $\calA$ is in $S_1^{\perp}$ and is outside of the span of the previous vectors (produced by $\calA$) that lie in $S_1^{\perp}$ (recall that $S_1$ is known to $\calA$ and $\calB$ and hence checking $S_1^{\perp}$ also takes at most $O(n^3)$ time).

Consider the following distribution $\calD_1$: sample $(T_{i, 1}, \ldots, T_{i, \ell+1}) \gets \calS_s$ i.i.d. $m$ times and for each set of superspaces, give $\calB$ an obfuscation of all membership programs $((\sfO_{T_{1, 1}}, \ldots, \sfO_{T_{1, \ell+1}}), \ldots, (\sfO_{T_{m, 1}}, \ldots, \sfO_{T_{m, \ell+1}}))$. By an identical argument as in Lemma \ref{lemma:info_anti_concentration}, we have that $D \geq t+1$ with probability at least $1-e^{-\Omega(n)}$ assuming $\frac{t}{\eps} \cdot \left(2^{t-s}+2^{-t+\ell}\right) \leq o(1)$.

We now execute $\calB$ on a different distribution $\calD_2$: sample $(T_1, \ldots, T_{\ell+1}) \gets \calS_s$ once, then sample $m$ i.i.d. obfuscations of the same circuits $(C_{T_1}, \ldots, C_{T_{\ell+1}})$, denoted by $((\sfO_{T_1}^{(1)}, \ldots, \sfO_{T_{\ell+1}}^{(1)}), \ldots, (\sfO_{T_1}^{(m)}, \ldots, \sfO_{T_{\ell+1}}^{(m)}))$. By Corollary \ref{corollary:superspace_indistinguishability_crypto}, 
    \[ \calD_1 \approx_{f(\secp) - m \cdot n \cdot \poly(\secp), \frac{m \cdot \poly(n)}{f(\secp)}} \calD_2. \]
By assumption, we have that $\calB$ runs in time $\leq f(\secp)- m \cdot n \cdot \poly(\secp)$. Thus, whenever we execute $\calB$ on a sample from $\calD_2$, with probability at least 
    \[ 1-e^{-\Omega(n)}-\frac{m \cdot \poly(n)}{f(\secp)} \geq 1-\frac{m \cdot \poly(n)}{f(\secp)} \geq \frac{1}{2},  \]
$\calB$ will compute $D \geq t+1$. By an averaging argument, this means that with probability at least $1/4$ over the sampling of $T_1, \ldots, T_{\ell+1}$, $\calB$ will compute $D \geq t+1$ on the appropriate input with probability at least $1/4$; we refer to such sets of superspaces as ``good'' sets. But when $D \geq t+1$, this means that $\calB$ has at least $t+1$ linearly independent vectors in $S_1^{\perp} \setminus \{0\}$; since $T_1^{\perp}$ has dimension $n-r-s = t$, $\calB$ produces at least one vector in $S_1^{\perp} \setminus T_1^{\perp}$. As $\calB$ uses $m$ i.i.d. executions of $\calA$, for every set of superspaces $(T_1, \ldots, T_{\ell+1})$ which is ``good'', when we prepare obfuscations of $T_i$, $\calA(\sfO_{T_1}, \ldots, \sfO_{T_{\ell+1}}) \in S_1^{\perp} \setminus T_1^{\perp}$ with probability at least $\frac{1}{4m}$.

Thus, over random $(T_1, \ldots, T_{\ell+1})$,
\begin{align*}
    \Pr[\calA(\sfO_{T_1}, \ldots, \sfO_{T_{\ell+1}}) \in S_1^{\perp} \setminus T_1^{\perp} : (\sfO_{T_1}, \ldots, \sfO_{T_{\ell+1}}) \gets \calO_{\secp, p, s}] &\geq \frac{1}{4} \cdot \frac{1}{4m}  \\
    &= \frac{\eps}{16n \cdot (t+1)},
\end{align*}
as desired.
\end{proof}

\section{Missing Proofs from Section \ref{sec:long_signatures_oracle} and \ref{sec:applications}}\label{appendix:missing_proofs_oracle}
\begin{proof}[Proof of Lemma \ref{lemma:bloating_dual}]
    We consider a series of hybrids, and show that the success probability in each pair of hybrids is statistically close.
    \begin{itemize}
        \item \textbf{$\mathsf{Hyb}_0$: The original execution of $\calA$.} By assumption, $\calA$ succeeds with probability $\eps$.
    
        \item \textbf{$\mathsf{Hyb}_1$: Using small-range distribution.} Consider the random function $F$ which samples the randomness used to compute the cosets $(\matA_\vecy, \vecb_\vecy)$ for every $\vecy \in \bbZ_2^r$; these cosets are then used in $\calP$, $\calP^{-1}$, and $\calD$. We now swap $F$ with a small-range version $F'$, which is sampled as follows: let $R := (300 \cdot q^3) \cdot \frac{2^7 \cdot n^2}{\eps}$. For every $\vecy \in \bbZ_2^r$, sample $\idx_\vecy \gets [R]$; for every $\idx \in [R]$, sample a coset $(\matA_{\idx}, \vecb_{\idx})$. Define $F'(\vecy) := (\matA_{\idx_\vecy}, \vecb_{\idx_\vecy})$.
        
        By Theorem A.6 of \cite{AGQY22}, every quantum algorithm making at most $q$ queries to $F$ or $F'$ has distinguishing advantage at most $\frac{300 \cdot q^3}{R} < \frac{\eps}{8}$, so the outputs of $\mathsf{Hyb}_0$ and $\mathsf{Hyb}_1$ have statistical distance at most $\frac{\eps}{8}$, and thus $\calA$ succeeds in this hybrid succeeds with probability at least $\frac{7\eps}{8}$.
    
        \item \textbf{$\mathsf{Hyb}_2$: Relaxing the dual oracle.} In $\mathsf{Hyb}_1$, on input $(j, \vecy, \vecv)$, the oracle $\calD$ accepted if and only if $\vecv \in \ColSpan(\matA_{\idx_\vecy}^{[j:n-r]})^{\perp}$. Consider the following modification. For each $\idx \in [R]$, sample a random $(r+s+\ell)$-dimensional superspace $T_{\idx, \ell+1}^{\perp}$ of $\ColSpan\left(\matA_\idx^{[\ell+1:n-r]}\right)^{\perp}$ by sampling $s$ random linearly independent vectors $v_1, \ldots, v_s$ such that 
            \[ \spn\{v_1, \ldots, v_s\} \cap \ColSpan(\matA_\idx^{[\ell+1:n-r]})^{\perp} = \{0\} \]
        and define 
            \[ T_{\idx, k}^{\perp} := \spn\{\ColSpan(\matA_{\idx}^{[k:n-r]})^{\perp}, v_1, \ldots, v_s\} \]
        for $k \in [\ell]$. Now, on input $(j, \vecy, \vecv)$, the oracle $\calD$ will instead accept if and only if $\vecv \in T_{\idx_\vecy, j}^{\perp}$.
    
        By Lemma \ref{lemma:subspace_hiding_info}, for each $\idx \in [R]$, changing $\calD$ is $O\left(\frac{q \cdot \ell \cdot s}{\sqrt{2^{n-r-s-\ell}}}\right)$-indistinguishable. Since we use the above indistinguishability $R$ times, we get $R \cdot O\left(\frac{q \cdot \ell \cdot s}{\sqrt{2^{n-r-s-\ell}}}\right)$-indistinguishability. Thus, if we denote $\calA$'s success probability in $\mathsf{Hyb_2}$ by $\eps_2$, we have that
        \begin{align*}
            \eps_2 &\geq \frac{7\eps}{8} - R \cdot O\left(\frac{q \cdot \ell \cdot s}{\sqrt{2^{n-r-s-\ell}}}\right) \geq \frac{7\eps}{8}-O\left(\frac{q^4 \cdot \ell \cdot s \cdot n^2 \cdot \frac{1}{\eps}}{\sqrt{2^{n-r-s-\ell}}}\right) \\
            &\geq \frac{7\eps}{8}-O\left(\frac{q^4 \cdot n^4 \cdot \frac{1}{\eps}}{\sqrt{2^{n-r-s-\ell}}}\right) \geq \frac{3\eps}{4},
        \end{align*}
        by assumption (2).
    
        \item \textbf{$\mathsf{Hyb}_3$: Changing the success predicate.} Recall that in $\mathsf{Hyb_2}$, we say $\calA$ is successful if it finds a collision, i.e. it produces a pair $x_0 \neq x_1$ such that $H(x_0) = H(x_1)$. We now say that $\calA$ is successful only if $H(x_0) = H(x_1)$ \emph{and} $\vecu_0 - \vecu_1 \notin T_{\idx_\vecy, 1}$. For every $\idx \in [R]$, let $\eps_2^{(\idx)}$ be the probability that $\calA$ finds $x_0 \neq x_1$ such that $y_0 = y_1 := \vecy$ and $\idx_\vecy = \idx$; by definition, $\sum_\idx \eps_2^{(\idx)} = \eps_2$. Observe that in this case, $\vecu_0 - \vecu_1 \in \ColSpan(\matA_{\idx_\vecy}) \setminus \{0\}$. Let $\delta^{(\idx)}$ be the probability that $\calA$ finds $x_0, x_1$ such that $y_0 = y_1 := \vecy$, $\idx_\vecy = \idx$, and $\vecu_0 - \vecu_1 \in T_{\idx_\vecy, 1} \setminus \{0\}$.

        Define $\eps_3^{(\idx)}$ to be the probability that $\calA$ finds $x_0, x_1$ such that $y_0 = y_1 := \vecy$, $\idx_\vecy = \idx$, and $\vecu_0 - \vecu_1 \notin T_{\idx_\vecy, 1}$. Let $L$ be the subset of indices $\idx \in [R]$ such that $\eps_2^{(\idx)} \geq \frac{\eps_2}{2R}$; this means that $\sum_{\idx \in L} \eps_2^{(\idx)} \geq \eps_2/2$. Our goal is to show that for every $\idx \in L$, $\eps_3^{(\idx)} \geq \frac{\eps_2^{(\idx)}}{16n^2}$. By definition, we have that $\eps_3^{(\idx)} \geq \eps_2^{(\idx)}-\delta^{(\idx)}$, so if $\delta^{(\idx)} \leq \eps_2^{(\idx)}/2$, then we are done. Thus, we assume that 
            \[ \delta^{(\idx)} > \frac{\eps_2^{(\idx)}}{2} \geq \frac{\eps_2}{4R} \geq \Omega\left(\frac{\eps^2}{q^3 \cdot n^2}\right). \] 
        Let $t := n-r-s \leq n-1$ and $m := \frac{n(t+1)}{\delta^{(\idx)}} \leq O\left(\frac{q^3 \cdot n^4}{\eps^2}\right)$. By assumptions (1) and (2), we have that
            \[ \frac{t}{\delta^{(\idx)}} \cdot \left(2^{t-s}+2^{-t+\ell}\right) = O\left(\frac{q^3 \cdot n^3}{\eps^2}\right) \cdot \left(2^{t-s}+2^{-t+\ell}\right) \leq o(1), \]
        and 
            \[ \frac{q \cdot \ell \cdot m^2 \cdot s}{2^{t/2}} = O\left(\frac{q^7 \cdot n^{10}/\eps^4}{2^{(t-1)/2}}\right) = O\left(\frac{q^7 \cdot n^{10}/\eps^4}{\sqrt{2^{n-r-s-\ell}}}\right) \leq o(1), \]
        so the conditions of Lemma \ref{lemma:info_anti_concentration} are fulfilled.\footnote{Strictly speaking, $\calA$ only has access to one oracle which, when restricted to inputs $(k, \cdot, \cdot)$, behaves as the oracle $\calO_{T_{\idx_\vecy, k}}$, but this is an even weaker model of access and so Lemma \ref{lemma:info_anti_concentration} readily applies.} Thus, we have that for $\idx \in L$, $\calA$ finds $x_0, x_1$ such that $y_0 = y_1 := y$ and $\vecu_0-\vecu_1 \in \ColSpan(\matA_{\idx_\vecy}) \setminus T_{\idx_\vecy, 1}$, and thus $\vecu_0-\vecu_1 \notin T_{\idx_\vecy, 1}$, with probability at least $\frac{\delta^{(\idx)}}{16n(t+1)} \geq \frac{\delta^{(\idx)}}{16n^2} \geq \frac{\eps_2^{(\idx)}}{32n^2}$.

        Therefore, the success probability of $\calA$ in this hybrid satisfies
            \[ \eps_3 = \sum_\idx \eps_3^{(\idx)} \geq \sum_{\idx \in L} \eps_3^{(\idx)} \geq \sum_{\idx \in L} \frac{\eps_2^{(j)}}{32n^2} \geq \frac{\eps_2}{64n^2} \geq \frac{3\eps}{256n^2}. \]
        
        \item \textbf{$\mathsf{Hyb}_{3.1}$: Using the randomness of $\Pi$.}    
        Recall that for every $\idx \in [R]$, the process of sampling $T_{\idx, k}^{\perp}$ involves sampling $s$ random linearly independent vectors $\vecv_{\idx, 1}, \ldots, \vecv_{\idx, s}$ and adding them to $S_{\idx, k}^{\perp}$. By Lemma \ref{lemma:sampling_superspaces_matrices}, we can equivalently sample $T_{\idx, k}^{\perp}$ by sampling a random full column-rank matrix $\matM_\idx \in \bbZ_2^{(n-r-\ell) \times (n-r-\ell)}$ and a random matrix $\matM'_\idx \in \bbZ_2^{(n-r-\ell) \times \ell}$ for each $\idx \in [R]$, computing $\overline{\matA}_\idx := \matA_\idx \cdot \begin{bmatrix} \matI_\ell & \matZero \\ \matM'_\idx & \matM_\idx \end{bmatrix}$ and setting 
            \[ T_{\idx, k}^{\perp} = \ColSpan\left(\begin{bmatrix} \overline{\matA}^{[k:\ell]} & \overline{\matA}^{[\ell+s+1:n-r]} \end{bmatrix}\right)^{\perp}. \]
        Since this new sampling process is statistically equivalent, the change is indistinguishable.

        Define the permutation $\Gamma$ over $\{0, 1\}^n$ as follows: for an input $\vecs \in \{0, 1\}^n$, take the left $r$ bits denoted $\vecy \in \bbZ_2^r$, compute $\idx_\vecy \in [R]$, and then apply matrix multiplication by $\matC_{\idx_\vecy} := \begin{bmatrix} \matI_\ell & \matZero \\ \matM'_{\idx_\vecy} & \matM_{\idx_\vecy} \end{bmatrix}$ to the remaining right $n-r$ bits. Since $\matC_{\idx_\vecy}$ is full-rank for all $\idx_\vecy$, $\Gamma$ is indeed a permutation. We now apply $\Gamma$ to the output of $\Pi$ inside the execution of a query to $\calP$, and apply $\Gamma^{-1}$ to the input of $\Pi^{-1}$ inside the execution of a query to $\calP^{-1}$. Since $\Pi$ is random and $\Gamma$ is fixed, the composition produces a random permutation as well and thus the success probability in $\mathsf{Hyb}_{3.1}$ is identical.

        \item \textbf{$\mathsf{Hyb}_{3.2}$: Moving $\Gamma$ into the coset computation.}
        In $\mathsf{Hyb}_{3.2}$, we stop applying $\Gamma$ and $\Gamma^{-1}$ and instead multiply the output of $\Pi$ by $\matC_\idx$ in $\calP$ and multiply the original input to $\Pi^{-1}$ in $\calP^{-1}$ by $\matC_\idx^{-1}$. By definition, the oracle distributions in $\mathsf{Hyb}_{3.1}$ and $\mathsf{Hyb}_{3.2}$ are identical, so the success probability of $\calA$ in $\mathsf{Hyb}_{3.2}$ is also $\eps_3$.
        
        \item \textbf{$\mathsf{Hyb}_4$: Using the randomness of $F'$.}
        In $\mathsf{Hyb}_4$, we stop multiplying by $\matC_\idx$ and $\matC_\idx^{-1}$ and directly define $T_{\idx, k}$ to be $\ColSpan\left(\begin{bmatrix} \matA_\idx^{[k:\ell]} & \matA_\idx^{[\ell+s+1:n-r]} \end{bmatrix}\right)$. Observe that for any fixed choice of $\Pi$, the distribution of $\matA_\idx = \begin{bmatrix} \matI_{\ell} & \matZero \\ \matK_\idx & \matL_\idx \end{bmatrix}$ has random matrices $\matK_\idx$ and random full column-rank matrices $\matL_\idx$, so we have that
        \begin{align*}
            \overline{\matA}_\idx := \matA_\idx \cdot \matC_\idx = \begin{bmatrix} \matI_{\ell} & \matZero \\ \matK_\idx & \matL_\idx \end{bmatrix} \cdot \begin{bmatrix} \matI_\ell & \matZero \\ \matM'_j & \matM_\idx \end{bmatrix} = \begin{bmatrix} \matI_{\ell} & \matZero \\ \matK_\idx + \matL_\idx \cdot \matM'_\idx & \matL_\idx \cdot \matM_\idx\end{bmatrix},
        \end{align*}
        which has the same distribution as $\matA_\idx$ (as both $\matL_\idx$ and $\matM_\idx$ are random full column-rank matrices and $\matK_\idx$ is a random matrix). Thus, the success probability $\eps_4$ of $\calA$ in $\mathsf{Hyb}_4$ is identical to the success probability of $\calA$ in $\mathsf{Hyb}_{3.2}$, which is simply $\eps_3$.
    
        \item \textbf{$\mathsf{Hyb}_5$: Rewinding small-range distributions.}
        We undo the process of sampling an $R$-small range distribution by swapping $F'$ back to a completely random function $F$. By an identical argument as before, $\calA$ succeeds in this hybrid with probability at least 
            \[ \eps_4 - \frac{300q^3}{R} = \eps_3 - \frac{300q^3}{R} \geq \frac{3\eps}{256n^2} - \frac{300q^3}{R} = \frac{3\eps}{256n^2} - \frac{\eps}{128n^2} = \frac{\eps}{256n^2}. \]
        We then observe that the success probability of $\calA$ in $\mathsf{Hyb}_5$ is exactly the probability we want to bound, which concludes the proof.
    \end{itemize}
\end{proof}

\begin{proof}[Proof of Lemma \ref{lemma:simulating_dual}]
    Given oracle access to $\overline{\calP}$ and $\overline{\calP}^{-1}$ which come from $(\overline{\calP}, \overline{\calP}^{-1}, \overline{\calD})$, $\calB$ acts as follows:
    \begin{itemize}
        \item Sample a random function $F'$ that outputs sufficient (polynomial in $r$) randomness and a random $n$-bit permutation $\Gamma$. On input $\vecy \in \{0, 1\}^r$, use the randomness of $F'(\vecy)$ to generate a random vector $\vecd_\vecy \in \bbZ_2^n$, random matrix $\matC_{\vecy, 1} \in \bbZ_2^{(n-\ell) \times \ell}$, and random full-rank matrix $\matC_{\vecy, 2} \in \bbZ_2^{(n-\ell) \times (n-\ell)}$. Denote by $\matC_{\vecy}$ the full-rank matrix $\begin{bmatrix} \matI_\ell & \matZero \\ \matC_{\vecy, 1} & \matC_{\vecy, 2} \end{bmatrix} \in \bbZ_2^{n \times n}$. This defines the following oracles:
        \item $(\vecy \in \bbZ_2^r, \vecu \in \bbZ_2^n) \gets \calP(\vecx \in \bbZ_2^n)$: 
        \begin{itemize}
            \item $(\overline{x} \in \{0, 1\}^{r+s}, \tilde{x} \in \bbZ_2^{n-r-s-\ell}, x' \in \bbZ_2^\ell) \gets \Gamma(\vecx)$.
            \item $(\vecy \in \bbZ_2^r, \overline{\vecu} \in \bbZ_2^{r+s}) \gets \overline{\calP}(\overline{x})$.
            \item $(\matC_\vecy \in \bbZ_2^{n \times n}, \vecd_\vecy \in \bbZ_2^n) \gets F'(\vecy)$.
            \item $\vecu \gets \matC_\vecy \cdot \begin{pmatrix} x' \\ \overline{\vecu} \\ \tilde{x} \end{pmatrix} + \vecd_\vecy$.
        \end{itemize}
        \item $(\vecx \in \bbZ_2^n \cup \{\bot\}) \gets \calP^{-1}(\vecy \in \bbZ_2^r, \vecu \in \bbZ_2^n)$:
        \begin{itemize}
            \item $(\matC_\vecy, \vecd_\vecy) \gets F'(\vecy)$.
            \item $\begin{pmatrix} x' \\ \overline{\vecu} \\ \tilde{x} \end{pmatrix} \gets \matC_\vecy^{-1} \cdot (\vecu-\vecd_\vecy)$.
            \item $(\overline{x} \in \bbZ_2^{r+s} \cup \{\bot\}) \gets \overline{\calP}^{-1}(\vecy, \overline{\vecu})$
            \item Output $\vecx \gets \Gamma^{-1}(\overline{x}, \tilde{x}, x')$ if $\overline{x} \neq \bot$, else output $\bot$.
        \end{itemize}
        \item $\{0, 1\} \gets \calD'(j \in [\ell+1], \vecy \in \bbZ_2^r, \vecv \in \bbZ_2^n)$:
        \begin{itemize}
            \item $(\matC_\vecy, \vecd_\vecy) \gets F'(\vecy)$.
            \item Output 1 iff $\vecv^T \cdot \begin{bmatrix}
                \matC_\vecy^{[j:\ell]} & \matC_\vecy^{[\ell+s+1:n-r]}
            \end{bmatrix} = 0^{n-r-s-j+1}$.
        \end{itemize}
    \end{itemize}
    Finally, $\calB$ runs $(x_0, x_1) \gets \calA^{\calP, \calP^{-1}, \calD'}$, $(\overline{x}_b, \tilde{x}_b, x'_b) \gets \Gamma(x_b)$ and outputs $(\overline{x}_0, \overline{x}_1)$. 
    
    Define 
    \begin{align*}
        \matA_\vecy &:= \matC_\vecy \cdot \begin{bmatrix} \matI_{\ell} & \matZero & \matZero \\ \matZero & \overline{\matA}_\vecy & \matZero \\ \matZero & \matZero & \matI_{n-r-s-\ell} \end{bmatrix} = \begin{bmatrix} \matI_{\ell} & \matZero \\ \matC_{\vecy, 1} & \matC_{\vecy, 2} \end{bmatrix} \cdot \begin{bmatrix} \matI_{\ell} & \matZero & \matZero \\ \matZero & \overline{\matA}_\vecy & \matZero \\ \matZero & \matZero & \matI_{n-r-s-\ell} \end{bmatrix} \\
        &= \begin{bmatrix} \matI_{\ell} & \matZero \\ \matC_{\vecy, 1} & \matC_{\vecy, 2} \cdot \begin{bmatrix} \overline{\matA}_\vecy & \matZero \\ \matZero & \matI_{n-r-s-\ell} \end{bmatrix} \end{bmatrix} \in \bbZ_2^{n \times (n-r)},
    \end{align*}
    where $\overline{\matA}_\vecy \in \bbZ_2^{(r+s) \times s}$ is the matrix arising from $\overline{\calP}$ and $\overline{\calP}^{-1}$. As $\matC_{\vecy, 2}$ and $\overline{\matA}_\vecy$ have full column-rank, so does $\matA_\vecy^{[\ell+1:n-r]}$; on the other hand, $\matC_{\vecy, 1}$ remains a random matrix; thus, $\matA_\vecy$ has full column-rank. Now, since $(\vecu_0 - \vecu_1) \notin \ColSpan\left(\begin{bmatrix}\matA_\vecy^{[1:\ell]} & \matA_\vecy^{[\ell+s+1:n-r]}\end{bmatrix}\right)$, this means that if we look at each of the two coordinates vectors $\vecw_0$, $\vecw_1$ (which each have $n-r$ bits) of $\vecu_0$ and $\vecu_1$, they differ somewhere between the $(\ell+1)$'th and $(\ell+s)$'th bit. However, $(\vecw_b)_{[\ell+1:\ell+s]}$ arises from applying the permutation $\overline{\Pi}$ on $\{0, 1\}^{r+s}$ defined by the oracles $\calP, \calP^{-1}$ to $\overline{x}_b$ and taking the last $s$ bits of the output. As $x_0, x_1$ have the same $\vecy$'s and thus so do $\overline{x}_0$ and $\overline{x}_1$, it must be the case that $\overline{x}_0 \neq \overline{x}_1$ in order for $\vecw_0$ to differ from $\vecw_1$ in the appropriate indices.

    If we define $\eps_{\calB}$ as the probability that $\calA$ outputs $x_0, x_1$ such that $y_0 = y_1 := \vecy$ and $(\vecu_0 - \vecu_1) \notin \ColSpan\left(\begin{bmatrix}\matA_\vecy^{[1:\ell]} & \matA_\vecy^{[\ell+s+1:n-r]}\end{bmatrix}\right)$, we wish to lower bound the probability of $\eps_{\calB}$. We do this through a series of hybrids.
    \begin{itemize}
        \item \textbf{$\mathsf{Hyb}_0$: The distribution $(\calP, \calP^{-1}, \calD') \gets \calB^{\overline{\calP}, \overline{\calP^{-1}}}$, as simulated by $\calB$.}
        
        This is the original execution, and thus has success probability $\eps_{\calB}$.
        \item \textbf{$\mathsf{Hyb}_1$: Removing the inner permutation $\overline{\Pi}$ by using $\Gamma$.}

        Instead of applying $\overline{\Pi}$ and $\Pi^{-1}$ inside $\overline{\calP}$ and $\overline{\calP}^{-1}$, respectively, only use $\Gamma$. This hybrid is statistically identical to the previous hybrid since a concatenation of a permutation with a random permutation $\Gamma$ is still a random permutation.

        \item \textbf{$\mathsf{Hyb}_2$: Taking $\matA_\vecy, \vecb_\vecy$ to be the direct output of $F$ by using $F'$.}

        Instead of sampling $\matC_\vecy$ and $\vecd_\vecy$ before computing $\matA_\vecy := \matC_\vecy \cdot \begin{bmatrix} \matI & \matZero & \matZero \\ \matZero & \overline{\matA}_\vecy & \matZero \\ \matZero & \matZero & \matI_{n-r-s} \end{bmatrix}$ and $\vecb_\vecy := \matC_\vecy \cdot \begin{pmatrix} \matZero \\ \overline{\vecb}_\vecy \\ \matZero \end{pmatrix} + \vecd_\vecy$, where $\overline{\matA}_\vecy, \overline{\vecb}_\vecy$ are defined by the output of the inner random function, we sample $\matA_\vecy$ and $\vecb_\vecy$ directly. We observe that by the randomness of $\vecd_\vecy$, $\matC_{\vecy, 1}$, $\matC_{\vecy, 2}$, and $\overline{\matA}_{\vecy}$, the joint distributions of matrices and offset vectors in both hybrids are identical. Therefore, the success probability of $\calA$ in this hybrid is also $\eps_{\calB}$.
        \begin{remark}\label{remark:coset_shift_distribution}
            Note that 
            \[ \vecb_\vecy-\vecd_\vecy = \matC_\vecy \cdot \begin{pmatrix} \matZero \\ \overline{\vecb}_\vecy \\ \matZero \end{pmatrix} = \begin{bmatrix} \matI_{\ell} & \matZero \\ \matC_{\vecy, 1} & \matC_{\vecy, 2} \end{bmatrix} \cdot \begin{pmatrix} \matZero \\ \overline{\vecb}_\vecy \\ \matZero \end{pmatrix} = \begin{pmatrix} \matZero \\ \matC_{\vecy, 2} \cdot \begin{pmatrix}\overline{\vecb}_\vecy \\ \matZero \end{pmatrix}\end{pmatrix} \]
            has only zeros in the first $\ell$ coordinates. Thus, if $\vecd_\vecy$ is distributed randomly, then $\vecb_\vecy$ is distributed randomly; if $\vecd_\vecy$ is uniformly random with the $\ell$'th coordinate set to 1, then so is $\vecb_\vecy$.
        \end{remark}
    \end{itemize}
    Since the distribution generated in $\mathsf{Hyb}_2$ is precisely $\calO'_{n, r, \ell, s}$, this means that $\eps_{\calB} = \eps$, as desired.
\end{proof}

\begin{proof}[Proof of Lemma \ref{lemma:incompressible_OSS_original_properties}]
The perfect correctness follows from the perfect correctness of Construction \ref{fig:oss-scheme-classical} and observing that the verification algorithm $\Ver$ in Construction \ref{fig:oss-scheme-classical} does not actually need the full capability of $\calP^{-1}$ and that having access to $\calD_0$ suffices.

The oracle $\calD_0$ can be implemented using $\calP^{-1}$. Thus, it suffices to show the strong signature security of Construction \ref{fig:oss-scheme-incompressible} with respect to the oracles $\calP$, $\calP^{-1}$, and $\calD$ (where we will abuse notation and also refer to this distribution of oracles by $\widetilde{\calO}_{n, r, \ell}$). 

As with Construction \ref{fig:oss-scheme-classical}, we can ``bloat'' the dual oracle. Let $\widetilde{\calO}'_{n, r, \ell, s}$ denote the following distribution over oracles $\calP$, $\calP^{-1}$, and $\calD'$. The oracles $\calP$ and $\calP^{-1}$ are defined identically to the oracles in $\widetilde{\calO}_{n, r, \ell}$. However, our new dual oracle $\calD'$ will be a relaxed version of the original dual oracle $\calD$:
    \[ \calD'(j, \vecy, \vecv) = \begin{cases} 1 & \text{if } \vecv^T \cdot \begin{bmatrix}
        \matA_\vecy^{[j:\ell]} & \matA_\vecy^{[\ell+1+s:n-r]}
    \end{bmatrix} = 0^{n-r-s-j+1} \text{ and } j \in [\ell+1], \\ 0 & \text{otherwise.} \end{cases} \]
The analog to Lemma \ref{lemma:bloating_dual} remains true for these sets of oracle distributions:
\begin{lemma}[Analogous Bloating Lemma]\label{lemma:bloating_dual_incompressible}
    Suppose there is an oracle-aided $q$-query quantum algorithm $\calA$ such that
        \[ \Pr[(y_0 = y_1) \land (x_0 \neq x_1):\begin{aligned} (\calP, \calP^{-1}, \calD) &\gets \widetilde{\calO}_{n, r, \ell} \\ (x_0, x_1) &\gets \calA^{\calP, \calP^{-1}, \calD} \\ (y_b, \vecu_b) &\gets \calP(x_b)\end{aligned}] \geq \eps. \]
    Then we have that
        \[ \Pr[\begin{array}{l} {\scriptstyle(y_0 = y_1 := \vecy) \hspace{5pt} \land} \\ {\scriptstyle(\vecu_0-\vecu_1) \notin \ColSpan}\left(\begin{bmatrix}{\scriptstyle \matA_\vecy^{[1:\ell]}} & {\scriptstyle \matA_\vecy^{[\ell+s+1:n-r]}}\end{bmatrix}\right) \end{array}:\begin{aligned} {\scriptstyle (\calP, \calP^{-1}, \calD')} &{\scriptstyle \gets \widetilde{\calO}'_{n, r, \ell, s}} \\ {\scriptstyle (x_0, x_1)} &{\scriptstyle \gets \calA^{\calP, \calP^{-1}, \calD'}} \\ {\scriptstyle (y_b, \vecu_b)} &{\scriptstyle \gets \calP(x_b)}\end{aligned}] \geq \frac{\eps}{256n^2}. \]
\end{lemma}
\begin{proof}
    The proof of Lemma \ref{lemma:bloating_dual_incompressible} is identical to the proof of Lemma \ref{lemma:bloating_dual}, since the oracle distribution of $\widetilde{\calO}_{n(\secp), r(\secp), \ell(\secp)}$ is the same as the distribution of $\calO_{n(\secp+1), r(\secp+1), \ell(\secp+1)}$ except for the coset shift vectors $\{\vecb_\vecy\}_{\vecy}$ whose distribution is not used in the proof.
\end{proof}
Just like with Construction \ref{fig:oss-scheme-classical}, we can entirely simulate access to a bloated dual given only access to the primal oracles. 
\begin{lemma}[Analogous Dual Simulation Lemma]\label{lemma:simulating_dual_incompressible}
    Suppose that there is an oracle aided $q$-query algorithm $\calA$ such that
        \[ \Pr[\begin{array}{l} {\scriptstyle(y_0 = y_1 := \vecy) \hspace{5pt} \land} \\ {\scriptstyle(\vecu_0-\vecu_1) \notin \ColSpan}\left(\begin{bmatrix}{\scriptstyle \matA_\vecy^{[1:\ell]}} & {\scriptstyle \matA_\vecy^{[\ell+s+1:n-r]}}\end{bmatrix}\right) \end{array}:\begin{aligned} {\scriptstyle (\calP, \calP^{-1}, \calD')} &{\scriptstyle \gets \widetilde{\calO}'_{n, r, \ell, s}} \\ {\scriptstyle (x_0, x_1)} &{\scriptstyle \gets \calA^{\calP, \calP^{-1}, \calD'}} \\ {\scriptstyle (y_b, \vecu_b)} &{\scriptstyle \gets \calP(x_b)}\end{aligned}] \geq \eps. \]
    Then there is an oracle aided $q$-query algorithm $\calB$ such that
        \[ \Pr[(\overline{y}_0 = \overline{y}_1) \land (\overline{x}_0 \neq \overline{x}_1):\begin{aligned} (\overline{\calP}, \overline{\calP}^{-1}, \overline{\calD}) &\gets \calO_{r+s, r, 0} \\ (\overline{x}_0, \overline{x}_1) &\gets \calB^{\overline{\calP}, \overline{\calP}^{-1}} \\ (\overline{y}_b, \overline{\vecu}_b) &\gets \overline{\calP}(\overline{x}_b)\end{aligned}] \geq \eps, \]
    where $s := 16 \cdot (\secp+1)$ and $r := s \cdot \secp$.
\end{lemma}
Note that the new distribution $\calO_{r+s, r, 0}$ is the same dual-free oracle distribution as in Section \ref{sec:long_signatures_oracle}, which uses fully random coset shift vectors.
\begin{proof}
The proof is identical to that of Lemma \ref{lemma:simulating_dual} with one difference: instead of sampling the random function $F'$ such that on input $\vecy \in \{0, 1\}^r$, $F'(\vecy)$ provides randomness to generate a random vector $\vecd_\vecy \in \bbZ_2^n$, $\vecd_\vecy$ will be sampled to be a random vector whose $\ell = (\secp+1)$'th coordinate is 1. It is not hard to verify that with this change, the rest of the proof goes through (see Remark \ref{remark:coset_shift_distribution} for more details).
\end{proof}

We can now apply Theorems \ref{thm:dual-free-oracle} and \ref{thm:coset_partition_collision_resistance} and conclude that Construction \ref{fig:oss-scheme-incompressible} has strong signature security (following the same outline as in Section \ref{sec:long_signatures_oracle} until Corollary \ref{corollary:strong_security_oracle}).
\end{proof}

\section{Missing Proofs from Section \ref{sec:long_signatures_plain}}\label{appendix:missing_proofs_plain}
The proofs of the following lemmas will closely track the general outline of \cite{SZ25}.
\begin{proof}[Proof of Lemma \ref{lemma:bloating_dual_crypto}]
    We include a proof of Lemma \ref{lemma:bloating_dual_crypto}, which very closely follows the proof of Lemma 54 in \cite{SZ25}, for completeness.
    We will consider a series of hybrids, and show that the success probability in each pair of hybrids is computationally close.
    \begin{itemize}
        \item \textbf{$\mathsf{Hyb}_0$: The original execution of $\calA$.}
        
        By definition, $\calA$ succeeds with probability $\eps$.
        \item \textbf{$\mathsf{Hyb}_1$: Preparing to switch to a small-range distribution.}
        
        Let $(\LFKeyGen, \LFF)$ be a $(f(\kappa), \frac{1}{f(\kappa)})$-secure lossy function with output length $m$ and lossiness parameter $w$. We would like to apply Lemma \ref{lemma:obfuscating_samplers}, so we will describe two PRFs and one ensemble of distributions (which will serve as both $\calD_0$ and $\calD_1$). The first PRF takes as input $\{0, 1\}^d$ and applies $\sfF(\klin, \cdot)$ to get randomness $r$. The second PRF takes as input $\{0, 1\}^d$, samples $\pk_{\LF} \gets \LFKeyGen(1^\kappa, 0, 1^w)$, applies $\LFF(\pk_{\LF}, \cdot)$ to the input (padded with $\kappa-d$ zeros), and then applies $\sfF(\klin', \cdot)$ to get randomness $r'$.

        We note that $\pk_{\LF}$ produces an associated injective function with probability $\geq 1-\negl(\kappa) = 1-\negl(\secp)$, so we assume this is the case in the following analysis.
        
        The first function is a puncturable PRF by definition. It is not hard to see that the second function is indeed a puncturable PRF: the key is simply $(\pk_{\LF}, \klin')$, and evaluation can be done by padding with zeros and successively applying $\LFF(\pk_{\LF}, \cdot)$ and $\sfF(\klin', \cdot)$. Puncturing is also relatively simple: for a set $S$, one evaluates the set $S$ on $\LFF(\pk_{\LF}, \cdot)$ to get a set of images $S'$ before puncturing $\klin'\{S'\} \gets \Punc(\klin', S')$. Correctness follows from the injectivity of $\LFF$ and the correctness of $\sfF$, while puncturing security follows from the injectivity of $\LFF$ and the puncturing security of $\sfF$ on inputs of length $m$.
        
        If we now define distributions $\calD = \calD_0 = \calD_1$ which take as input $\vecy \in \{0, 1\}^d$ and randomness $r$ and construct a coset $(\matA_\vecy, \vecb_\vecy)$, it is easy to see that in $\mathsf{Hyb}_0$, the circuit $E_0(\vecy) := \calD(\vecy; \sfF(\klin, \vecy))$ is used black-box by $P$, $P^{-1}$, and $D$ (and all three circuits are obfuscated).
        
        In $\mathsf{Hyb}_1$, we will instead use the circuit $E_1(\vecy) := \calD(\vecy; \sfF(\klin', \LFF(\pk_{\LF}, \vecy)))$, where $\pk_{\LF}$ and $\klin'$ are freshly sampled. It follows by Lemma \ref{lemma:obfuscating_samplers} that the output of $\mathsf{Hyb}_0$ and $\mathsf{Hyb}_1$ are $(f(\kappa)-\poly(\kappa), O(|\calX|/f(\kappa)))$-computationally indistinguishable, where $\calX$ is the set of all possible values of $\vecy$, which has size $2^r$. We incur some loss due to assuming $\LFF$ is inejctive, and thus it follows by Assumptions (1) and (3) that the success probability of $\calA$ in this hybrid is 
            \[ \eps_1 \geq \eps - O\left(\frac{2^r}{f(\kappa)}\right)-\negl(\secp) \geq \frac{31\eps}{32}. \]
        \item \textbf{$\mathsf{Hyb}_2$: Switching to a small number of cosets.}
        
        $E_1$ now samples a lossy key $\pk'_{\LF} \gets \LFKeyGen(1^{\kappa}, 1, 1^w)$ instead of an injective key $\pk_{\LF} \gets \LFKeyGen(1^{\kappa}, 0, 1^w)$. By the security of the lossy function, the output of this hybrid is $(f(\kappa), \frac{1}{f(\kappa)})$-computationally indistinguishable from the previous hybrid. Thus, Assumption (1) implies that the success probability of $\calA$ in this hybrid is
            \[ \eps_2 \geq \eps_1 - \frac{1}{f(\kappa)} \geq \frac{15\eps}{16}. \]
        We will refer to the image of the lossy function $\LFF(\pk'_{\LF}, \cdot)$ by $\calX$. Note that in this hybrid, the coset $\matA_\vecy, \vecb_\vecy$ is also an efficiently computable function of $\vecx = \LFF(\pk'_{\LF}, \vecy) \in \calX$ (and is the same for all $\vecy$'s which map to the same $\vecx$), so we will refer to these cosets by $\matA_\vecx, \vecb_\vecx$. 
        \item \textbf{$\mathsf{Hyb}_3$: Using obfuscated circuits inside $D$.} 
        
        Recall that in $\mathsf{Hyb}_2$, $D(j, \vecy, \vecv)$ computes $\vecx$ before applying $\sfF(\klin, \vecx)$ to get pseudorandomness $r_\vecx$ which is used to generate a coset $(\matA_\vecx, \vecb_\vecx)$; $D(j, \vecy, \vecv)$ then checks membership in the dual of $\ColSpan(\matA_\vecx^{[j:n-r]}) := S_{\vecx, j}$.
        
        In $\mathsf{Hyb}_3$, we additionally sample another puncturable PRF key $k_S$. After computing $\vecx$, we apply $\sfF(k_S, \vecx)$ to obtain additional pseudorandomness $r'_\vecx$; using $r'_\vecx$, we generate obfuscations $\sfO_{S_{\vecx, j}^{\perp}} \gets \IO(1^{\kappa}, 1^{p}, C_{S_{\vecx, j}^{\perp}})$ of the membership circuits for $S_{\vecx, j}^{\perp}$. $D$ now uses $\sfO_{S_{\vecx, j}^{\perp}}$ instead of $C_{S_{\vecx, j}^{\perp}}$ to decide dual membership. By the correctness of the inner obfuscation scheme, $D$ remains functionally equivalent, so the two hybrids are $(f(\kappa), \frac{\ell}{f(\kappa)})$-computationally close. Thus, by Assumption (1), the success probability of $\calA$ in this hybrid is
            \[ \eps_3 \geq \eps_2 - \frac{\ell}{f(\kappa)} \geq \frac{29\eps}{32}. \]
        \item \textbf{$\mathsf{Hyb}_4$: Relaxing dual verification inside $D$.}
        
        Once again, we would like to apply Lemma \ref{lemma:obfuscating_samplers}, so we now describe the two ensembles of distributions we want to use; we will use one puncturable PRF $\sfF$. The first distribution $\calD_{S, \klin}$ takes as input $\vecx$ and randomness $r$ and computes $S_\vecx$ from $\sfF(\klin, \vecx)$ before using $r$ to generate obfuscations $\sfO_{S_{\vecx, j}^{\perp}} \gets \IO(1^{\kappa}, 1^{p}, C_{S_{\vecx, j}^{\perp}})$. The second distribution $\calD_{T, \klin}$ takes as input $\vecx$ and randomness $r$ and computes $S_\vecx$ from $\sfF(\klin, \vecx)$ before using $r$ to sample superspaces $T_{\vecx, j}^{\perp}$ (with $s$ more dimensions) of $S_{\vecx, j}^{\perp}$ and generate obfuscations $\sfO_{T_{\vecx, j}^{\perp}} \gets \IO(1^{\kappa}, 1^{p}, C_{T_{\vecx, j}^{\perp}})$.
        
        Specifically, the first part of $r$ will be used to generate a random full column-rank matrix $\matM_\vecx \in \bbZ_2^{(n-r-\ell) \times (n-r-\ell)}$ and a random matrix $\matM'_\vecx \in \bbZ_2^{(n-r-\ell) \times \ell}$. This defines matrices $\matC_\vecx := \begin{bmatrix} \matI_{\ell} & \matZero \\ \matM'_\vecx & \matM_\vecx \end{bmatrix} \in \bbZ_2^{(n-r) \times (n-r)}$ and $\overline{\matA}_\vecx := \matA_\vecx \cdot \matC_\vecx \in \bbZ_2^{n \times (n-r)}$. We can now define 
            \[ T_{\vecx, j}^{\perp} := \ColSpan\left(\begin{bmatrix} \overline{\matA}_\vecx^{[j:\ell]} & \overline{\matA}_\vecx^{[\ell+s+1:n-r]} \end{bmatrix}\right)^{\perp}. \]
        
        By Lemmas \ref{lemma:sampling_superspaces_matrices} and \ref{lemma:subspace_hiding_crypto}, the distributions $\calD_{S, \klin}$ and $\calD_{T, \klin}$ are $(f(n-r-s-\ell)-\poly(n), \poly(n)/f(n-r-s-\ell))$-computationally indistinguishable (even though $S_{\vecx, j}$ are known).
        
        Since $|\calX| \leq 2^w$ with probability $1-\negl(\secp)$, in what follows, we assume this is the case. Consider the two circuits $E_{S, \klin}(\vecx) := \calD_{S, \klin}(\vecx, \sfF(k_S, \vecx))$ and $E_{T, \klin}(\vecx) := \calD_{T, \klin}(\vecx, \sfF(k_T, \vecx))$ for uniformly sampled keys $k_S, k_T$. Observe that $E_{S, \klin}$ is used black-box in $D$ in $\mathsf{Hyb}_3$, and $D$ will be obfuscated. Thus, if we switch to $E_{T, \klin}$, by Lemma \ref{lemma:obfuscating_samplers}, this is $(f(n-r-s-\ell)-\poly(n), 2^w \cdot \poly(n)/f(n-r-s-\ell))$-computationally indistinguishable. We conclude that $\mathsf{Hyb}_3$ and $\mathsf{Hyb}_4$ are $(f(n-r-s-\ell)-\poly(n), 1/f(\kappa)+2^w \cdot \poly(n)/f(n-r-s-\ell))$-computationally indistinguishable. 
        
        Therefore, after incurring some loss from the assumption that $|\calX| \leq 2^w$, we have that by Assumptions (1) and (3), the success probability of $\calA$ in this hybrid is
            \[ \eps_4 \geq \eps_3 - \negl(\secp)-\frac{2^w \cdot \poly(n)}{f(n-r-s-\ell)} \geq \frac{7\eps}{8}. \] 
        \item \textbf{$\mathsf{Hyb}_5$: Changing the success predicate.}
        
        Recall that in $\mathsf{Hyb_4}$, we say $\calA$ is successful if it finds a collision, i.e. it produces a pair $x_0 \neq x_1$ such that $y_0 = y_1 := \vecy$. We now say that $\calA$ is successful only if $y_0 = y_1 := \vecy$ \emph{and} $\vecu_0 - \vecu_1 \notin T_{\vecx, 1}$. As $|\calX| \leq 2^w$ with probability $1-\negl(\secp)$, in what follows, we assume this is the case.
        
        For every $\vecx \in \calX$, let $\eps_4^{(\vecx)}$ be the probability that $\calA$ finds $x_0 \neq x_1$ such that $y_0 = y_1 := \vecy$ and $\vecx = \LFF(\pk_{\LF}, \vecy)$; by definition, $\sum_{\vecx \in \calX} \eps_4^{(\vecx)} = \eps_4$. Observe that when a collision occurs, we have $\vecu_0 - \vecu_1 \in \ColSpan(\matA_{\vecx}) \setminus \{0\} = S_{\vecx, 1} \setminus \{0\}$. Let $\delta^{(\vecx)}$ be the probability that $\calA$ finds $x_0, x_1$ such that $y_0 = y_1 := \vecy$, $\vecx = \LFF(\pk_{\LF}, \vecy)$, and $\vecu_0 - \vecu_1 \in T_{\vecx, 1} \setminus \{0\}$.

        Define $\eps_5^{(\vecx)}$ to be the probability that $\calA$ finds $x_0, x_1$ such that $y_0 = y_1 := \vecy$, $\vecx = \LFF(\pk_{\LF}, \vecy)$, and $\vecu_0 - \vecu_1 \notin T_{\vecx, 1}$. Let $L$ be the subset of $\calX$ such that $\eps_4^{(\vecx)} \geq \frac{\eps_4}{2|\calX|}$; this means that $\sum_{\vecx \in L} \eps_4^{(\vecx)} \geq \eps_4/2$. Our goal is to show that for every $\vecx \in L$, $\eps_5^{(\vecx)} \geq \frac{\eps_4^{(\vecx)}}{16n^2}$. By definition, we have that $\eps_5^{(\vecx)} \geq \eps_4^{(\vecx)}-\delta^{(\vecx)}$, so if $\delta^{(\vecx)} \leq \eps_4^{(\vecx)}/2$, then we are done. Thus, we assume that 
            \[ \delta^{(\vecx)} > \frac{\eps_4^{(\vecx)}}{2} \geq \frac{\eps_4}{4|\calX|} \geq \Omega\left(\frac{\eps}{2^w}\right). \]
        Let $t := n-r-s \leq n-1$ and $m := \frac{n(t+1)}{\delta^{(\vecx)}} \leq O\left(\frac{n^2 \cdot 2^w}{\eps}\right)$. By Assumption (2), we have that
            \[ \frac{t}{\delta^{(\vecx)}} \cdot \left(2^{t-s}+2^{-t+\ell}\right) = O\left(\frac{2^w \cdot t}{\eps}\right) \cdot \left(2^{t-s}+2^{-t+\ell}\right) \leq o(1), \]
        and by Assumption (3),
            \[ \frac{m \cdot \poly(n) \cdot T_{\calA}}{f(n-r-s-\ell)} = \frac{2^w \cdot \poly(n) \cdot T_{\calA}/\eps}{f(n-r-s-\ell)} \leq o(1), \]
        so the conditions of Lemma \ref{lemma:crypto_anti_concentration} are fulfilled. Thus, it follows from Lemma \ref{lemma:crypto_anti_concentration} and the security of the iO and puncturable PRF that for $\vecx \in L$,
            \[ \eps_5^{(\vecx)} \geq \frac{\delta^{(\vecx)}}{16n(t+1)}-O\left(\frac{1}{f(\kappa)}\right) \geq \frac{\delta^{(\vecx)}}{16n^2}-O\left(\frac{1}{f(\kappa)}\right) \geq \frac{\eps_4^{(\vecx)}}{32n^2}-O\left(\frac{1}{f(\kappa)}\right). \]
        Therefore, after incurring some loss from the assumption that $|\calX| \leq 2^w$, we have that by Assumption (3), the success probability of $\calA$ in this hybrid is
        \begin{align*}
            \eps_5 &= \sum_{\vecx \in \calX} \eps_5^{(\vecx)}-\negl(\secp) \geq \left(\sum_{\vecx \in L} \eps_5^{(\vecx)}\right)-\negl(\secp) \\
            &\geq \sum_{\vecx \in L} \left[\frac{\eps_4^{(\vecx)}}{32n^2}-O\left(\frac{1}{f(\kappa)}\right)\right]-\negl(\secp) \geq \frac{\eps_4/2}{32n^2}-O\left(\frac{|\calX|}{f(\kappa)}\right)-\negl(\secp) \\
            &\geq \frac{7\eps}{512n^2}-O\left(\frac{2^w}{f(\kappa)}\right)-\negl(\secp) \geq \frac{3\eps}{256n^2}.
        \end{align*}
        \item \textbf{$\mathsf{Hyb}_6$: Using the security of the P-PRP.}
        
        Sample all components except $\kin$ for the initial permutation, thus defining subspaces $S_{\vecx, \ell+1} \subseteq \ldots \subseteq S_{\vecx, 1}$ and $T_{\vecx, \ell+1} \subseteq \ldots \subseteq T_{\vecx, 1}$.
        
        Now consider the following permutation $\Gamma$ over $\{0, 1\}^n$. Denote by $h_0$ the first $r$ bits and $h_1$ the last $n-r$ bits of the input to $\Gamma$. Recall that in $P$, $P^{-1}$, and $D$, the value of $\vecx$ (and more specifically the matrices $\matM_\vecx$ and $\matM'_\vecx$ which are used to compute $\matC_\vecx$) can be written as an efficient function of $H(x) \in \{0, 1\}^r$. Thus, we define $\Gamma$ as the (efficient) permutation which takes $h_0$, computes $\matM_{h_0}$, $\matM'_{h_0}$, and $\matC_{h_0}$, interprets $h_1$ as a vector in $\bbZ_2^{n-r}$, and multiplies $h_1$ by $\matC_{h_0}$. Note that this means we will now use the PRF key $k_T$ in $P$ and $P^{-1}$ as well, because it is needed to compute $\matM_{h_0}$ and $\matM'_{h_0}$ and hence to compute $\Gamma$.

        For every value $h_0 \in \{0, 1\}^r$, multiplying by $\matC_{h_0}$ is an affine transformation, which is $(2^{n-r} \cdot (n-r)^2, \poly(n-r))$-decomposable by Lemma \ref{lemma:decomposable_perms}. Thus, $\Gamma$ applies a controlled decomposable permutation and is itself a $(2^r \cdot 2^{n-r} \cdot (n-r)^2, \poly(n-r) + \poly\log(2^n)) = (2^n \cdot n^2, \poly(n)) = (2^{\poly(\secp)}, \poly(\secp)) = (2^{\poly(\kappa)}, \poly(\kappa))$-decomposable permutation by Lemma \ref{lemma:decomposable_perms}. 

        Thus, if we switch to applying $\Gamma$ to the output of $\Pi$ (and $\Gamma^{-1}$ to the input of $\Pi^{-1}$), since the circuits $P$ and $P^{-1}$ which use $\Pi$ and $\Pi^{-1}$ are obfuscated, this change is $(f(\kappa), O(\frac{1}{f(\kappa)}))$-computationally indistinguishable by Lemma \ref{lemma:PRPs}.
        \item \textbf{$\mathsf{Hyb}_7$: Using the security of outside iO.}
        
        For every $h \in \bbZ_2^r$, we make the following change to $P$, $P^{-1}$, and $D$. Instead of composing the permutation $\Gamma$ to the output of $\Pi$ (by using $\kin^{\Gamma, 1}$), we switch to using just $\kin$ which applies $\Pi$ as-is. However, after generating $(\matA_\vecx, \vecb_\vecx)$, we will compute $\matM_\vecx$, $\matM'_\vecx$, and $\matC_\vecx$ to get $\overline{\matA}_\vecx := \matA_\vecx \cdot \matC_\vecx$. $P$ and $P^{-1}$ will now use $(\overline{\matA}_\vecx, \vecb_\vecx)$ instead of $(\matA_\vecx, \vecb_\vecx)$. Additionally, in $D$, instead of obfuscating the membership check circuits for $T_{\vecx, i}^{\perp}$, we go back to using $\overline{\matA}_\vecx$ to directly check for membership.
        
        Since $P$, $P^{-1}$, and $D$ remain functionally equivalent, by the security of the outer iO scheme that obfuscates $P$, $P^{-1}$, and $D$, this change is $(f(\kappa), O(\frac{1}{f(\kappa)}))$-computationally indistinguishable from $\mathsf{Hyb}_6$. Thus, by Assumption (1), the success probability of $\calA$ in this hybrid is
            \[ \eps_7 \geq \eps_6 - O\left(\frac{1}{f(\kappa)}\right) \geq \eps_5 - O\left(\frac{1}{f(\kappa)}\right) \geq \frac{\eps}{128n^2}. \]
        \item \textbf{$\mathsf{Hyb}_8$: Using the security of the puncturable PRF.}
        
        Recall the process in $\mathsf{Hyb}_7$ for generating our cosets in all three circuits $P$, $P^{-1}$, and $D$: given $\vecy$ we compute $\vecx$ with the lossy function, and then apply a puncturable PRF with two i.i.d keys, $\klin$ and $k_T$ to obtain $(\matA_\vecx, \vecb_\vecx)$ and the matrix $\matC_\vecx$. We then compute $\overline{\matA}_\vecx := \matA_\vecx \cdot \matC_\vecx$ and use $(\overline{\matA}_\vecx, \vecb_\vecx)$ in all circuits. Our goal is to apply Lemma \ref{lemma:obfuscating_samplers}, so we will describe two distributions and two puncturable PRFs.

        The first distribution $\calD_2$ takes as input $\vecx$ and randomness $(r_1, r_2)$, uses $r_1$ to generate $(\matA_\vecx, \vecb_\vecx)$ and $r_2$ to generate $\matC_\vecx$, computes $\overline{\matA}_\vecx := \matA_\vecx \cdot \matC_\vecx$, and outputs $(\overline{\matA}_\vecx, \vecb_\vecx)$. The second distribution $\calD_3$ takes as input $\vecx$ and randomness $r_1$ and uses $r_1$ to generate $(\matA_\vecx, \vecb_\vecx)$. By an identical argument as in the proof of Lemma \ref{lemma:bloating_dual}, it is easy to see that $\calD_2$ and $\calD_3$ are statistically equivalent. 

        Consider the function $\sfF'$ which takes as input $\vecx$ and outputs $(r_1, r_2) := (\sfF(k_1, \vecx), \sfF(k_2, \vecx))$ for randomly sampled keys $k_1, k_2$. We claim that $\sfF'$ is a puncturable PRF whose keys are simply a pair of keys for $\sfF$. Puncturing $\sfF'$ is simple: $\Punc'(k = (k_1, k_2), S) = (\Punc(k_1, S), \Punc(k_2, S))$. The correctness of $\sfF'$ follows directly from the correctness of $\sfF$, and it is not hard to see that $\sfF'$ has $(f(\kappa)-\poly(\kappa), O(\frac{1}{f(\kappa)}))$-puncturing security by the security of $\sfF$.

        As $|\calX| \leq 2^w$ with probability $1-\negl(\secp)$, in what follows, we assume this is the case. We now define the circuits $E_2(\vecx) = \calD_2(\vecx; \sfF'((\klin, k_T), \vecx))$ and $E_3(\vecx) = \calD_3(\vecx; \sfF(\klin', \vecx))$ for freshly sampled keys $\klin, k_T, \klin'$. Observing that all three circuits in $\mathsf{Hyb}_7$ use $E_2$ black-box, we see that Lemma \ref{lemma:obfuscating_samplers} implies that switching to $E_3$ in all three circuits is $(f(\kappa)-\poly(\kappa), O(\frac{2^w}{f(\kappa)}))$-computationally indistinguishable. 
        
        Therefore, after incurring some loss from the assumption that $|\calX| \leq 2^w$, we have that by Assumptions (1) and (3), the success probability of $\calA$ in $\mathsf{Hyb}_8$ is
            \[ \eps_8 \geq \eps_7 - O\left(\frac{2^w}{f(\kappa)}\right)-\negl(\secp) \geq \frac{\eps}{128n^2}-\frac{\eps}{512n^2} = \frac{3\eps}{512n^2}. \]
        \item \textbf{$\mathsf{Hyb}_9$: Switching back to injective mode of the lossy function.}
        
        We make the same change from $\mathsf{Hyb}_1$ to $\mathsf{Hyb}_2$, but in the other direction. Recall that in the previous hybrid $\mathsf{Hyb}_8$, on input $\vecy$, we compute $\vecx \gets \LFF(\pk_{\LF}, \vecy)$ and use $\vecx$ to compute all other components. We now sample an injective key $\pk'_{\LF} \gets \LFKeyGen(1^{\kappa}, 0, 1^w)$ instead of the lossy key $\pk_{\LF}$. By the security of the lossy function, the output of this hybrid is $(f(\kappa), \frac{1}{f(\kappa)})$-computationally indistinguishable from $\mathsf{Hyb}_8$. Thus, Assumption (1) implies that the success probability of $\calA$ in this hybrid is
            \[ \eps_9 \geq \eps_8 - \frac{1}{f(\kappa)} \geq \frac{\eps}{256n^2}. \]
        \item \textbf{$\mathsf{Hyb}_{10}$: Stopping use of the lossy function.}
        
        We make the same change from $\mathsf{Hyb}_0$ to $\mathsf{Hyb}_1$, but in the other direction. Instead of applying $\LFF$ to $\vecy || 0^{\kappa-d}$ to get $\vecx$ and then applying $\sfF$ to $\vecx$ to get randomness, we directly apply $\sfF$ (with a different key) to $\vecy$ to get our randomness. By an identical argument as before, this hybrid is $(f(\kappa)-\poly(\kappa), O(2^r/f(\kappa)))$-computationally indistinguishable from $\mathsf{Hyb}_9$, so by Assumptions (1) and (3), the success probability of $\calA$ in this hybrid is
            \[ \eps_{10} \geq \eps_9 - O\left(\frac{2^r}{f(\kappa)}\right)-\negl(\secp) \geq \frac{\eps}{512n^2}. \]
    \end{itemize}
    We conclude by noting that $\mathsf{Hyb}_{10}$ is exactly the distribution $(\calP, \calP^{-1}, \calD') \gets \widetilde{\Setup}(1^{\secp}, n, r, \ell, s)$. 
\end{proof}

\begin{proof}[Proof of Lemma \ref{lemma:simulating_dual_crypto}]
    Given $\overline{\calP}$ and $\overline{\calP}^{-1}$ from $(\overline{\calP}, \overline{\calP}^{-1}, \overline{\calD}) \gets \widetilde{\Setup}(1^{\secp}, r+s, r, 0, 0)$ and the code of $\calA$, $\calB$ acts as follows:
    \begin{itemize}
        \item Sample a puncturable PRF key $k_{\sfF}$ that outputs a sufficient (polynomial in $\kappa$) number of bits and a P-PRP key $k_{\Gamma}$ for a P-PRP on $\{0, 1\}^n$. On input $\vecy \in \{0, 1\}^d$, use the pseudorandomness of $\sfF(k_{\sfF}, \vecy)$ to generate a vector $\vecd_\vecy \in \bbZ_2^n$, matrix $\matC_{\vecy, 1} \in \bbZ_2^{(n-\ell) \times \ell}$, and full-rank matrix $\matC_{\vecy, 2} \in \bbZ_2^{(n-\ell) \times (n-\ell)}$. Denote by $\matC_{\vecy}$ the full-rank matrix $\begin{bmatrix} \matI_\ell & \matZero \\ \matC_{\vecy, 1} & \matC_{\vecy, 2} \end{bmatrix} \in \bbZ_2^{n \times n}$; we write as shorthand $(\matC_{\vecy}, \vecd_\vecy) \gets \sfF(k_{\sfF}, \vecy)$ to refer to this process. This defines the following circuits:
        \item $(\vecy \in \bbZ_2^d, \vecu \in \bbZ_2^n) \gets P(x \in \bbZ_2^n)$: 
        \begin{itemize}
            \item $(\overline{x} \in \{0, 1\}^{r+s}, \tilde{x} \in \bbZ_2^{n-r-s-\ell}, x' \in \bbZ_2^\ell) \gets \Pi(k_{\Gamma}, x)$.
            \item $(\vecy \in \bbZ_2^d, \overline{\vecu} \in \bbZ_2^{r+s}) \gets \overline{\calP}(\overline{x})$.
            \item $(\matC_\vecy \in \bbZ_2^{n \times n}, \vecd_\vecy \in \bbZ_2^n) \gets \sfF(k_{\sfF}, \vecy)$.
            \item $\vecu \gets \matC_\vecy \cdot \begin{pmatrix} x' \\ \overline{\vecu} \\ \tilde{x} \end{pmatrix} + \vecd_\vecy$.
        \end{itemize}
        \item $(x \in \bbZ_2^n \cup \{\bot\}) \gets P^{-1}(\vecy \in \bbZ_2^d, \vecu \in \bbZ_2^n)$:
        \begin{itemize}
            \item $(\matC_\vecy, \vecd_\vecy) \gets \sfF(k_{\sfF}, \vecy)$.
            \item $\begin{pmatrix} x' \\ \overline{\vecu} \\ \tilde{x} \end{pmatrix} \gets \matC_\vecy^{-1} \cdot (\vecu-\vecd_\vecy)$.
            \item $(\overline{x} \in \bbZ_2^{r+s} \cup \{\bot\}) \gets \overline{\calP}^{-1}(\vecy, \overline{\vecu})$
            \item Output $x \gets \Pi^{-1}(k_{\Gamma}, (\overline{x}, \tilde{x}, x'))$ if $\overline{x} \neq \bot$, else output $\bot$.
        \end{itemize}
        \item $\{0, 1\} \gets D'(j \in [\ell+1], \vecy \in \bbZ_2^d, \vecv \in \bbZ_2^n)$:
        \begin{itemize}
            \item $(\matC_\vecy, \vecd_\vecy) \gets \sfF(k_{\sfF}, \vecy)$.
            \item Output 1 iff $\vecv^T \cdot \begin{bmatrix}
                \matC_\vecy^{[j:\ell]} & \matC_\vecy^{[\ell+s+1:n-r]}
            \end{bmatrix} = 0^{n-r-s-j+1}$.
        \end{itemize}
        \item Generate obfuscations $\calP \gets \IO(1^{\kappa}, 1^p, P)$, $\calP^{-1} \gets \IO(1^{\kappa}, 1^p, P^{-1})$, $\calD' \gets \IO(1^{\kappa}, 1^p, D')$.
    \end{itemize}
    Finally, $\calB$ runs $(x_0, x_1) \gets \calA(\calP, \calP^{-1}, \calD')$, $(\overline{x}_b, \tilde{x}_b, x'_b) \gets \Pi(k_{\Gamma}, x_b)$ and outputs $(\overline{x}_0, \overline{x}_1)$. 

    Define 
    \begin{align*}
        \matA_\vecy &:= \matC_\vecy \cdot \begin{bmatrix} \matI_{\ell} & \matZero & \matZero \\ \matZero & \overline{\matA}_\vecy & \matZero \\ \matZero & \matZero & \matI_{n-r-s-\ell} \end{bmatrix} = \begin{bmatrix} \matI_{\ell} & \matZero \\ \matC_{\vecy, 1} & \matC_{\vecy, 2} \end{bmatrix} \cdot \begin{bmatrix} \matI_{\ell} & \matZero & \matZero \\ \matZero & \overline{\matA}_\vecy & \matZero \\ \matZero & \matZero & \matI_{n-r-s-\ell} \end{bmatrix} \\
        &= \begin{bmatrix} \matI_{\ell} & \matZero \\ \matC_{\vecy, 1} & \matC_{\vecy, 2} \cdot \begin{bmatrix} \overline{\matA}_\vecy & \matZero \\ \matZero & \matI_{n-r-s-\ell} \end{bmatrix} \end{bmatrix} \in \bbZ_2^{n \times (n-r)},
    \end{align*}
    where $\overline{\matA}_\vecy \in \bbZ_2^{(r+s) \times s}$ is the matrix arising from $\overline{\calP}$ and $\overline{\calP}^{-1}$. As $\matC_{\vecy, 2}$ and $\overline{\matA}_\vecy$ have full column-rank, so does $\matA_\vecy^{[\ell+1:n-r]}$; on the other hand, $\matC_{\vecy, 1}$ remains a random matrix; thus, $\matA_\vecy$ has full column-rank. Now, since $(\vecu_0 - \vecu_1) \notin \ColSpan\left(\begin{bmatrix}\matA_\vecy^{[1:\ell]} & \matA_\vecy^{[\ell+s+1:n-r]}\end{bmatrix}\right)$, this means that if we look at each of the two coordinates vectors $\vecw_0$, $\vecw_1$ (which each have $n-r$ bits) of $\vecu_0$ and $\vecu_1$, they differ somewhere between the $(\ell+1)$'th and $(\ell+s)$'th bit. However, $(\vecw_b)_{[\ell+1:\ell+s]}$ arises from applying the permutation $\overline{\Pi}$ on $\{0, 1\}^{r+s}$ (defined by $\calP, \calP^{-1}$ to $\overline{x}_b$) and taking the last $s$ bits of the output. As $x_0, x_1$ have the same $\vecy$'s and thus so do $\overline{x}_0$ and $\overline{x}_1$, it must be the case that $\overline{x}_0 \neq \overline{x}_1$ in order for $\vecw_0$ to differ from $\vecw_1$ in the appropriate indices.

    If we define $\eps_{\calB}$ as the probability that $\calA$ outputs $x_0, x_1$ such that $y_0 = y_1 := \vecy$ and $(\vecu_0 - \vecu_1) \notin \ColSpan\left(\begin{bmatrix}\matA_\vecy^{[1:\ell]} & \matA_\vecy^{[\ell+s+1:n-r]}\end{bmatrix}\right)$, we wish to lower bound the probability of $\eps_{\calB}$. We do this through a series of hybrids.
    \begin{itemize}
        \item \textbf{$\mathsf{Hyb}_0$: The distribution $(\calP, \calP^{-1}, \calD') \gets \calB(\overline{\calP}, \overline{\calP^{-1}})$, as simulated by $\calB$.}
        
        This is the original execution, and thus has success probability $\eps_{\calB}$.
        \item \textbf{$\mathsf{Hyb}_1$: Removing the inner obfuscation.}
        
        In this hybrid, we remove the obfuscations from $\overline{\calP}$ and $\overline{\calP}^{-1}$ and use the underlying plain circuits $\overline{P}$ and $\overline{P}^{-1}$. By the correctness of the inner iO scheme, we have not changed the functionality of the final circuits $P$, $P^{-1}$, $D$, so the security of the outer iO scheme guarantees that this change is $(f(\kappa), O(\frac{1}{f(\kappa)}))$-computationally indistinguishable.
        \item \textbf{$\mathsf{Hyb}_2$: Removing the inner permutation $\overline{\Pi}_{\mathsf{in}}$ by using the security of the P-PRP.}
        
        Let $\overline{\Pi}_{\mathsf{in}}$ be the first P-PRP inside $\overline{P}$. In the previous hybrid, we apply the $n$-bit P-PRP $\Pi(k_{\Gamma}, \cdot)$ to $x$ before applying the inner permutation $\overline{\Pi}_{\mathsf{in}}(\overline{k}_{\mathsf{in}}, \cdot)$ to the first $r+s$ output bits of $\Pi(k_{\Gamma}, x)$. The inner permutation can also be thought of as a permutation $\Gamma'$ on $\{0, 1\}^n$ which acts as the identity on the last $n-r-s$ bits and applies $\overline{\Pi}_{\mathsf{in}}(\overline{k}_{\mathsf{in}}, \cdot)$ to the first $r+s$ bits. 
        
        By Theorem \ref{theorem:OP_PRPs}, the inner P-PRP $\overline{\Pi}_{\mathsf{in}}(\overline{k}_{\mathsf{in}}, \cdot)$ is $(2^{\poly(\kappa)}, \poly(\kappa))$-decomposable. This implies that $\Gamma'$ is also $(2^{\poly(\kappa)}, \poly(\kappa))$-decomposable.

        Since the circuits $P$ and $P^{-1}$ which apply $\Pi$ and $\Pi^{-1}$ are obfuscated by the outer iO scheme, if we stop applying $\Gamma'$ to the output of $\Pi$ (and $(\Gamma')^{-1}$ to the input of $\Pi$), this change is $(f(\kappa), O(\frac{1}{f(\kappa)}))$-computationally indistinguishable by Lemma \ref{lemma:PRPs}.
        \item \textbf{$\mathsf{Hyb}_3$: Taking $\matA_\vecy$ to be the direct output of the PRF $\sfF$.}
        
        In the previous hybrid, we computed $\matA_\vecy := \matC_\vecy \cdot \begin{bmatrix} \matI & \matZero & \matZero \\ \matZero & \overline{\matA}_\vecy & \matZero \\ \matZero & \matZero & \matI_{n-r-s} \end{bmatrix}$ and $\vecb_\vecy := \matC_\vecy \cdot \begin{pmatrix} \matZero \\ \overline{\vecb}_\vecy \\ \matZero \end{pmatrix} + \vecd_\vecy$, where $\overline{\matA}_\vecy, \overline{\vecb}_\vecy$ are defined by the output of the inner PRF $\overline{\sfF}(\overline{\klin}, \vecy)$ and $\matC_\vecy, \vecd_\vecy$ are computed from the output of $\sfF(k_{\sfF}, \vecy)$.

        Our goal will be to apply Lemma \ref{lemma:obfuscating_samplers}, so we begin by describing two distributions and two puncturable PRFs.

        The first distribution $\calD_0$ takes as input $\vecy$ and randomness $(r_1, r_2)$, uses $r_1$ to generate $(\overline{\matA}_\vecy, \overline{\vecb}_\vecy)$ and $r_2$ to generate $(\matC_\vecy, \vecd_\vecy)$, computes $\matA_\vecy := \matC_\vecy \cdot \begin{bmatrix} \matI & \matZero & \matZero \\ \matZero & \overline{\matA}_\vecy & \matZero \\ \matZero & \matZero & \matI_{n-r-s} \end{bmatrix}$ and $\vecb_\vecy := \matC_\vecy \cdot \begin{pmatrix} \matZero \\ \overline{\vecb}_\vecy \\ \matZero \end{pmatrix} + \vecd_\vecy$, and outputs $(\matA_\vecy, \vecb_\vecy)$. The second distribution $\calD_1$ takes as input $\vecy$ and randomness $r_1$ and uses $r_1$ to generate $(\matA_\vecy, \vecb_\vecy)$. $\calD_2$ and $\calD_3$ are statistically equivalent by the fact that $\matC_{\vecy, 1}$ is a random matrix, $\matC_{\vecy, 2}$ and $\overline{\matA}_\vecy$ are random full column-rank matrices, and $\vecd_\vecy$ is a random vector. 

        Now consider the function $\sfF'$ which takes as input $\vecy$ and outputs $(r_1, r_2) := (\overline{\sfF}(k_1, \vecy), \sfF(k_2, \vecy))$ for randomly sampled keys $k_1, k_2$. By an essentially identical argument as in the proof of Lemma \ref{lemma:bloating_dual_crypto}, we have that $\sfF'$ is a puncturable PRF.

        We now define the circuits $E_0(\vecy) = \calD_0(\vecy; \sfF'((\overline{\klin}, k_{\sfF}), \vecy))$ and $E_1(\vecy) = \calD_1(\vecy; \sfF(k'_{\sfF}, \vecy))$ for freshly sampled keys $\overline{\klin}, k_{\sfF}, k'_{\sfF}$. 
        
        Observing that the circuit $E_0$ is only used black-box in $P$, $P^{-1}$, and $D$, Lemma \ref{lemma:obfuscating_samplers} implies that switching to $E_1$ is $(f(\kappa)-\poly(\kappa), O(\frac{2^r}{f(\kappa)}))$-computationally indistinguishable. 

        \item \textbf{$\mathsf{Hyb}_4$: Moving the second permutation outside $\overline{P}$ and $\overline{P}^{-1}$.}
        
        Recall that in the previous hybrid, the second inner permutation $\overline{\Pi}^{-1}(\kout, \cdot)$, which acts on $\{0, 1\}^d$ to generate $\vecy$, is used in $\overline{P}$ and $\overline{P}^{-1}$. We simply move the application of the permutation outside of $\overline{P}$ and $\overline{P}^{-1}$. This does not change the functionality of $P$, $P^{-1}$, or $D$, so this change is $(f(\kappa), O(\frac{1}{f(\kappa)}))$-computationally indistinguishable by the security of the outer iO scheme.
        \item \textbf{$\mathsf{Hyb}_5$: Dropping $\overline{P}$ and $\overline{P}^{-1}$.} 
        
        Having moved the permutation $\overline{\Pi}^{-1}(\kout, \cdot)$ to the outside of $\overline{P}$ and $\overline{P}^{-1}$, we observe that $P$, $P^{-1}$, and $D$ never use the output of $\overline{P}$ and $\overline{P}^{-1}$, so we can simply generate $P$, $P^{-1}$, and $D$ without using $\overline{P}$ and $\overline{P}^{-1}$ at all. This does not change the functionality of $P$, $P^{-1}$, or $D$, so this change is $(f(\kappa), O(\frac{1}{f(\kappa)}))$-computationally indistinguishable by the security of the outer iO scheme.
    \end{itemize}
    It is easy to see that the final distribution in $\mathsf{Hyb}_5$ is precisely the output distribution of $\widetilde{\Setup}(1^{\secp}, n, r, \ell, s)$, and the outputs of $\mathsf{Hyb}_0$ and $\mathsf{Hyb}_5$ are $(f(\kappa)-\poly(\kappa), O(\frac{2^r}{f(\kappa)}))$-computationally indistinguishable. By hypothesis, we have that the success probability of $\calA$ in the final hybrid is $\eps$, and thus we have by the lemma's assumptions that $\eps_B \geq \eps - O\left(\frac{2^r}{f(\kappa)}\right) \geq \eps/2$, as desired.
\end{proof}

\section{Improved Analysis of the \texorpdfstring{\cite{SZ25}}{} Distinguisher}\label{appendix:more_details}

\newcommand{\aux}{\mathsf{aux}}
\newcommand{\Measure}{\mathsf{Measure}}
In this section, we show that the distinguisher described in \cite{SZ25} does in fact also succeed in distinguishing between the case where the preimage state has not been measured at all and the case that (only) the first bit of the preimage (or indeed, any given bit) has been measured. Recall that \cite{SZ25} showed the weaker statement that the distinguisher succeeds in distinguishing between the case where the preimage state has not been measured at all and the case that {\em all the bits} of the preimage have been measured.

We first recall the construction of \cite{SZ25}.
\construction{One-Shot Signature Scheme, \cite{SZ25}}{OriginalOSS}{
    Let $\secp \in \bbN$ be the statistical security parameter. Define $s := 16 \cdot \secp$, $r := s \cdot (\secp-1)$, and $n := r+\frac{3}{2} \cdot s$.
    
    Let $\Pi: \{0, 1\}^n \to \{0, 1\}^n$ be a random permutation and let $F : \{0, 1\}^r \to \{0, 1\}^{n \cdot (n-r+1)}$ a random function. Let $H(x)$ denote the first $r$ output bits of $\Pi(x)$, and $J(x)$ denote the remaining $n-r$ bits, which are interpreted as a vector in $\bbZ_2^{n-r}$. For each $\vecy \in \{0, 1\}^r$, let $\matA_\vecy \in \bbZ_2^{n \times (n-r)}$ be a random matrix with full column-rank, and $\vecb_\vecy \in \bbZ_2^n$ be uniformly random, both are generated by the output randomness of $F(\vecy)$.
    
    Then, we let $\calP: \{0, 1\}^n \to (\{0, 1\}^r \times \bbZ_2^n)$, $\calP^{-1}: (\{0, 1\}^r \times \bbZ_2^n) \to \{0, 1\}^n \cup \{ \bot \}$, $\calD: (\{0, 1\}^r \times \bbZ_2^n) \to \{0, 1\}$ be the following oracles:
    \begin{itemize}
        \item $\calP(x) = (\vecy, \matA_\vecy \cdot J(x) + \vecb_\vecy)$ where $\vecy = H(x)$
        \item $\calP^{-1}(\vecy, \vecu) = \begin{cases} \Pi^{-1}(\vecy, \vecz) & \exists \vecz \in \bbZ_2^{n-r} \text{ such that } \matA_\vecy \cdot \vecz + \vecb_\vecy = \vecu \\ \bot & \text{otherwise} \end{cases}$
        \item $\calD(\vecy, \vecv) = \begin{cases} 1 & \text{if } \vecv^{T} \cdot \matA_\vecy = 0^{n-r} \\ 0 & \text{otherwise} \end{cases}$
    \end{itemize}
    We denote the above distribution of oracles by $\calO_{n, r}$. We define our hash function as $H$, which can be easily computed by querying $\calP$, considering only the first $r$ bits and discarding the second output.
}

\begin{lemma}
For a given security parameter $\secp$, let $\calO_{n, r} := \calO_{n(\secp), r(\secp)}$ be the oracle distribution of Construction \ref{construction:OriginalOSS}.

There exists a pair of $\QPT$ algorithms $(\calS, \calD)$ such that for all $\secp \in \bbN$,
\renewcommand{\arraystretch}{0.75}
\begin{align*}
    {\scriptstyle \abs{\Pr[{\scriptstyle \calD(\ket*{\psi_{y, b}}\!, \aux) = 1:} \begin{array}{r} {\scriptstyle (\calP, \calP^{-1}, \calD) \gets \calO_{n, r}} \\ {\scriptstyle (\ket*{\psi}, \aux) \gets \calS^{(\calP, \calP^{-1}, \calD)}} \\ {\scriptstyle \ket*{\psi_{y, b}} \gets \Measure_{H, 1}(\ket*{\psi})} \end{array}] -
    \Pr[{\scriptstyle \calD(\ket*{\psi_y}\!, \aux) = 1:} \begin{array}{r} {\scriptstyle (\calP, \calP^{-1}, \calD) \gets \calO_{n, r}} \\ {\scriptstyle (\ket*{\psi}, \aux) \gets \calS^{(\calP, \calP^{-1}, \calD)}} \\ {\scriptstyle \ket*{\psi_y} \gets \Measure_{H}}(\ket*{\psi}) \end{array}]}} \geq \frac{1}{4}.
\end{align*}
Here, $\ket*{\psi_{y, b}} \gets \Measure_{H, 1}(\ket*{\psi})$ means to measure the first bit of $\ket*{\psi}$ in the computational basis after measuring $H^{(\calP, \calP^{-1}, \calD)}(\cdot)$, resulting in the outcome $(b, y)$, while $\ket*{\psi_y} \gets \Measure_{H}$ means to measure $H^{(\calP, \calP^{-1}, \calD)}(\cdot)$, resulting in outcome $y$. Then the state $\ket*{\psi}$ collapses to $\ket*{\psi_y}$, which contains some superposition of preimages of $y$.
\end{lemma}
\begin{proof}
    $\calS_H$ computes a uniform superposition $\ket{+}^{\otimes n}$ and outputs it as $\ket{\psi}$ with no auxiliary information. 

    $\calD_H$, given an unknown $n$-qubit state $\ket{\phi} := \sum_x \alpha_x \ket{x}$, acts as follows:
    \begin{enumerate}
        \item Compute $\calP$ to obtain the state
            \[ \sum_x \alpha_x \ket{x} \ket{\vecy_x} \ket*{\matA_\vecy \cdot \vecw_x + \vecb_\vecy}. \]
        \item Apply $\calP^{-1}$ to uncompute the input register holding $x$ to obtain the state
            \[ \sum_x \alpha_x \ket{\vecy_x} \ket*{\matA_\vecy \cdot \vecw_x + \vecb_\vecy}.\]
        \item Apply $H^{\otimes n}$ to the rightmost register (which contains coset vectors) to obtain the state
            \[ \sum_x \alpha_x \ket{\vecy_x} (H^{\otimes n} \ket*{\matA_\vecy \cdot \vecw_x + \vecb_\vecy}). \]
        \item Apply $\calD(\cdot, \cdot)$ before measuring and returning the output bit register.
    \end{enumerate}
    The case where just $H$ is measured is the same as in \cite{SZ25}: $\calD_H$ receives the state $\sum_{x:H(x)=y} \ket{x}$ for some $y$, and after the second step $\calD_H$ holds the state
        \[ \ket{y} \sum_{x: H(x) = y} \ket*{\matA_\vecy \cdot \vecw_x + \vecb_\vecy}. \]
    Since applying the QFT to a uniform superposition over a coset yields a uniform superposition over the dual subspace (with a phase), it follows that $\calD_H$ outputs 1 with probability 1.

    In the case where both $H$ and the first preimage bit is measured, $\calD_H$ receives the state $\sum_{x:H(x)=y, x_1 = b} \ket{x}$ for some hash $y$ and bit $b$. Define the set $S_{y, b} = \{\matA_\vecy \cdot \vecw_x + \vecb_\vecy\}_{\vecw_x: x_1 = b} \subseteq \{\matA_\vecy \cdot \vecw_x + \vecb_\vecy\}_{\vecw_x \in \{0, 1\}^{n-r}}$. At the end of the second step, $\calD_H$ holds the state
        \[ \frac{1}{\sqrt{|S_{y, b}|}} \ket{y} \otimes \sum_{\vecs \in S_{y, b}} \ket*{\vecs}. \]
    After applying the QFT $H^{\otimes n}$, $\calD_H$ holds the state
        \[ \frac{1}{\sqrt{2^n \cdot |S_{y, b}|}} \ket{y} \otimes \sum_{\vecz \in \bbZ_2^n} \left(\sum_{\vecs \in S_{y, b}} (-1)^{\vecs \cdot \vecz}\right) \ket{\vecz}. \]
    Thus, the probability of step 4 outputting 1 is precisely
    \begin{align*}
        \frac{1}{2^n \cdot |S_{y, b}|} \sum_{\vecz \in \matA_\vecy^{\perp}} \left(\sum_{\vecs \in S_{y, b}} (-1)^{\vecs \cdot \vecz}\right)^2 &= \frac{1}{2^n \cdot |S_{y, b}|} \sum_{\vecz \in \matA_\vecy^{\perp}} \left(\sum_{\vecs \in S_{y, b}} (-1)^{\vecb_\vecy \cdot \vecz}\right)^2 \\
        &= \frac{1}{2^n \cdot |S_{y, b}|} \sum_{\vecz \in \matA_\vecy^{\perp}} |S_{y, b}|^2 = \frac{2^r \cdot |S_{y, b}|^2}{2^n \cdot |S_{y, b}|} = \frac{|S_{y, b}|}{2^{n-r}}.
    \end{align*}
    If we fix a permutation $\Pi$, we know that each hash value $y$ is equally likely to occur, and conditioned on $\Pi$ and $y$, the outcome $b$ will be measured with probability $\frac{|S_{y, b}|}{2^{n-r}}$. Thus, the probability that $\calD_H$ outputs 1 is precisely
    \begin{align*}
        \bbE_{\Pi, y}\left[\frac{|S_{y, 0}|^2}{2^{2(n-r)}}+\frac{|S_{y, 1}|^2}{2^{2(n-r)}}\right] &= \bbE_{\Pi, y}\left[\frac{|S_{y, 0}|^2+(2^{n-r}-|S_{y, 0}|)^2}{2^{2(n-r)}}\right] \\
        &= \bbE_{\Pi, y}\left[\frac{2|S_{y, 0}|^2+2^{2(n-r)}-2^{n-r+1}|S_{y, 0}|}{2^{2(n-r)}}\right] \\
        &= \bbE_{\Pi, y}\left[\frac{2|S_{y, 0}|^2}{2^{2(n-r)}}\right] + \frac{2^{2(n-r)}-2^{n-r+1} \cdot 2^{n-r-1}}{2^{2(n-r)}} \\
        &= 2\bbE_{\Pi, y}\left[\left(\frac{|S_{y, 0}|}{2^{n-r}}\right)^2\right] \\
        &= 2\left[\left(\frac{2^{n-r-1}}{2^{n-r}}\right)^2+\frac{2^{n-r}}{2^{2(n-r)}} \cdot \frac{2^{n-1}}{2^n} \cdot \frac{2^n-2^{n-1}}{2^n} \cdot \frac{2^n-2^{n-r}}{2^n-1}\right] \\
        &= \frac{1}{2}+\frac{1}{2^{n-r+1}} \cdot \frac{2^n-2^{n-r}}{2^n-1} \leq \frac{1}{2}+\frac{1}{2^{n-r+1}} \leq \frac{3}{4},
    \end{align*}
    since $|S_{y, 0}|$ is a hypergeometric variable.
    
    We conclude that the distinguishing advantage of $(\calS_H, \calD_H)$ is at least $1-\frac{3}{4} = \frac{1}{4}$, as desired.
\end{proof}
\end{document}